\documentclass{elsarticle}
\usepackage[normalem]{ulem}
\usepackage{amsmath}
\usepackage{enumerate}
\usepackage{amsfonts}
\usepackage{yfonts}
\usepackage{hyperref}
\usepackage{subfigure}
\usepackage{psfrag}

\usepackage{epsfig}
\usepackage[latin1]{inputenc}
\usepackage{float}
\usepackage{graphicx}
\usepackage{cancel}
\usepackage{mathrsfs}
\usepackage{amssymb}
\usepackage{amsfonts}
\usepackage{amsmath}
\usepackage{slashed}
\usepackage{graphicx}
\usepackage{bm}
\usepackage{color}
\newcommand{\be}{\begin{equation}}
\newcommand{\ee}{\end{equation}}
\newcommand{\bea}{\begin{eqnarray}}
\newcommand{\eea}{\end{eqnarray}}
\newcommand{\bwt}{\begin{widetext}}
\newcommand{\ewt}{\end{widetext}}

\newcommand{\nn}{\nonumber}

\newcommand{\bi}{\begin{itemize}}
\newcommand{\ei}{\end{itemize}}

\begin{document}
\begin{frontmatter}

\title {Thermalization of holographic Wilson loops  in spacetimes with spatial anisotropy}
\author[1]{D. S. Ageev}
\ead{ageev@mi.ras.ru}
\author[1]{I. Ya. Aref'eva}
\ead{arefeva@mi.ras.ru}
\author[2,3]{A. A. Golubtsova}
\ead{golubtsova@theor.jinr.ru}
\author[4]{E. Gourgoulhon}
\ead{eric.gourgoulhon@obspm.fr}

\address[1]{Steklov Mathematical Institute, Russian Academy of Sciences,
Gubkina str. 8, 119991, Moscow, Russia}
\address[2]{Bogoliubov Laboratory of Theoretical Physics, JINR, 141980,  Dubna, Moscow region, Russia}
\address[3]{Dubna State University, Universitetskaya str. 19, 141980, Dubna, Russia}
\address[4]{Laboratoire Univers et Th\'{e}ories, Observatoire de Paris,
CNRS, Universit\'e Paris Diderot, Universit\'e PSL,
5 place Jules Janssen, 92190 Meudon, France}

\begin{abstract}
In this paper, we study behaviour of Wilson loops in the boost-invariant  nonequilibrium anisotropic quark-gluon plasma produced in heavy-ion collisions
within the holographic approach. We describe the thermalization studying the evolution of the Vaidya metric in the boost-invariant and spatially anisotropic background. 
To probe the system during this process we calculate rectangular Wilson loops oriented in different spatial directions. 
We find that anisotropic effects  are more visible for the Wilson loops lying in the transversal plane
 unlike the Wilson loops with partially longitudinal orientation.  In particular, we observe that the Wilson loops can  thermalizes  first unlike to the order of the isotropic model. We see that Wilson loops on transversal contours have the shortest thermalization time.
We also calculate the string tension and the pseudopotential at different temperatures for the static quark-gluon plasma. We show that the pseudopotential related to the configuration on the transversal plane  has the screened Cornell form.
We also show that the jet-quenching parameter  related with the average of the light-like Wilson loop  exhibits  the dependence  on orientations.

\end{abstract}
\end{frontmatter}

\tableofcontents

\section{Introduction}\label{Sec:intro}

Wilson loops are known to play a key role as fundamental probes of gauge theories, in particular QCD.
Owing to Wilson loops one can define many important quantities, for instance,
 we can derive the potential of a quark-antiquark interaction from the expectation value of the space-time rectangular Wilson loop.
In the lattice QCD the Wilson loops are the prime observables and their expectation values are defined non-pertur\-batively \cite{Wilson:1974sk}. 
One can also determine expectation values of Wilson loops in the framework of perturbative QCD after suitable renormalization \cite{Arefeva:1980zd}.

At the same time,  expectation values of Wilson loops are used to characterize properties of the quark-gluon plasma produced in heavy ion collisions (HIC). 
With the help of Wilson loops one can perform the analysis of radiative parton energy loss, quarkonium suppression, jet quenching, etc.\cite{Solana}.

In this paper, we investigate the behaviour of Wilson loops in the nonequilibrium spatially anisotropic quark-gluon plasma (QGP) produced in heavy-ion collisions
in the framework of the holographic correspondence. It is widely appreciated that the holography, or  the gauge/gravity duality, provides a powerful tool for studying dynamics of the strong coupling system, in particular,  
the QGP formed in heavy-ion collisions \cite{Solana, DeWolf, IA}.  
 Main idea of this approach is using natural prescriptions of the generalized AdS/CFT correspondence 
 to recover non-perturbative QCD phenomena, particularly,  non-perturbative
 vacuum phenomena, finite temperature,  high-dense  and non-zero chemical potential phenomena.
 In this strategy,  fitting parameters are ones specifying the form of the 5-dimensional metric. 
 The 5-dimensional background
 is supposed to be a solution of Einstein equations with a suitable matter content, not necessary related with string theory.
 
According to the holographic approach,  creation of the quark-gluon plasma in HIC can be represented here as a collision of  shock waves in the 5-dimensional bulk, in which a black hole is formed \cite{GPY}-\cite{APP}.  After the collision the shocks slowly decay, leaving the plasma described by hydrodynamics. 
 The formation of the black hole can be also described by
 an infalling shell  \cite{Vad} propagating in the 5-dimensional bulk. 
A gravitational collapse of the thin shell to the black hole provides also a gravitational dual description of a more general class of thermalization processes \cite{1103.2683}-\cite{1211.6041}. Note, that the holographic approach  is convenient to incorporate anisotropic properties of the QGP created in heavy-ion collisions \cite{1312.2285,Giataganas:2013lga}.
 
The holographic approach has been  widely used to study Wilson loops in different settings. Expectation values of Wilson loops  within the gauge/gravity duality have been calculated for  the strongly coupled  ${\cal N}=4$ super Yang-Mills theory \cite{JMM2,9803135,9803137,Sonnenschein:1999if}.
The string dual description of the real QCD  is unknown in spite of a lot of performed effort to find it
\cite{0306018,Mateos,SS}.  However, suitable "bottom-up" holographic QCD models matching with experimental and lattice results have been proposed in \cite{HW,Karch2006,Andreev2006ct, White2007tu, GKMN,Pirner,HHY,GKMMN, Ageev}.  Various physical quantities, in particular, expectation values of Wilson loops, have been calculated holographically.   Wilson loops in static anisotropic backgrounds,  static non-relativistic background, Lifshitz backgrounds  and backgrounds with hyperscaling violation have been examined in   \cite{Giataganas:2013lga}, \cite{ALR}-\cite{1406.6389} and refs therein. As it is known following the holographic dictionary, gravity duals of light-like Wilson loops can be used for calculation of the jet quenching parameter \cite{LRW,LRW2} controlling the description of medium-induced energy loss for partons in QCD. Holographic evaluation of the jet-quenching parameter for anisotropic QGP was considered in \cite{1203.0561,Rebhan:2012bw, Giataganas:2013lga}.

The special feature of this paper is that we consider spatially anisotropic backgrounds,
  which have also boost invariance
  \begin{equation}\label{Lif-like0}
ds^{2} =\frac{R^2}{z^2}\left(-dt^{2} + dx^{2} + (\frac{aR}{z})^{2/\nu-2}(dy^2_1 + dy^2_2) +  d z^{2}\right).
\end{equation}
These backgrounds are characterized  by the scale parameter $R$ and anisotropic parameters $a$ and $\nu$.

 This spacetime is in accordance with the geometry of HIC, where one has a selected direction -- the axis of ions collisions. As has been shown before \cite{AG}, to fit the experimental form of the dependence of total multiplicity on energy, obtained at LHC,  one  can assume that the holographic model 
has  boost invariance and a spatial anisotropy controlled by the so-called critical exponent. More precisely, it has been shown in \cite{AG} that the critical exponent should be taken $\nu=4.45$.
 The metrics of the form \eqref{Lif-like0} are dual to  models with so called Lifshitz-like fixed points \cite{TAYLOR,ALT}. 
 These metrics are also occurs as the IR limit for the anisotropic background suggested for studies of QGP in \cite{MT2}.
It's also worth to be noted that the anizotropic background \eqref{Lif-like0} differs from the  Lifshitz metrics \cite{KLM} by the anisotropic scaling of spatial coordinates.
  The Vaidya shell in the background \eqref{Lif-like0} has been found  in \cite{AGG} and used to estimate 
the thermalization time of the 2-point correlators and the  holographic entanglement entropy in HIC
  \cite{IYaA,AGG}. The holographic model  with the shell in (\ref{Lif-like0}) was used to explore the behaviour of the quark-antiquark potential in \cite{Ali-Akbari:2017} with the method developed previously for in the AdS-Vaidya background \cite{Ali-Akbari:2015ooa}.
    
  As already mentioned, the information about processes during HIC can be read off from the expectation values of Wilson loops.
Thus, it is natural to study the behaviour of Wilson loops during the HIC within the same holographic model we used to fit the energy dependence of total multiplicity. In particular,  it is reasonable to investigate thermalization of Wilson loops  and their behaviour in the end of the thermalization process. By thermalization of Wilson loop we mean the thermalization of its expectation value.

In the present work we calculate holographically Wilson loops in the backgrounds  \eqref{Lif-like0} with the Vaidya shell and black brane. We consider spatial and light-like  configurations, which represent rectangles with two infinitely long sides and two sides of  finite lengths. By virtue of the  metric \eqref{Lif-like0} possessing a spatial ani\-so\-tropy, the expectation values of Wilson loops depend on the orientation of the corresponding configuration.
Further, we will study potentials and its evolution during the thermalization process. 
We show that the order of thermalization in our background is the following: first thermalizes two-point correlator, then Wilson loops and the last thermalization occurs for the entanglement entropy. Our anisotropy reduces the thermalization time for non-local observables as compared to the isotropic case. We also find an analytical representation for  the 
holographic  light-like  Wilson loops in the background \eqref{Lif-like0} and  derive the jet quenching parameter.

The paper is organized as the following. In Sec.~\ref{Sec:setup} we briefly remind the holographic description of Wilson loops, gravitational backgrounds and the notations. In  Sec.~\ref{Sect:WL-ind} we analyze the static Wilson loops as well as calculate pseudopotentials, string tensions for different orientations of Wilson loop and the jet quenching parameter. 
In Sec.~\ref{Sect:5.2} we study the nonequilibrium dynamics of the same oriented Wilson loops and present the results. In Sec.~\ref{Sect:6} the thermalization time of Wilson loops is estimated
and we compare it with the thermalization times of the entanglement entropy and two-point correlation functions. We conclude with a discussion of our results and further directions.

\section{Set Up}\label{Sec:setup}

\subsection{Wilson  loops}\label{Sect:WL}
In this work we consider rectangular Wilson loops.
As already noticed, Wilson loops contain the information about the force between quarks.
Following the holographic approach \cite{JMM2,9803135} the expectation value of the Wilson loop in the fundamental representation calculated on the gravity side reads as:
\begin{eqnarray}\label{7.1}
W[C]=\langle  {\mbox{Tr}}_F \,e^{i\oint_{C} dx_\mu A_\mu}\rangle  = e^{- S_{string}[C]},
\end{eqnarray}
 where $C$ in a contour on the boundary.   More precisely, we mean that  the contour is at the "regularized
 boundary" $z=z_0$. On the gauge side we suppose to deal with the Wilson loop in the anisotropic gauge theory, i.e. theory on the 4-dimensional flat anisotropic background,
 \be\label{4-anis}
 ds_4^2=-dt^{2} + dx^{2} + (\frac{aR}{z_0})^{2/\nu-2}(dy^2_1 + dy^2_2).
\ee
The regularized version of this theory corresponds to the gauge theory on the anisotropic lattice \cite{Arefeva:1993hi}, where spacing in the longitudinal and transversal directions, 
$a_\parallel$ and $a_\perp$,  are different so that $a_\parallel/a_{\perp}=(aR/z_0)^{1-1/\nu}$. One can develop  renormalizations in this theory and also the renormalization of the Wilson loops in the analogy with the isotropic theory.  $F$ means the fundamental representation (we will omit this symbol in what follows),  $S_{string}$ is the minimal action of the string  hanging from 
 the contour $C$ in the bulk. 
 The Nambu-Goto action  can be represented as
\be\label{7.1b}
S_{string} = \frac{1}{2\pi \alpha'}\int d\sigma^{1} d\sigma^{2}\sqrt{-\det (h_{\alpha \beta})},\ee
with $ h_{\alpha\beta} = g_{MN} \partial_{\alpha}X^{M}\partial_{\beta}X^{N}$.
In (\ref{7.1b})  $\sigma^{1}$, $\sigma^{2}$ parametrize the worldsheet, $g_{MN}$ is the background metric, $M, N =1,\ldots,5$ and
 $X^{M}=X^{M}(\sigma^{1}, \sigma^{2})$ specify the  string worldsheet.

 The pseudopotential of the interquark interaction can be extracted from the rectangular spatial Wilson loop of size \footnote{Note that we take $\ell<L$  for real calculations, where  $\ell$ is large, but not infinite.} $\ell\times L$,  for large $L$
 \be\label{SpWL}
 W(\ell,L)  = \langle {\mbox {Tr}}\, e^{i\oint_{\ell\times L} dx_\mu A_\mu} \rangle = e^{- {\cal V}(\ell)L},
\ee
 and its defines the so called pseudopotential $\cal V$. We note, that for large $L$  the behavior of $W(\ell,L)$ is different from the behaviour of the time-like one.
Then the pseudopotential can be straightforwardly extracted from the string action \eqref{7.1b} as follows
\be\label{VTension}
{\cal V}(\ell)=\frac{S_{string}}{L}.
\ee
As it is known from the QCD lattice calculations the spatial Wilson loops obey the area law at all
 temperature, i.e.
 the spatial string tension $\sigma_s$ is given by 
 \be\label{sp-tension}
 \sigma_s=\lim _{\ell\to\infty} \frac{{\cal V}(\ell) }{\ell}.
 \ee
The quantity $\sigma_{s}$ differs from the usual string tension which is defined from time-like Wilson-loops. 
By virtue to the non-Abelian Stokes formula equal time spatial Wilson loops \cite{Arefeva:1979dp} are related with the spatial components of the energy-momen\-tum tensor
and by this reason $\sigma_s$ is also called the magnetic string tension. Spatial Wilson loops have been studied on the lattice
\cite{Bali:1993tz,Petreczky:2012rq}, analytically \cite{Simonov}, and also within the gauge/gravity duality \cite{Andreev:2006eh,Alanen:2009ej}.

\subsection{The boost invariant anisotropic  metrics}\label{Sect:2.2}

We will study spatial Wilson loops in gravity backgrounds with spatial anisotropy,
given by \eqref{Lif-like0},
where the critical exponent  $\nu$  controls the deviation from isotropic case. In \cite {AGG} we have called  this metric  as an Lifshitz-like metrics. In this paper to avoid a misleading with  the Lifshitz metrics we mainly call the metric \eqref{Lif-like0} as the boost invariant spatial anisotropic metic.   Note, that in \eqref{Lif-like0} we used the standard dimensional coordinates. In what follows we take $R^2=2\pi \alpha '$, $a=1$ and use the dimensionless  coordinates 
$\tilde t= t/(2\pi \alpha')^{1/2}$, $\tilde z= z/(2\pi \alpha')^{1/2}$ etc., and we remove tilde on the top of coordinates. So, the metric take the form
 \begin{equation}\label{Lif-like-RL}
ds^{2} =2\pi\alpha'\left( \frac{-dt^{2} + dx^{2}}{z^{2}} + \frac{dy^{2}_{1} + dy^{2}_{2}}{z^{2/\nu}}  + \frac{d z^{2}}{ z^2 }\right).
\end{equation} 

One can see that the background (\ref{Lif-like-RL}) with $\nu =1$ comes to be the 5-dimensional AdS spacetime.
As already mentioned, the choice of this metric is motivated by the fact that holographic estimations of the total multiplicity performed 
in this background reproduce the experimental dependence of the multiplicity on the energy \cite{AG}.

The non-zero temperature generalization of (\ref{Lif-like0})  was constructed in \cite{AGG}\footnote{The computations have been checked with SageManifolds, which
is an extension of the free computer algebra system SageMath. The
corresponding worksheets are publicly available at the following links:\\
{\footnotesize
\url{https://cloud.sagemath.com/3edbca82-97d6-41b3-9b6f-d83ea06fc1e9/raw/Lifshitz_black_brane.html}\\
\url{https://cloud.sagemath.com/3edbca82-97d6-41b3-9b6f-d83ea06fc1e9/raw/Vaidya-Lifshitz.html}}}:
\begin{equation}\label{Ll-bh}
ds^{2} =2\pi\alpha'\left( \frac{-f( z)dt^{2} + dx^{2}}{z^{2}} + \frac{dy^{2}_{1} + dy^{2}_{2}}{z^{2/\nu}}  + \frac{d z^{2}}{ z^2 f(z)}\right),
\end{equation}
with the blackening function
\begin{equation} \label{bh-f}
f = 1- m z^{2 + 2/\nu}.
\end{equation}
For $\nu = 1$ the background (\ref{Ll-bh}) with (\ref{bh-f}) represents the metric of the AdS black brane.

This background (\ref{Ll-bh})-(\ref{bh-f}) describes holographically the anisotropic media  on the boundary with the  temperature
corresponding to  the Hawking temperature of the black brane:
\be\label{temperature}
T=\frac{1}{\pi}\frac{ (\nu +1) }{2\nu }\,m^{\frac{\nu }{2 \nu +2}} .
\ee

To study the thermalization process,  corresponding to the black brane formation in the dual language, we will  use the Vaidya generalization of solution \eqref{Lif-like0} 

  \cite{AGG}:
\be\label{Vaidya-LL}
ds^{2} = 2\pi\alpha'\left(- \frac{f(v,z)dv^2  + 2 dv d z-dx^{2}}{z^2}+\frac{dy^{2}_{1} + dy^{2}_{2}}{ {z}^{2/\nu}}
\right)
\ee
with
\bea\label{Vaidya-f}
f = 1- m(v) z^{2 + 2/\nu}.
\eea
The metric (\ref{Vaidya-LL}) has been written in ingoing Eddington-Finkel\-stein coordinates $(v,r)$. 
The  function $m(v)$ in (\ref{Vaidya-f}) defines the thickness of the shell smoothly interpolating between the zero-temperature (\ref{Lif-like0}) at $v=-\infty$ and black brane backgrounds (\ref{Ll-bh}) at $v=\infty$.

 We choose the following form for the function $f$ 
 \be\label{f}
f(z,v) = 1- \frac{m}{2}\left(1+\tanh \frac{v}{\alpha}\right)z^{\frac{2}{\nu}+2},\ee
where $\alpha$ is a parameter. For the calculations in this paper we keep $\alpha=0.2$.

\section{Spatial Wilson loops in a time-independent background}\label{Sect:WL-ind}

In this work we consider rectangular Wilson loops in the static background \eqref{Ll-bh} located in the spatial planes $xy_{1}$(or $xy_{2}$) and $y_{1}y_{2}$.
One can delineate the following possible configurations:
\begin{itemize}
\item   a rectangular loop in the $xy_{1}$ (or $xy_{2}$) plane  with a short side of the
length $\ell_{x}$ in the longitudinal $x$ direction  and a long side of the length $L_{y_{1}}$ along the transversal $y_1$  direction,
so that 
\begin{eqnarray}\label{C1xy1}
x \in [0, \ell _x], \quad y_{1} \in [0, L_{y_1}],\,\,\,\,\,\,\,\ell _x<L_{y_1};
\end{eqnarray}
\item a rectangular loop in  the $xy_1$ plane  with a short side of the
length $\ell_{y_{1}}$ in the transversal $y_1$ direction  and a long side of the length $L_{x}$ along the longitudinal $x$  direction, 
\begin{eqnarray}\label{C2xy1}
x \in [0, L_{x}], \quad y_{1} \in [0, \ell_{y_{1}}],\,\,\,\,\,\,\,\ell _{y_{1}}<L_x;
\end{eqnarray}
\item a rectangular loop in the transversal $y_1y_2$  plane  with a short side of the
length $\ell_{y_{1}}$ in one of  transversal  directions (say $y_1$)  and a long side of the length $L_{y_{2}}$ along the other  transversal   direction
$y_2$, namely
\begin{eqnarray}\label{C3y1y2}
y_{1} \in [0, \ell_{y_{1}}], \quad y_{1} \in [0, L_{y_{2}}],\,\,\,\,\,\,\,\ell_{y_{1}}< L_{y_{2}}.
\end{eqnarray}
\end{itemize}

In this section we perform all calculations in the  static spacetime (\ref{Ll-bh}) with (\ref{bh-f}) (using Eddington-Finkelstein coordinates).

\subsection{Wilson loops on the $xy_1$-plane}

\subsubsection{Rectangular strip infinite along the $y_1$-direction}

We start from the rectangular Wilson loop in the $xy_1$-plane assuming that the large extent is oriented in the $y_1$-direction (see (\ref{C1xy1})). We parametrize the world-sheet of the string in the following way
$\sigma^{1} = x, \quad \sigma^{2} = y_{1}$, assuming  $v = v(x)$, $z = z(x)$ and   boundary conditions: $z(\pm \ell_{x} /2)=0, z(0)=z_*, v(0) =v_*, z'(0)= 0,  v'(0)=0$.

 Taking into account  (\ref{7.1b}) and \eqref{Vaidya-LL}  with the stationary $f$ given by \eqref{bh-f}, the Nambu-Goto action can be presented as
\begin{eqnarray}
S_{x,y_{1(\infty)}} =   \int dy_{1} dx \frac{1}{z^{1/\nu}}\sqrt{\left(\frac{1}{z^{2}} - \frac{1}{z^{2}}f v'\,^{2} - \frac{2}{z^{2}}v'z'\right)},  
\label{7.2}
\end{eqnarray}
where it is supposed $\prime \equiv \frac{d}{d x}$. The subscript $x,y_{1(\infty)}$ in the LHS of (\ref{7.2}) indicates the orientation of the loop contour.

The action \eqref{7.2} on the  time-independent string configurations  after division on 
  the length  of the Wilson loop in the $y_1$-direction  can be rewritten in the form
 \begin{eqnarray}
\frac{S_{x,y_{1(\infty)}}}{L_{y_1}} =  \int \, \frac{ dx}{z^{1+1/\nu}\sqrt{f}}\sqrt{f(z) +z'\,^2}.  
\label{7.2a}
\end{eqnarray}
The RHS of \eqref{7.2a} defines a dynamical system  that has the first integral
\be
\label{J}
J=\frac{ 1}{z^{1+1/\nu}}\frac{\sqrt{f(z)}}{\sqrt{f(z) +z'\,^2}}.\ee
Note, that the same is true for the action in the form \eqref{7.2}. Indeed,  it defines the dynamical system with two degrees of freedom, $z=z(x)$ and $v=v(x)$, that for the stationary case has two integral of motions and excluding $v$
due to the conservation law one comes back to  \eqref{J}.

Since we have the symmetric boundary conditions, the top of the configuration $z=z(x)$ is at $x=0$, i.e. $z_*=z(0)$
and $z'(0)=0$, and $z_*$ is related with the first integral, 
\be\label{zstar}
\frac{ 1}{z^{1+1/\nu}}\frac{\sqrt{f(z)}}{\sqrt{f(z) +z'\,^2}}=\frac{ 1}{z_*^{1+1/\nu}}\ee
Equation \eqref{zstar} and the boundary condition $z(\pm \ell _x/2)=0$ give us the relation between the top point $z_*$
and the length
\be\label{7.3da}
\ell_x= 2  \int _{z_0} ^{z_*}\frac{dz}{\sqrt{f(z)\left(\left(\frac{z_*}{z}\right)^{2 + 2/\nu}-1\right)}},
\ee
So the  Nambu-Goto action (\ref{7.2}) can be rewritten in the form
\be\label{S1}\frac{S_{x,y_{1(\infty)}}}{L_{y_1}} =  2\int ^{z_*}_{z_0}\, \frac{ dz}{z^{1+1/\nu}\sqrt{f(z)}}\frac{1}{\sqrt{1-\left(\frac{z}{z_*}\right)^{2+2/\nu}}}.  
\ee
Here we use the regularized boundary conditions $z(\pm \ell_{x} /2)=z_0$. The action \eqref{S1} has a divergent when $z_0\to 0$
and one can present it in the form 
\bea\nn
\frac{S_{x,y_{1(\infty)}}}{2L_{y_{1}}}  &= &\frac{1}{z^{1/\nu}_{*}}\int^{1}_{z_0/z_*}\frac{dw}{w^{1 +1/\nu}}\Big[\frac{1}{\sqrt{f(z_{*}w)\left(1 - w^{2 + 2/\nu}\right) }}-1  \Big] + \frac{1}{z^{1/\nu}_{*}}\int^{1}_{z_0/z_*}\frac{dw}{w^{1+1/\nu}}.\\\label{7.4a}
\eea
Subtracting  the divergent part $\frac{\nu}{z_0^{1/\nu}}$ one gets 
the renormalized  Nambu-Goto action
\be
\frac{S_{x,y_{1(\infty)},ren}}{2L_{y_{1}}}  = \frac{1}{z^{1/\nu}_{*}}\int^{1}_{0}\frac{dw}{w^{1 +1/\nu}}\Big[\frac{1}{\sqrt{f(z_{*}w)\left(1 - w^{2 + 2/\nu}\right) }}-1  \Big] - \frac{\nu}{z^{1/\nu}_{*}}.\label{7.4a}
\ee
The length \eqref{7.3da} between the ends of the string admits the removing of regularization $z_0\to 0$ without renormalization
\begin{eqnarray}\label{7.3d}
\ell_x= 2 z_{*} \int _{0} ^1\frac{w^{1 + 1/\nu}\,dw}{\sqrt{f(z_{*}w)(1- w^{2 + 2/\nu})}}.
\end{eqnarray}

Then pseudopotential ${\cal V}_{x,y_{1(\infty)}}$  is given by:

\be\label{7.4a-V}
{\cal V}_{x,y_{1(\infty)}}=\frac{S_{x,y_{1(\infty)},ren}}{L_{y_{1}}}.
\ee

\begin{figure*}
\centering \begin{picture}(175,240)
\put(-90,-0) {\includegraphics[width=5.cm]{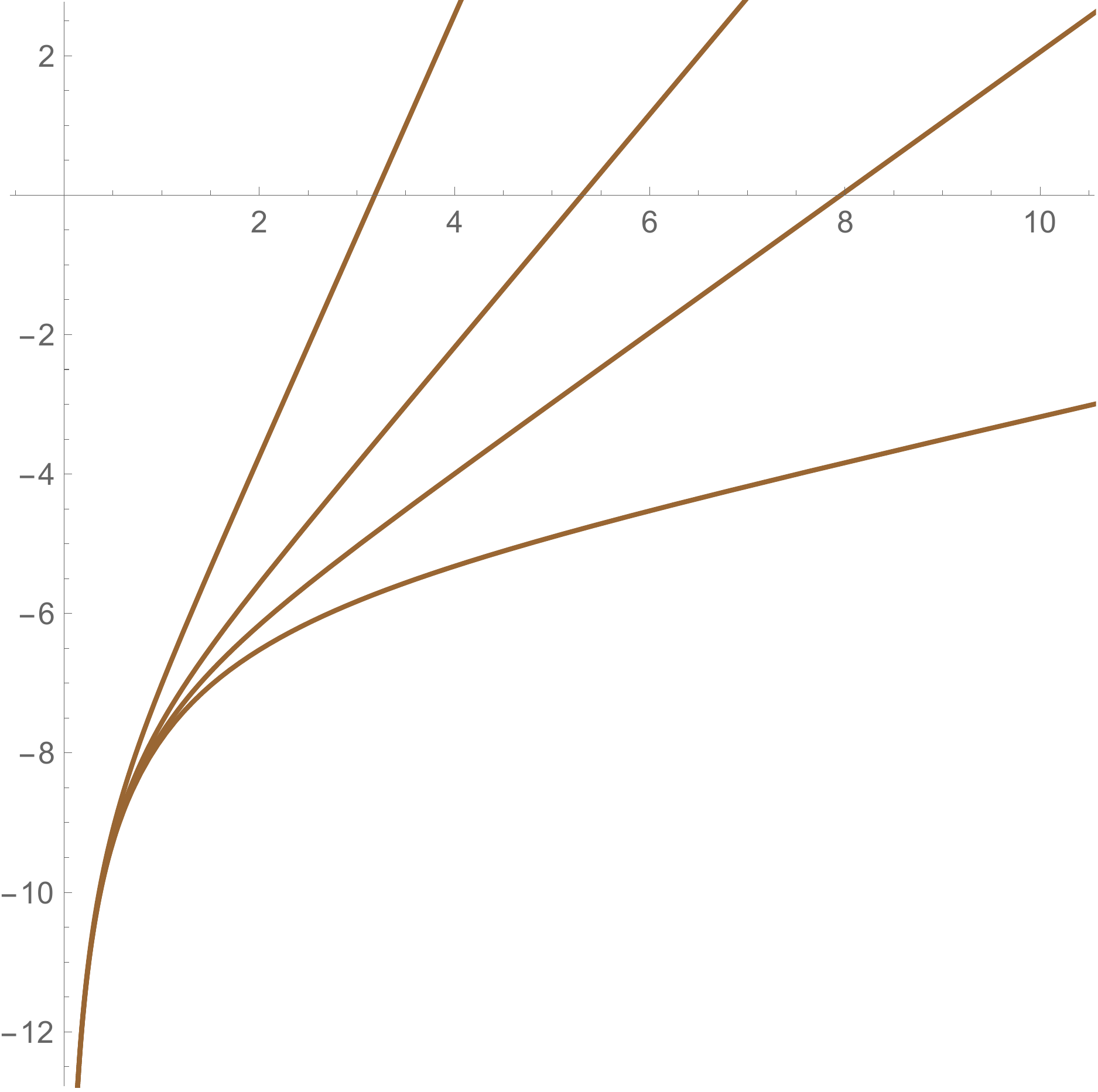}(a)}
 \put(-90,150){$\mathcal{V}_{x,y_{1(\infty)}} $}
 \put(60,110){$\ell_x$}
\put(105,5){\includegraphics[width=5.cm]{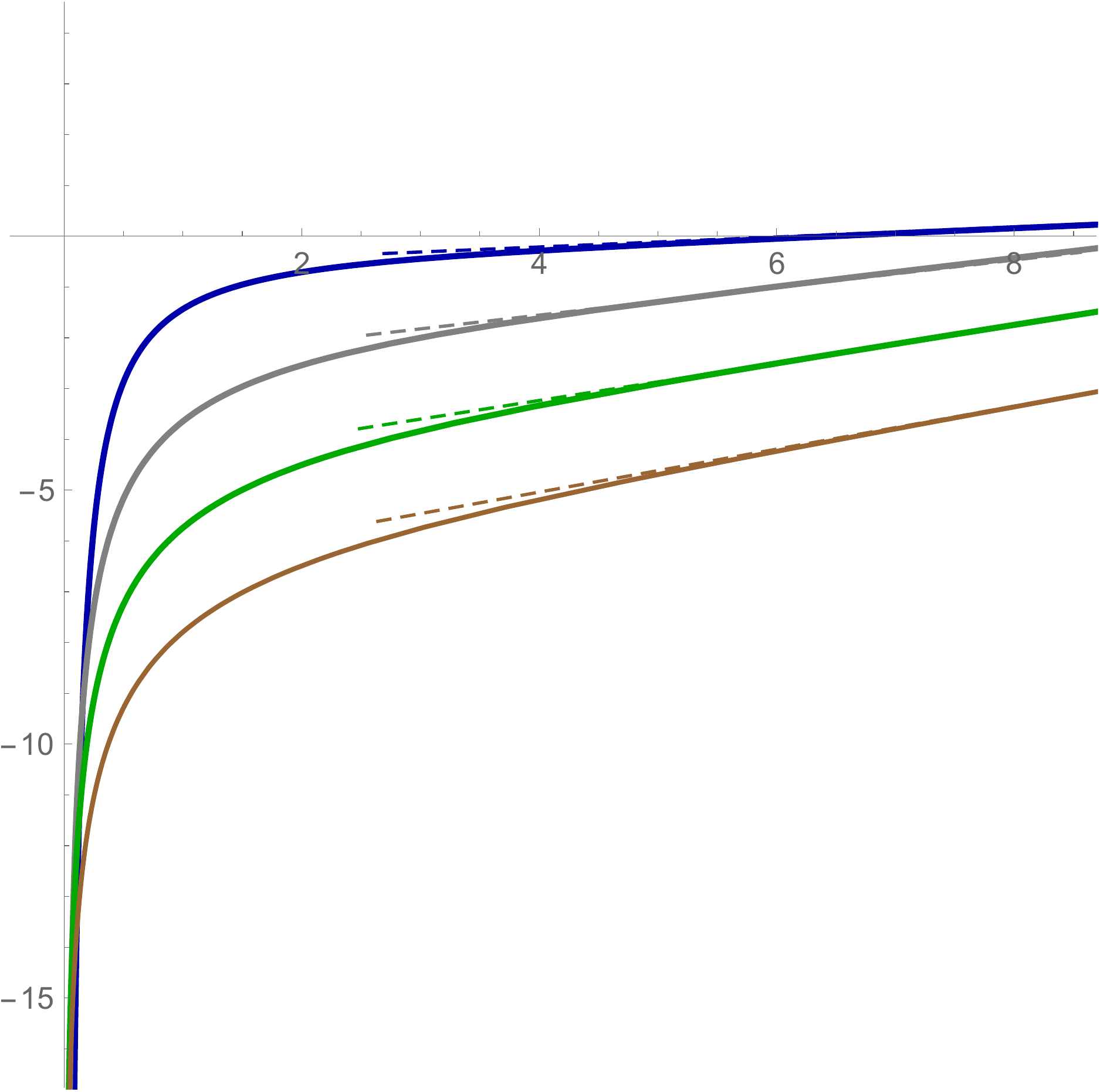}(b)}
\put(145,5){\includegraphics[width=3.cm]{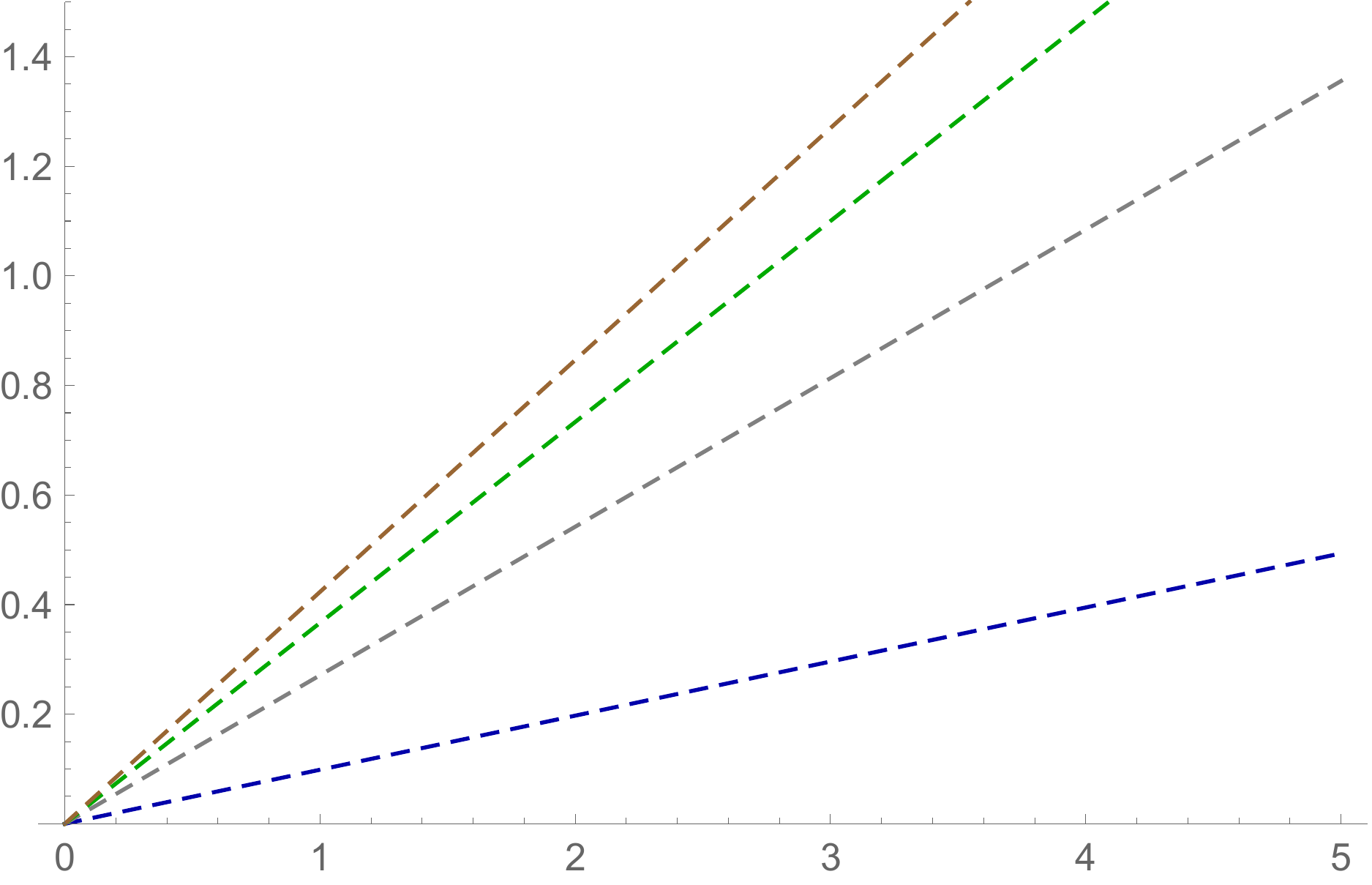}}
 \put(100,155){$\mathcal{V}_{x,y_{1(\infty)}} (\nu)$}
 \put(260,110){$\ell_x$}
 \end{picture}
 \caption{a) The pseudopotential ${\cal V}_{x,y_{1(\infty)}}$ corresponding to the action  ${S_{x,y_{1(\infty)}}}$ (\ref{7.4a}) as a function of $\ell_x$ (\ref{7.3d}) for $\nu=4$. We take the temperature  $T=0.08,0.2, 0.3, 0.5$  (from down to top).  b)  The behavior of the pseudopotential corresponding to (\ref{7.4a}) for $\nu =1,2,3,4$ (blue, gray, green and brown, respectively) at $T =0.1$. The dashed lines show the asymptotics at large $\ell_x$ given be \eqref{VxLarge}
  with \eqref{VLarge}. The inset plot zooms the slops of dashed lines. }
 \label{fig:11}
\end{figure*}
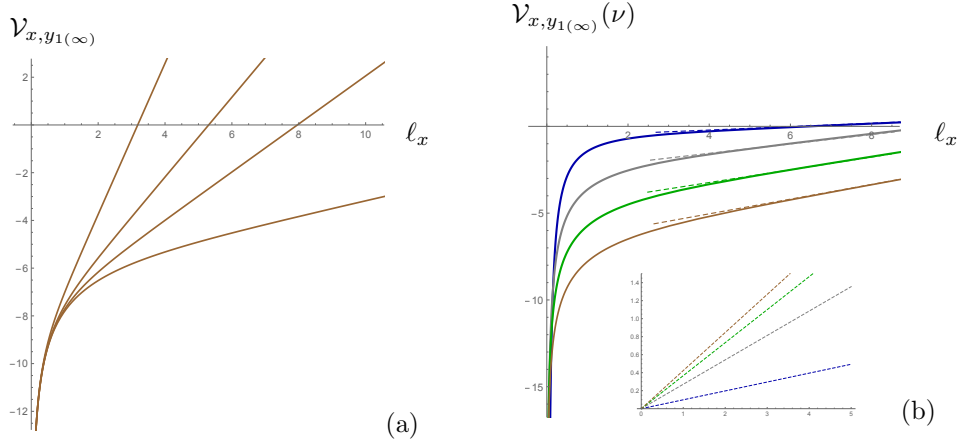

In Fig.\ref{fig:11} we present the dependence of the pseudopotential  ${\cal V}_{x,y_{1(\infty)}}$  (\ref{7.4a-V}) on the length (\ref{7.3d}). 
We see that for small $\ell_{x}$ the pseudopotential has the Coulomb part deformed by the critical exponent, thus
\begin{eqnarray}
\label{IR1}
\mathcal{V}_{x,y_{1(\infty)}} (\ell_x,\nu)\underset{\ell_x\sim 0}{\sim} -\frac{\mathcal{C}_1(\nu)}{\ell_x^{1/\nu}},
\end{eqnarray}
where $\mathcal{C}_1$ is some constant dependent on $\nu$.   Putting $m=0$ in \eqref{7.4a} and 
\eqref{7.3d}, we get the leading term \eqref{IR1}, while taking corrections on $m$ one has for $\nu=4$
\bea\nn
{\cal V}_{x,y_{1(\infty)}}  (\ell_x,4)&=& -\frac{7.80}{\ell_x^{1/4}}\left( 1-0.012 \,m\,\ell_x^{5/2}
   +{\cal O}\left(m^2\,\ell_x^{5}\right) \right).\\
   \eea
The asymptotics for arbitrary  $\nu$ are presented in Appendix \ref{App:B}.
 One should note that for small enough $\ell_{x}$  the behaviour of $\mathcal{V}_{x,y_{1(\infty)}}$ extracted from (\ref{7.4a}) for all $\nu$ has a form of the deformed Coulomb law with the power equal to $1/\nu$ reproducing 
the  Coulomb  behaviour of the pseudopotential in the case   $\nu =1$ (the AdS case).

For large distances $\ell_x$ the pseudopotential  ${\cal V}_{x,y_{1(\infty)}}$ behaves as a linearly increasing function
\be
\label{sigma1}
\mathcal{V}_{x,y_{1(\infty)}} (\ell_x,\nu)\underset{\ell_x\to\infty}{\sim} \sigma_{s,1}(\nu)\,\ell_x.\ee
To see this behaviour let us note that to get the large $\ell_x$ we have to take $z_*$ near $z_h$, or $z_*/z_h=w_0\to 1$, and in this case the denominator in the integrands in the RHS of \eqref{7.3d}  and 
 \eqref{7.4a} behaves as
\be
\sqrt{f(z_{*}w)(1- w^{2 + 2/\nu})}\approx  \frac{2 (\nu +1)}{\nu }(1-w),\ee
and we get the $\log$-behaviour for $\ell_x$ and ${\cal V}_{x,y_{1(\infty)}} $
\bea
\label{7.4aa}
\ell_x&\sim& - 2\frac{\nu z_*  }{ (\nu +1)}\log(1-w_0),\\
{\cal V}_{x,y_{1(\infty)}}  &\sim &-\frac{2}{z^{1/\nu}_{*}} \frac{\nu   }{ (\nu +1)}\log(1-w_0),\label{7.4aaa}
\eea
which lead us to the asymptotic
\bea\label{VxLarge}
{\cal V}_{x,y_{1(\infty)}}   &\underset{\ell_x\to \infty}{\sim }&\sigma_{s,1}(\nu)\,\ell_x,\eea
where  
\be
\label{VLarge}
\sigma_{s,1}(\nu)= 1/z^{1+1/\nu}_h=(\frac{2\pi  T}{1+1/\nu})^{1+1/\nu}.
\ee

From Fig.\ref{fig:11}.b we  see that  the asymptotics of a ${\cal V}_{x,y_{1(\infty)}}$ 
 at large $\ell_x$ is linear with the slope  given by formula \eqref{VLarge}. 
The formula \eqref{VLarge} exhibits the dependence on $\nu$ shown in the zoomed inset of Fig.\ref{fig:11}.b. 
  
   The string tension  \eqref{VLarge} can be also seen by dimensional analysis  keeping the anisotropic parameter $a$ in \eqref{Lif-like0}. Indeed, since we are working with metric \eqref{Ll-bh} and we can assign a factor $A$ to $t,x$ and $z$ and factor $A^{1/\nu}$ to $y1$ and $y_2$. Since the action $S$ is invariant under this rescaling we have that $\sigma_{1,s}\sim \frac{S}{\ell_x L_{y_1}}\sim A^{-(1+1/\nu)}\sim T^{1+1/\nu}.$

\subsubsection{ Rectangular strip infinite along the $x$-direction}

Another possible configuration in the $xy_1$ plane is the rectangular Wilson loop whose contour
 is infinite along the $x$-direction and has a finite size along the $y_1$-direction (see (\ref{C2xy1})). 
We specify this type of the configuration by the subscript $y_{1},\,x_{(\infty)}$.
 For the parameterization we take $v = v(y_{1})$, $z = z(y_{1})$ with boundary conditions $z(\pm \ell_{y_{1}}/2)=0$. 
 By virtue to this assumption the Nambu-Goto action (\ref{7.1b}) reads 
\begin{eqnarray}
S_{y_{1},\,x_{(\infty)}} = \int\,  \frac{dy_{1} dx}{z}\,\sqrt{\left(\frac{1}{z^{2/\nu}} - \frac{1}{z^{2}}f (v')^{2} - \frac{2}{z^{2}}v'z'\right)}, \label{7.5-2}
\end{eqnarray}
where it is supposed $\prime\equiv \frac{d}{d y_{1}}$.
Similar to the previous case this action can be rewritten as
\begin{eqnarray}
\frac{S_{y_{1},\,x_{(\infty)}}}{L_x} = \int\,  \frac{ dy_1}{z^2\sqrt{f}}\,\sqrt{fz^{2-2/\nu}+z'\,^2}, \label{7.5-2a}
\end{eqnarray}
and the first integral is related with the top point as 
\be
\frac{\sqrt{f(z)} z^{-2/\nu }}{\sqrt{f(z)
   z^{2-\frac{2}{\nu }}+z'\,^2}}=\frac{ 1}{
   z_*^{1+1/\nu }}.\ee
Due to this relation we get the expression for 
 the action
\bea\label{S2r}
\frac{S_{y_{1},\,x_{(\infty)}}}{L_{x}} &=&\frac{2}{z_{*}}\int^{1}_{z_{0}/z_{*}}\frac{dw}{w^{2}}\frac{1}{\sqrt{f(z_{*}w)\left(1 - w^{2 + 2/\nu}\right) }},
\label{S2r}
\eea and the length
\begin{eqnarray}\label{7.6}
\ell_{y_1} = 2 z^{1/\nu}_{*}\int_{z_{0}/z_{*}}^{1} \frac{w^{2/\nu}dw}{\sqrt{f(z_{*}w)\left(1 -  w^{2 +2/\nu}\right)}}.
\end{eqnarray}
where $z_{0}$  is the regularization. Similar to the previous case one can remove regularization in  \eqref{S2r} and in \eqref{7.6} after renormalization
\bea
\frac{S_{y_{1},\,x_{(\infty)},\,ren}}{2 L_{x}} &=&\frac{1}{z_{*}}\int^{1}_{0}\frac{dw}{w^{2}}\left[\frac{1}{\sqrt{f(z_{*}w)\left(1 - w^{2 + 2/\nu}\right) }} - 1\right] - \frac{1}{z_{*}}.
\label{7.6a}
\eea
The pseudopotential ${\cal V}_{y_{1},x_{(\infty)}}$ is  related to \eqref{7.6a} as
\be\label{7.6a-V}
{\cal V}_{y_{1},\,x_{(\infty)}}=\frac{S_{y_{1},\,x_{(\infty)},\,ren} }{L_x}.
\ee

\begin{figure*}
\centering \begin{picture}(175,220)
\put(-90,-0) {\includegraphics[width=5.5cm]{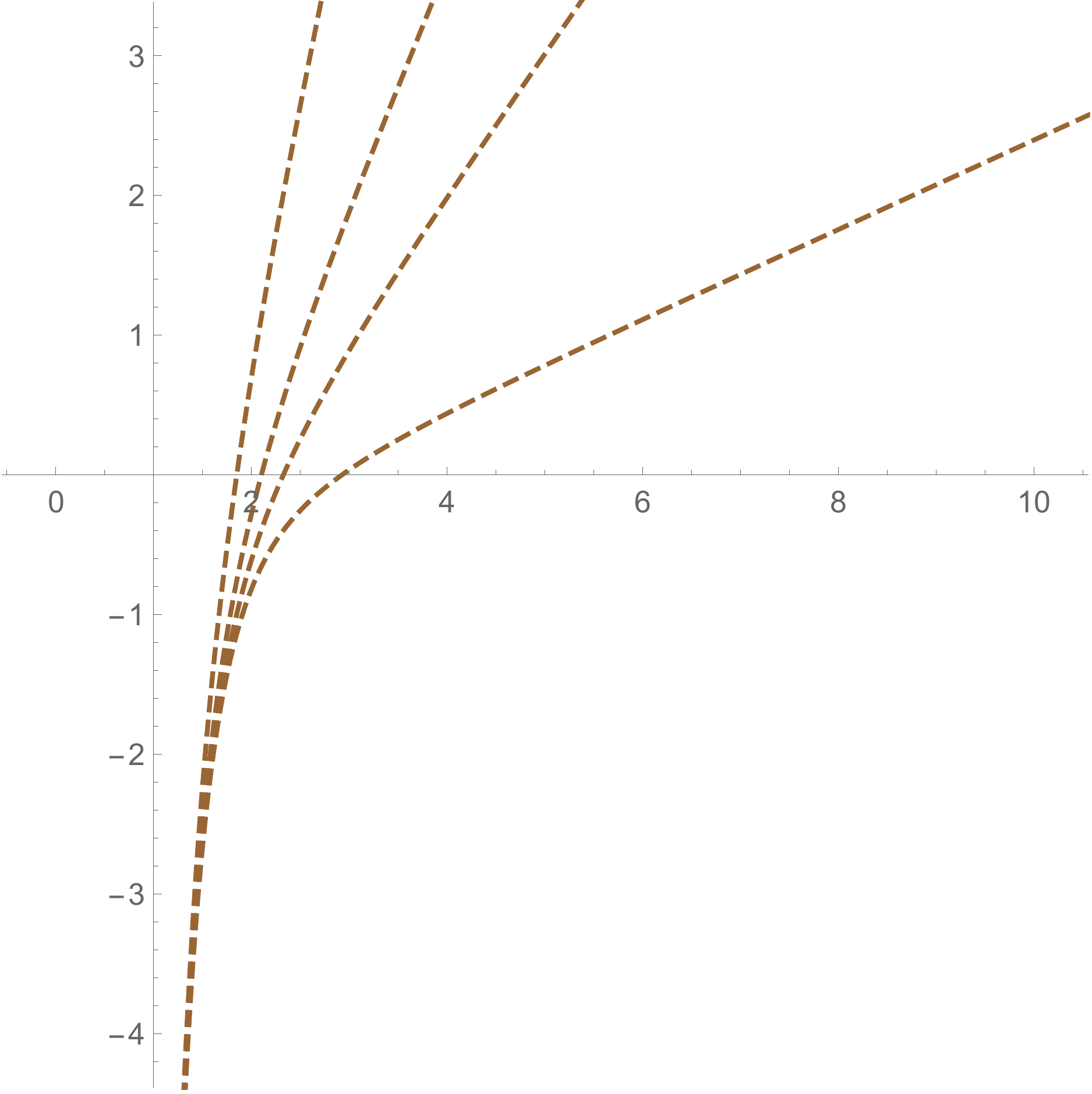}(a)}
 \put(-90,160){ ${\cal V}_{y_{1},\,x_{(\infty)}}$}
 \put(73,90){$\ell_{y_1}$}
\put(105,-0){\includegraphics[width=5.5cm]{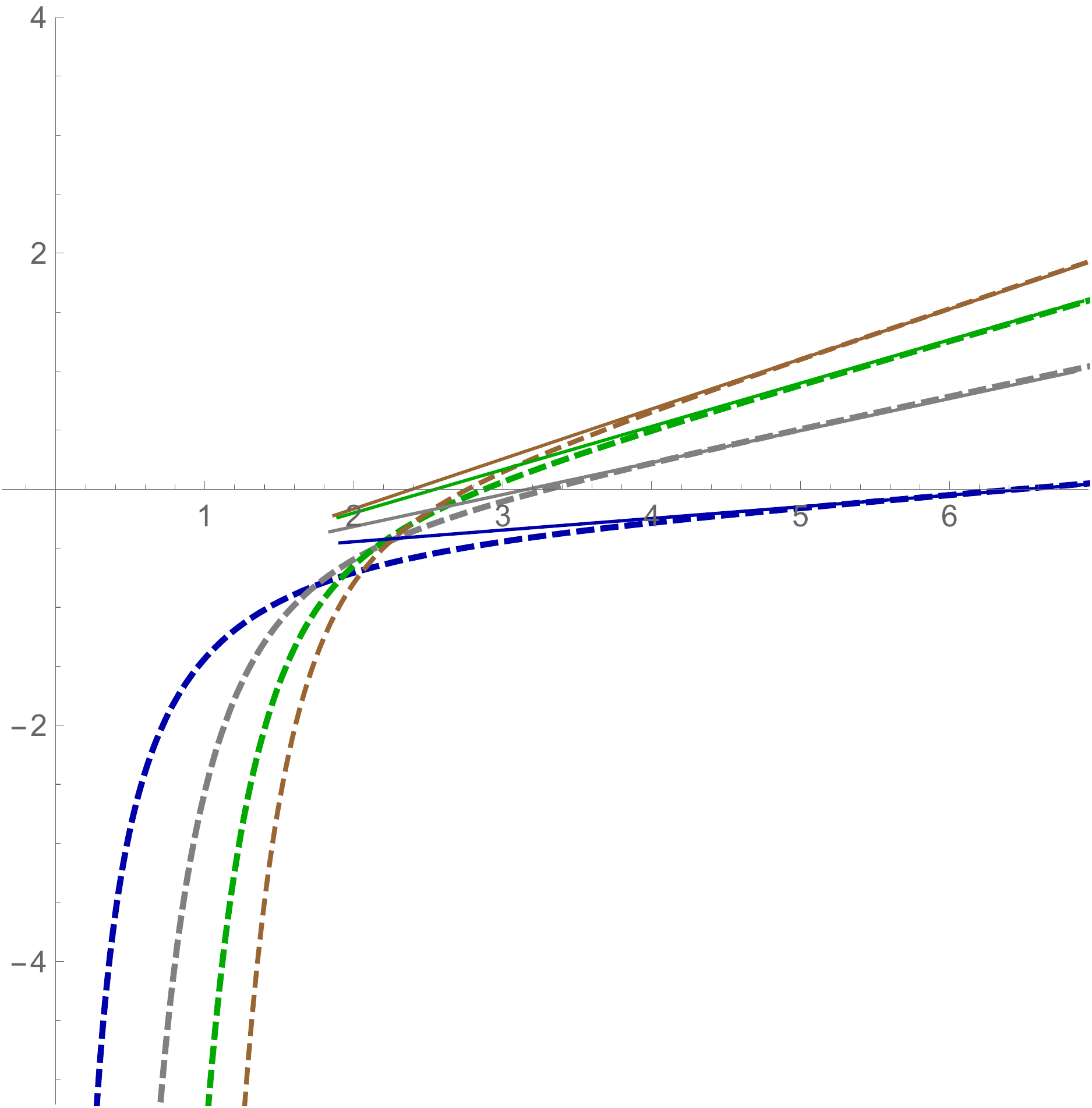}(b)}
\put(165,5){\includegraphics[width=3.cm]{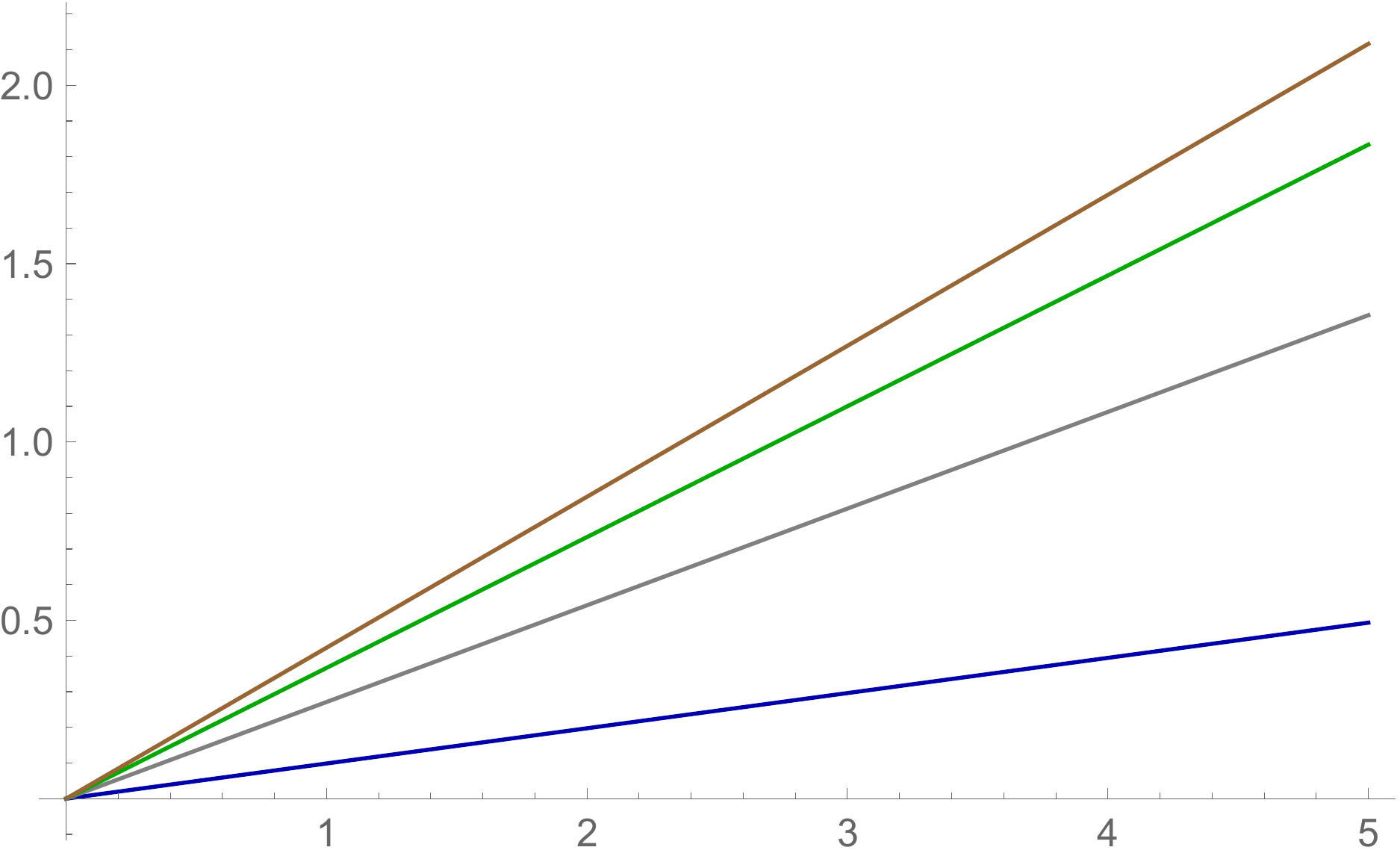}}
 \put(100,165){ ${\cal V}_{y_{1},\,x_{(\infty)}}(\nu)$}
 \put(265,85){$\ell_{y_1}$}
 \end{picture}
 \caption{a) The pseudopotential ${\cal V}_{y_{1},\,x_{(\infty)}}$ extracted from action (\ref{7.6a}) as a function of $\ell_{y_1}$ for $\nu =4$. We take $T=0.08,0.2, 0.3, 0.5$  from down to top.  (q.6.1)b) The behaviour of ${\cal V}_{y_{1},\,x_{(\infty)}}$  for $\nu =1,2,3,4$ (blue, gray, green and brown, respectively) at $T =0.1$. The solid thin lines show the asymptotic of ${\cal V}_{y_{1},\,x_{(\infty)}}$ for large $\ell_y$ given by formula \eqref{sigma2} for different  $\nu$.
 The inset plot zooms the slops of solid thin lines.}\label{fig:12}
\end{figure*}
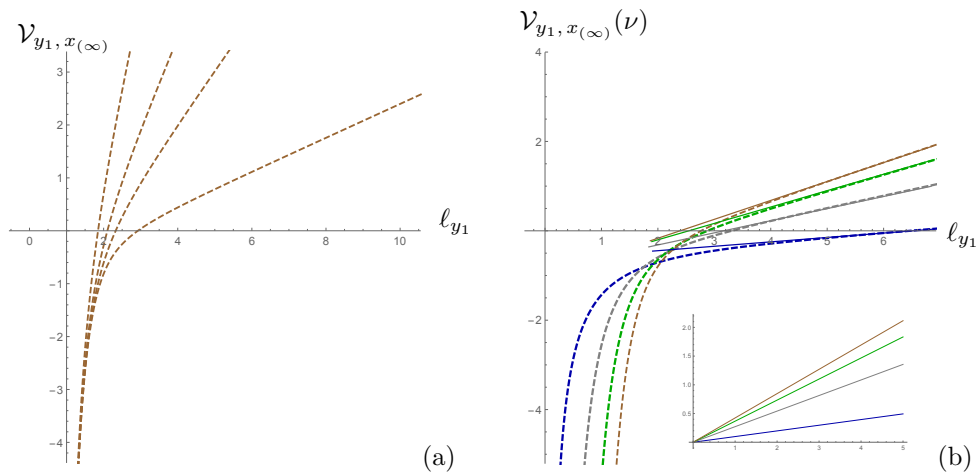

In Fig.~\ref{fig:12} we show the dependence of the pseudopotential extracted from the action \eqref{7.6a} 
on the length $\ell_{y_{1}}$ for different values of the temperature and the dynamical exponent. Now the pseudopotential has a power-law dependence on $\nu$ for small $\ell_{y_1}$, so that 
\begin{eqnarray}
\label{IR2}
\mathcal{V}_{y_{1},\,x_{(\infty)}} \underset{\ell_{y_1} \to 0}{\sim}  -\frac{\mathcal{C}_2(\nu)}{\ell_{y_1}^{\nu}},
\end{eqnarray}
with some constant $\mathcal{C}_2$ dependent on $\nu$.   One gets asymptotics \eqref{IR2} with $m=0$ in \eqref{7.6a} and 
\eqref{7.6}, and taking corrections on $m$ we obtain for $\nu=4$
\bea
{\cal V}_{y_{1},\,x_{(\infty)}}&=&-\frac{13.5}{\ell_{y_{1}}^4}\left(1-0.00031m \ell_{y_{1}}^{10}+{\cal{O}}(m^2\ell_{y_{1}}^{20} )\right).\nn\\
\eea

However, for large distances the pseudopotential represents a linear function of $\ell_{y_{1}}$ again
\be
\label{sigma2}
\mathcal{V}_{y_{1},\,x_{(\infty)}} (\ell_{y_{1}},\nu)\underset{\ell_{y_1}\to\infty}{\sim} \sigma_{s,2}(\nu)\,\ell_{y_{1}}
\ee
 and using estimation, similar to \eqref{7.4aa} and \eqref{7.4aaa} we  get 
 \be
\sigma_{s,2}(\nu)= \frac{1}{z_h^{1+1/\nu}} =(\frac{2\pi T}{1+1/\nu} )^{1+1/\nu},\ee
that is also in agreement with the dimensional analysis.
 
From Fig.~\ref{fig:12}.b we also see that the dependence of $\mathcal{V}_{y_{1},\,x_{(\infty)}} (\ell_{y_1},\nu)$ on $\ell_{y_{1}}$ at large length is linear for all the dynamical exponent $\nu$ with  the slops  slightly deviating from the $AdS$ case ($\nu=1$).

\subsection{Wilson loop on the $y_1y_2$-plane}\label{Sect:5.1.2}

Now we come to the spatial rectangular Wilson loop located on  the $y_1y_2$-plane.
Let us assume that the loop contour is infinite along the $y_2$-direction and has the finite extent of the length $\ell_{y_{1}}$ in  the $y_1$-direction (see (\ref{C3y1y2})). We specify this type of the orientation
by  the subscript $y_{1},\,y_{2(\infty)}$,  choosing only transversal coordinates for the parameterization of the worldsheet 
$\sigma^{1}  = y_{1}, \quad \sigma^{2} = y_{2}$.

 Taking into account, that $z(y_{1})$ satisfying  $ z(\pm \ell_{y_1}/2)=0$  one can represent the string action (\ref{7.1b}) in the following form
 \begin{eqnarray}\label{7.7a-2}
S_{y_{1},\,y_{2(\infty)}} = \int  \frac{dy_{1} dy_{2}}{z^{\frac1{\nu}}}\sqrt{\frac{1}{z^{\frac2{\nu}}} - \frac{f v^{\prime2}}{z^{2}} - \frac{2v'z'}{z^{2}}}, 
\end{eqnarray}
where it is supposed $\prime \equiv \frac{d}{d y_{1}}$.
Similar to the previous cases this action can be rewritten as
 \begin{eqnarray}\label{7.7a-2}
\frac{S_{y_{1},\,y_{2(\infty)}} }{L_{y_2}}= \int  
\frac{dy }{z^{1+1/\nu} \sqrt{f}}\sqrt{f z^{2-2/\nu} +z'\,^2}, 
\end{eqnarray}
and the first integral is related with $z_*$ point as
\be
\frac{\sqrt{f(z)} z^{1-\frac{3}{\nu }}}{\sqrt{f(z)
   z^{2-\frac{2}{\nu }}+z'\,^2}}=\frac{1}{z_*^{2/\nu}}.\ee
   Due to this relation we get the expression for 
 the action
\bea
\frac{S_{y_{1},\,y_{2(\infty)}}}{2L_{y_{2}}}& = &\int^{z_*}_{z_{0}} \frac{dz}{z^{1+1/\nu}}\frac{1}{\sqrt{f(z)\left(1 - \left(\frac{z}{z_*}\right)^{4/\nu}\right)}},
\label{7.6ab}
\eea
and  the length 
\begin{eqnarray}\label{7.7im}
\ell_{y_1} = \frac{2}{z^{2/\nu}_{*}}\int _{z_0}^{z_*}\frac{dz}{z^{1 - 3/\nu}\sqrt{f(z)\left(1 - \left(\frac{z}{z_*}\right)^{4/\nu}\right)}},
\end{eqnarray}
where $z_0$ is the regularization. Similar to the previous case one can remove regularization in \eqref{7.7im} directly 
and in \eqref{7.6ab} after renormalization.
The renormalized action (\ref{7.6ab}) in terms of the $w$-variable takes the form
\bea\label{Sy1y2c}
\frac{S_{y_{1},\,y_{2(\infty)},ren}}{2L_{y_{2}}}= \frac{1}{z^{1/\nu}_{*}}\int^{1}_{z_{0}/z_{*}} \frac{dw}{w^{1+1/\nu}}\Big[\frac{1}{\sqrt{f(z_{*}w)\left(1 - w^{4/\nu}\right)}}- 1\Big] - \frac{\nu}{z^{1/\nu}_{*}}.\label{7.7k}
\eea
The relation for the length is given by
\begin{eqnarray}\label{7.7i}
\ell _{y_1}= 2z^{1/\nu}_{*}\int _0^1\frac{dw}{w^{1 - 3/\nu}\sqrt{f(z_{*}w)\left(1 - w^{4/\nu}\right)}}.
\end{eqnarray}

Finally, the pseudopotential ${\cal V}_{y_{1},\,y_{2(\infty)}}$ extracted from \eqref{7.7k} reads as:
\be\label{7.7a-V}
{\cal V}_{y_{1},\,y_{2(\infty)}}=\frac{S_{y_{1},\,y_{2(\infty)}} }{L_{y_{2}}}.
\ee

\begin{figure*}
\centering \begin{picture}(175,220)
\put(-90,-0) {\includegraphics[width=5.5cm]{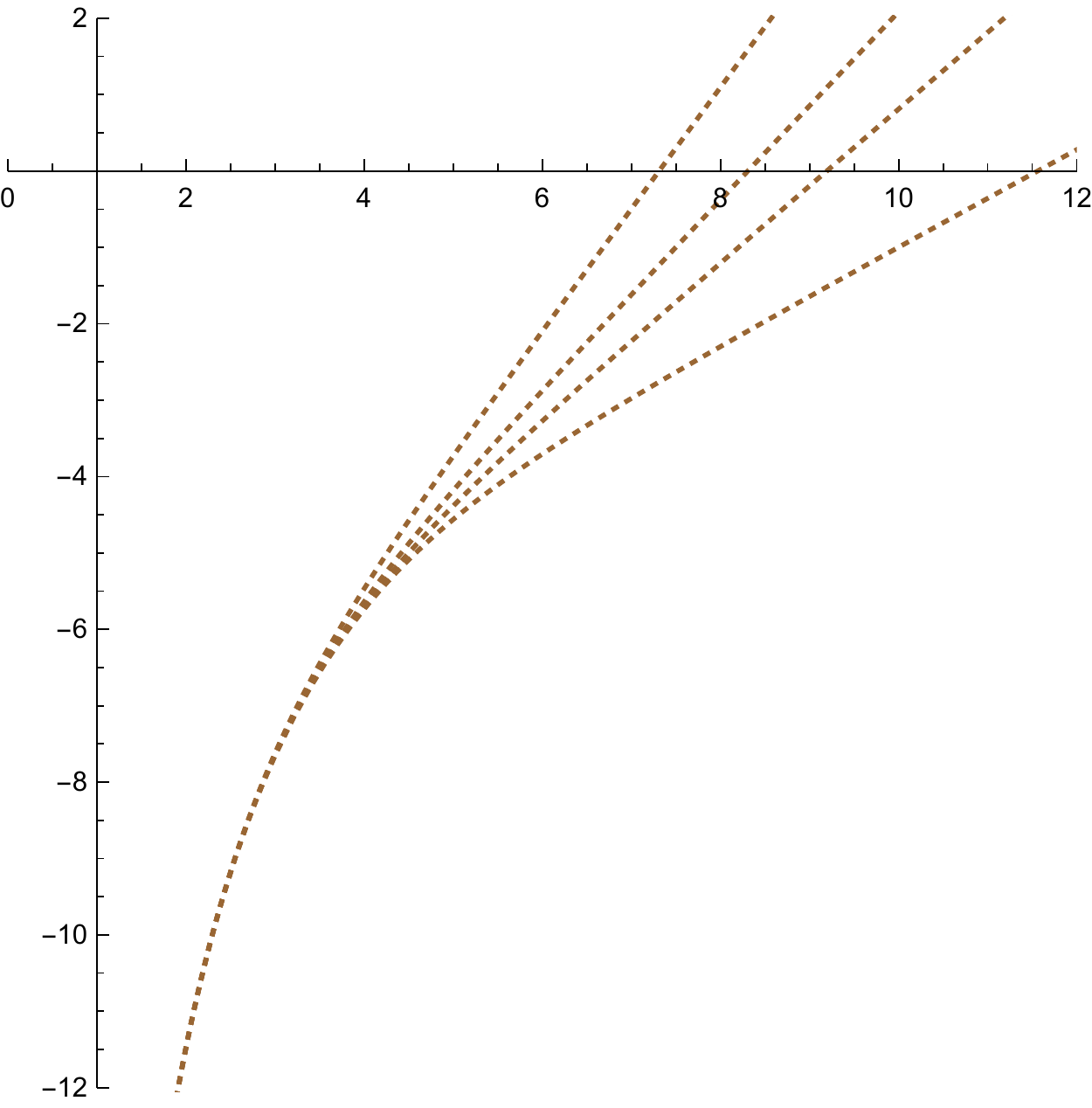}(a)}
 \put(-90,160){ ${\cal V}_{y_{1},\,y_{2(\infty)}}$}
 \put(75,130){$\ell_{y_1}$}
\put(105,20){\includegraphics[width=5.5cm]{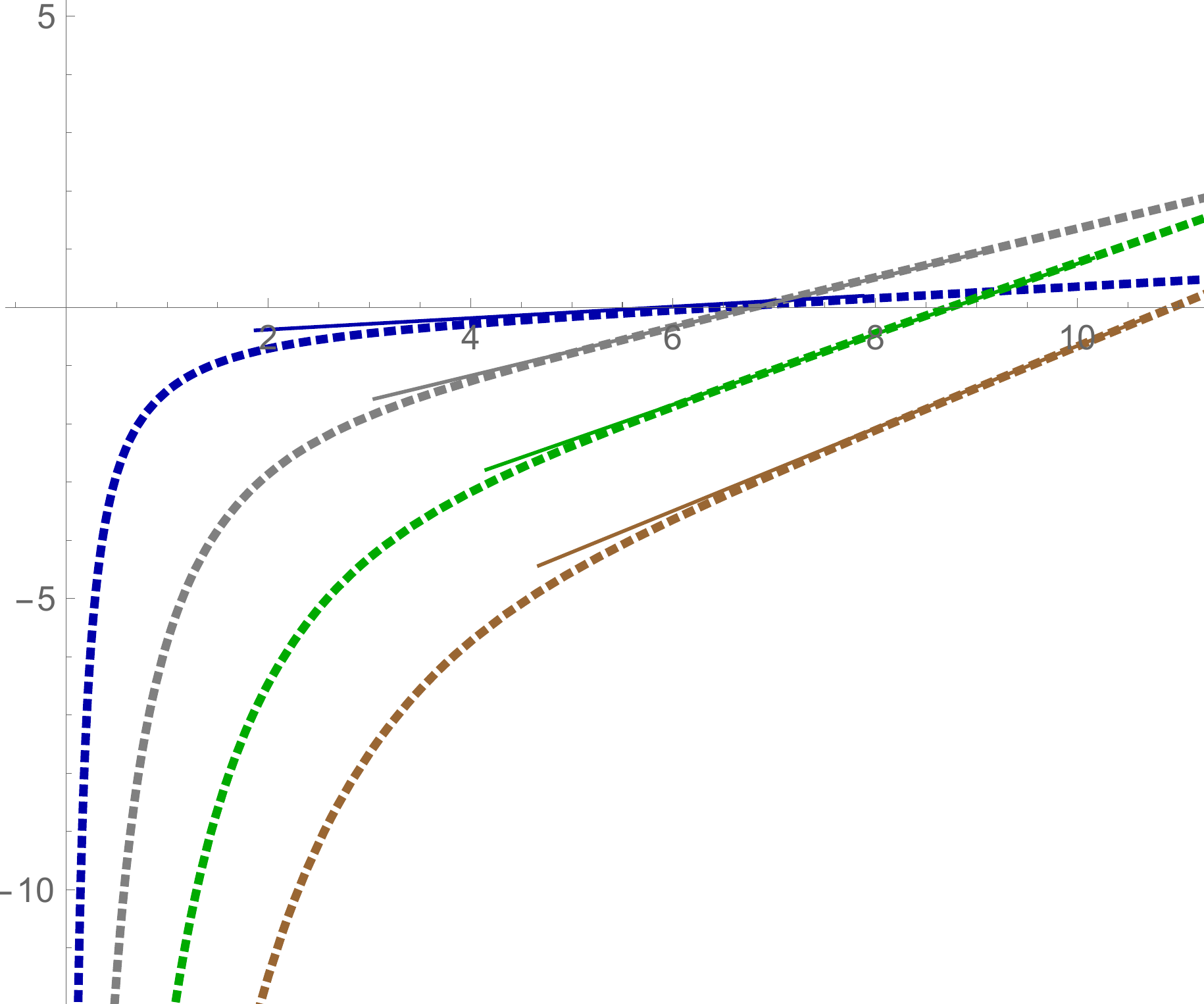}(b)}
\put(165,5){\includegraphics[width=3.cm]{WL-2-inset.pdf}}
 \put(100,160){ ${\cal V}_{y_{1},\,y_{2(\infty)}}(\nu)$}
 \put(265,110){$\ell_{y_1}$}
 \end{picture}
 \caption{a) The pseudopotential ${\cal V}_{y_{1},\,y_{2(\infty)}}$ extracted from action (\ref{Sy1y2c}) as a function of $\ell_{y_{1}}$ for $\nu =4$. We take $T=0.08,0.2, 0.3, 0.5$  from down to top.  The behaviour of ${\cal V}_{y_{1},\,x_{(\infty)}}$ corresponding to (\ref{Sy1y2c}) for $\nu =1,2,3,4$ (from left to right, respectively)  at $T =0.1$.
 The solid lines show the asymptotics given by formula \eqref{sigma3} with \eqref{sigma3m}.
 The inset plot zooms the slops of solid thin lines.}\label{fig:13}
\end{figure*}

In Fig.\ref{fig:13} we display the behaviour of the pseudopotential  (\ref{7.7a-V}) on the length (\ref{7.7i}). 

It is easy to see that the behavior of $\mathcal{V}_{y_{1},\,y_{2(\infty)}}$ in Fig.\ref{fig:13} is rather different from two previous cases.  From Fig.\ref{fig:13} (b) we observe, that now the dependence on $\nu$  is driven by some constant $\mathcal{C}_3$ relying on $\nu$. It should be noted that the pseudopotentials strongly deviate from the $AdS$  case ($\nu=1$) both in the UV  and the IR regions of $\ell_{y_{1}}$. 
Thus, one can write for small $\ell_{y_{1}}$
\begin{eqnarray}\label{Vy1y2c3}
\mathcal{V}_{y_{1},\,y_{2(\infty)}} (\ell_{y_1},\nu)\underset{\ell_{y_1}\to 0}{\sim} -\frac{\mathcal{C}_3(\nu)}{\ell_{y_1}},
\end{eqnarray}
where $\mathcal{C}_3$ is some constant dependent on $\nu$.  For $\nu=4$ one can write down (\ref{Vy1y2c3}) with the corrections on $m$ in the following form
 \bea\nn
{\cal V}_{y_{1},\,y_{2(\infty)}}(\ell_{y_{1}},4)=  -\frac{23.0}{\ell_{y_1}}\left(1 - 0.741\cdot 10^{-9}m\,l_{y_1}^{10}+{\cal O}\left(m^2\ell_{y_1}^{20}\right)\right).
\end{eqnarray}

For large $\ell_{y_{1}}$ we have 
\be
\label{sigma3}
\mathcal{V}_{y_{1},\,y_{2(\infty)}} (\ell_{y_1},\nu)\underset{\ell_{y_1}\to\infty}{\sim} \sigma_{s,3}(\nu)\,\ell_{y_1}.
\ee
 As in the previous cases the large $\ell_{y_1}$ behaviour is provided by the pole near to $w\sim 1$ in \eqref{7.7i} and we get, compare with \eqref{7.4aa} and \eqref{7.4aaa},
\be\label{sigma3m}
 \sigma_{s,3}(\nu)=  \frac{1}{z_h^{2/\nu}}=\left(\frac{2\pi T}{ 1+1/\nu}\right)^{2/\nu}.
\ee\\
Looking  at formulas (\ref{Vy1y2c3})-(\ref{sigma3}) one can conclude that the pseudopotential corresponding  to the configuration on the transversal plane  reproduces the  form of the Cornell potential.

\subsection{Spatial string tension dependence on the orientation}

It is interesting to analyze the behavior of the spatial string tension $\sigma_s$ (\ref{sp-tension}) for different orientations of the Wilson loop, i.e. the behaviour of $\sigma_{s,i}, i=1,2,3$ given by \eqref{sigma1}, \eqref{sigma2} and \eqref{sigma3}. The  temperature dependences of the spatial string tension in the confining background,
which reproduces the Cornell potential \cite{Andreev2006ct}, and the deconfining one
 have been studied in \cite{Andreev:2006eh} and  \cite{Andreev:2007rx}, respectively. In \cite{Andreev:2008tv} the universal behaviour of the spatial string tension for multiquark configurations was found.  In the  AdS/QCD model  \cite{Andreev2006ct} the string tension dependence matching lattice data was found in \cite{Alanen:2009ej}. 
 
\begin{figure*}
\centering
\centering \begin{picture}(185,150)
\put(-90,0){\includegraphics[width=5.5cm]{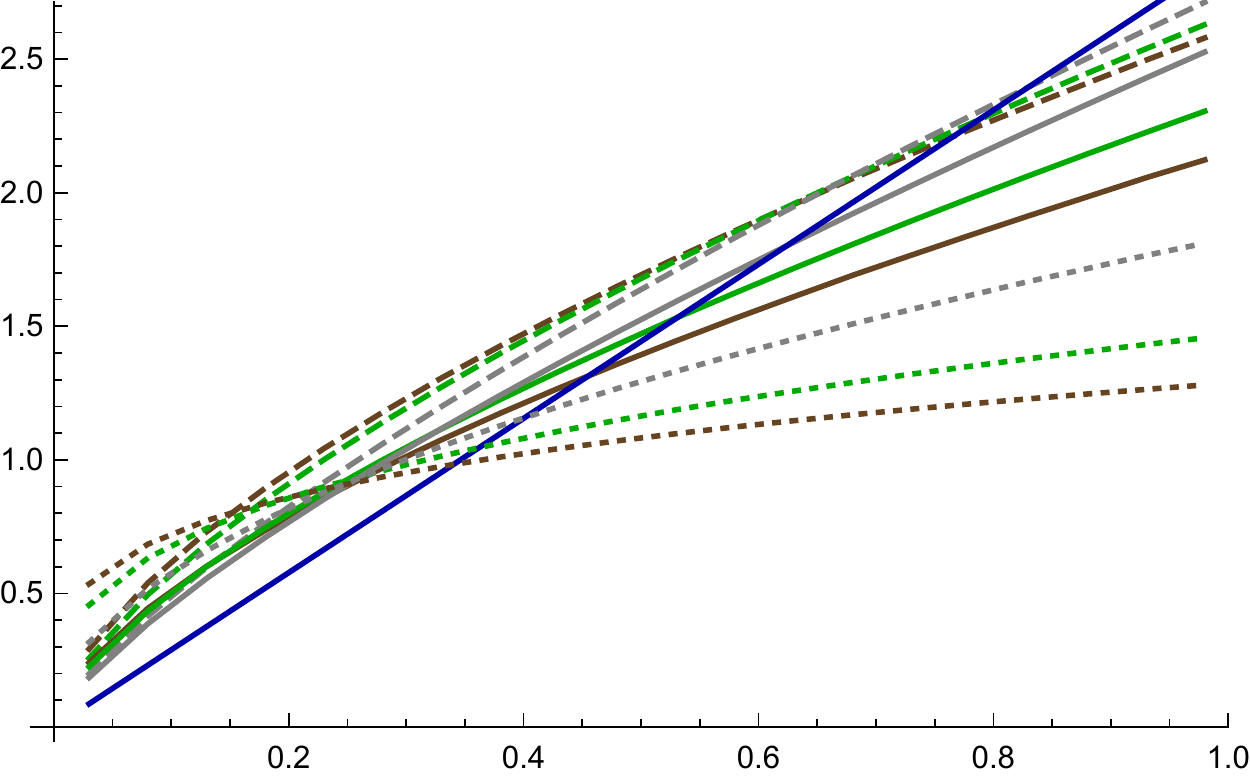}\,\,\,\,\,\,(a)}
 \put(-90,100){ $\sigma_s^{1/2}$}
 \put(75,10){ $T$}
\put(100,0){\includegraphics[width=5.5cm]{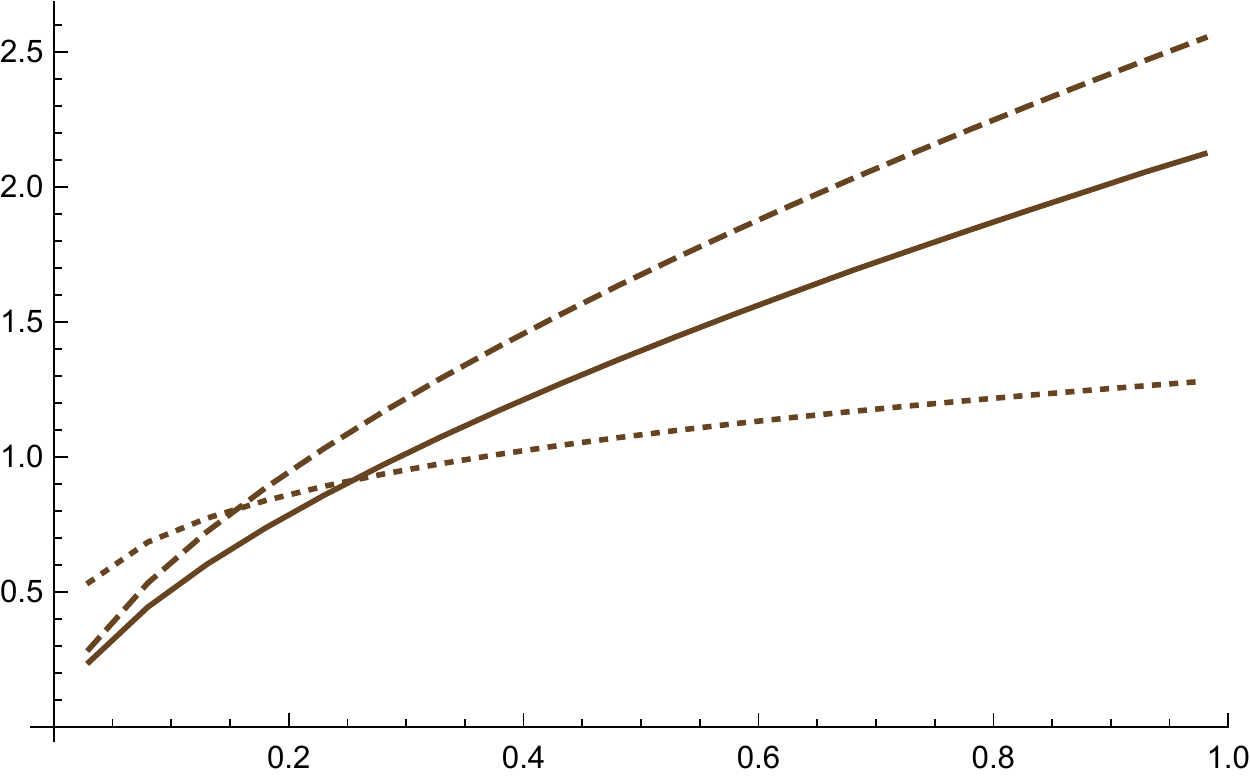}$\,\,\,\,\,\,$(b)}
 \put(260,10){ $T$} 
  \put(100,100){ $\sigma_s^{1/2}$}\end{picture}
  \caption{The dependence of the spatial string tension $\sqrt{\sigma_s}$ on orientation and temperature. The solid lines corresponds to
 the rectangular Wilson loop with a short extent in the $x$-direction, while the dashed lines correspond to a short extent in the $y$-direction.
The dotted lines  correspond to the rectangular Wilson loop in the transversal $y_1y_2$-plane. (a) Blue line corresponds to $\nu=1$, gray lines correspond to $\nu=2$, green lines correspond to $\nu=3$ and the brown ones correspond to $\nu =4$. (b) The spatial string tension $\sqrt{\sigma_s}$ for different  orientations for $\nu=4$.}
        \label{Sp-sigmaTnu4}
  \end{figure*}
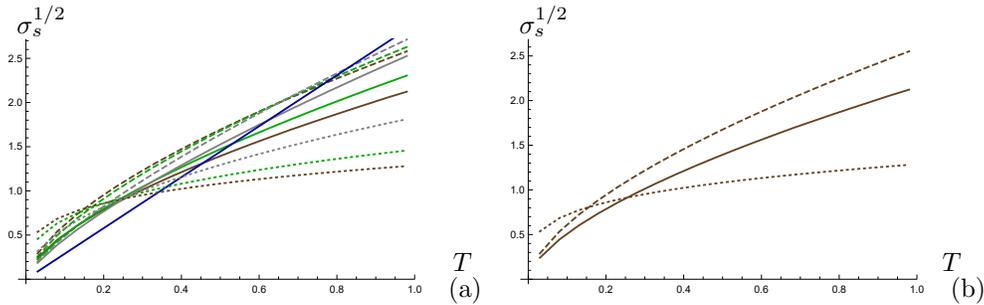

In Fig.\ref{Sp-sigmaTnu4} (the left panel)  we present the dependence of the spatial string tension $\sqrt{\sigma_s}$  
as a function of $T$ for all cases of the orientation and for $\nu=1,2,3,4$. We see, that for the configurations located on the $xy_{1}$-plane (partially longitudinal orientations shown by solid and dashed lines) the temperature dependence of the string tension for different $\nu$ are rather similar. The deviations of solid lines from dashed ones increase with increasing $T$. We also see that the  string tension corresponding to the Wilson loop in the $y_{1}y_{2}$-plane (the totally transversal orientation shown by the dotted lines) differs from the  behaviour of the Wilson loop including the longitudinal direction, showing less dependence on the temperature with  increasing $\nu$.  All these plots indicate that the structure of  chromomagnetic fields in our holographic model has strong dependence  on its orientation. In the right panel of Fig.\ref{Sp-sigmaTnu4} the spatial string tension $\sqrt{\sigma_s}$ for different orientations  for $\nu=4$ is presented.

We note that the behaviour of magnetic Wilson loops in heavy ion collisions
was worked out in \cite{Dumitru:2013koh,Dumitru:2014nka}  and is put in the context of the color glass condensate model. The universal behavior of a large magnetic Wilson loop was found  to have a nontrivial power-law dependence on the loop area.  They have also argued 
that in contrast to usual Coulomb phase behaviour, magnetic flux does not propagates uniformly in the transverse plane, but instead, it is concentrated in small domains.  In our work the Coulomb phases of pseudopotentials are modified for orientations different from the transversal one.

\subsection{The holographic light-like Wilson loops and jet-quenching}

One of the important characterizations of heavy-ions collisions is the jet quenching.
The jet-quenching parameter $\hat{q}$ introduced in \cite{Baier} is related with the average of the light-like Wilson loop in the adjoint representation
\cite{JQ}
\bea\label{WA}
W_A(C) = e^{-\frac{1}{4\sqrt 2}\hat{q} L_- \ell^{2}_{y_{1}}},
\eea
where $C$ is a rectangular contour with large extension $L_-$ in a light-like direction and small extension $\ell_{y_{1}}$ in a transversal one.
In the holographic approach $W_A(C)$ is equal to the classical string action $S$ of a string worldsheet  configuration stretched on 
  the contour $C$  on the boundary of the holographic background \cite{LRW}
\bea
W_A(C)=e^{2iS}.
\eea

In this section we focus on holographic light-like Wilson loops in the black brane background (\ref{Ll-bh}). Choosing  different light-like directions  we obtain the dependence of the jet quenching parameter  $\hat{q}$ on orientations.

Let us calculate the holographic light-like Wilson loop for the light-like coordinates related with the longitudinal direction, $x^{\pm} = \frac{t \pm x}{\sqrt{2}}$. The metric (\ref{Ll-bh}) in these light-cone coordinates takes the form
\bea\label{Q1}
ds^2=G_{--}(dx_+^2+dx_-^2)+G_{-+}dx_-dx_+ + G_{y_1y_1}dy_1^2+ G_{y_2y_2}dy_2^2+G_{zz}dz^2,
\eea
where
\bea\nn
&&G_{--}=\frac{1 - f(z)}{2z^{2}},\quad G_{-+}=\frac{1 + f(z)}{2z^{2}},\quad G_{y_1y_1}=G_{y_2y_2}=\frac{1}{z^{2/\nu}},\\
\label{Q1a}&& G_{zz}=\frac{1}{z^2f(z)}.
\eea
 After introducing the parametrization for the string worldsheet by coordinates
$\tau$ and  $\sigma$ such that
\begin{equation}
\tau = x^{-},\,\,\,\, \sigma=y_1,\,\,\,\,   z=z(y_1),
\end{equation}
 the string action takes the form
\bea\label{SLL2}
S = iL^{-} \int_{-\ell_{y_{1}}/2}^{\ell_{y_{1}}/2} d y_1\sqrt{G_{--} (G_{y_1y_1}+z'^2G_{zz})},
\eea
 where $i$ in front of the integral comes from the $\sqrt{-\det \gamma}$, since we deal with the Lorentz signature (here $\gamma$ is the induced metric on the string).
Note, that $G_{--}$ is positive definite function when $z<z_h$.
For small $\ell_{y_1}$ from this expression one gets \cite{1202.4436}
\bea\label{q-G}
\hat q^{-1}= \frac{1}{\sqrt{2}}\int_0^{z_h}\frac{ \sqrt{G_{zz}}}{
   \sqrt{G_{--}}G_{y_1y_1}} dz.
\eea
Evaluating the integral in \eqref{q-G} explicitly we get
\bea \label{qqq}
\hat q=-\frac{2^{\frac{2}{\nu }+2} \nu ^{\frac{\nu +2}{\nu }}  \pi ^{\frac{2}{\nu }-\frac{1}{2}}  \Gamma \left(-\frac{\nu }{2 \nu +2}\right)}{(\nu +1)^{\frac{2 (\nu +1)}{\nu }}\Gamma \left(1+\frac{1}{2 \nu +2}\right)}T^{\frac{\nu
   +2}{\nu }}.
\eea
  For $\nu=1$ formula \eqref{qqq} reproduces the $T^3$ dependence of the jet quenching parameter in the isotropic quark-gluon plasma \cite{LRW}, while in the case of $\nu>1$ the dependence on the temperature of the jet quenching parameter is caused by the anisotropic parameter.

Let us consider now  the contour $C$ with the light-like direction $y_1^{\pm} =\frac{ t \pm y_{1}}{\sqrt{2}}$. The metric 
in these coordinates has the form \eqref{Q1} with slightly different coefficients as compare to \eqref{Q1a}
\bea\label{Q2}
&&G_{--}=\frac{1}{2}\left(\frac{1}{z^{2/\nu}}-\frac{f}{z^{2}}\right),\quad G_{-+}=-\left(\frac{1}{z^{2/\nu} + \frac{f}{z^{2}}}\right),\quad\\&&\nn G_{x x}=\frac{1}{z^2},\quad G_{y_2y_2}=\frac{1}{z^{2/\nu}},\quad G_{zz}=\frac{1}{z^2f(z)}.
\eea
The small side of the contour  $C$ can be oriented in $x$ or  $y_2$ directions. In both cases the corresponding string actions have the form \eqref{SLL2} and contain the metric coefficient $G_{--}$ as in \eqref{Q2}. This  $G_{--}$   is non-positive definite below horizon $z_h$, $0<z<z_h$, for $\nu>1$. This makes the string action \eqref{SLL2} complex   that one can interpret as a suppression  of the jet quenching parameter.

\section{Spatial Wilson loops in a time-dependent background}
\label{Sect:5.2}

Now we move to consider the thermalization of rectangular Wilson loops in the Vaidya background (\ref{Vaidya-LL})-(\ref{f}), 
which describes collapsing geometry in the special anizotropic  spacetime \eqref{Lif-like0}.  
We proceed in a similar manner as in the static case studying three possible configurations of spatial Wilson loops.

\begin{figure}[h!]\centering
 \includegraphics[width=4.7cm]{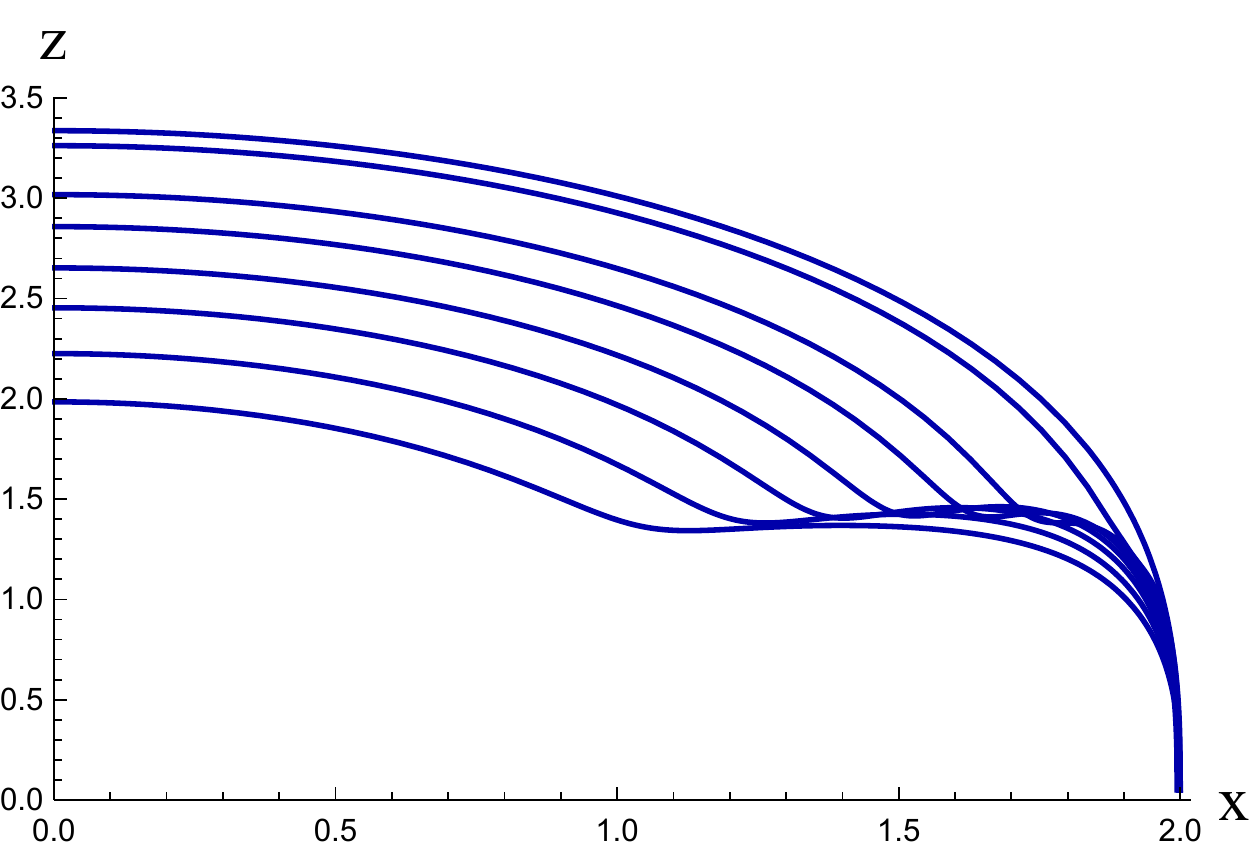}\,\,\,\,\,\,\,\,\,\,\,\,\,\,\,\,\,\,
  \includegraphics[width=4.7cm]{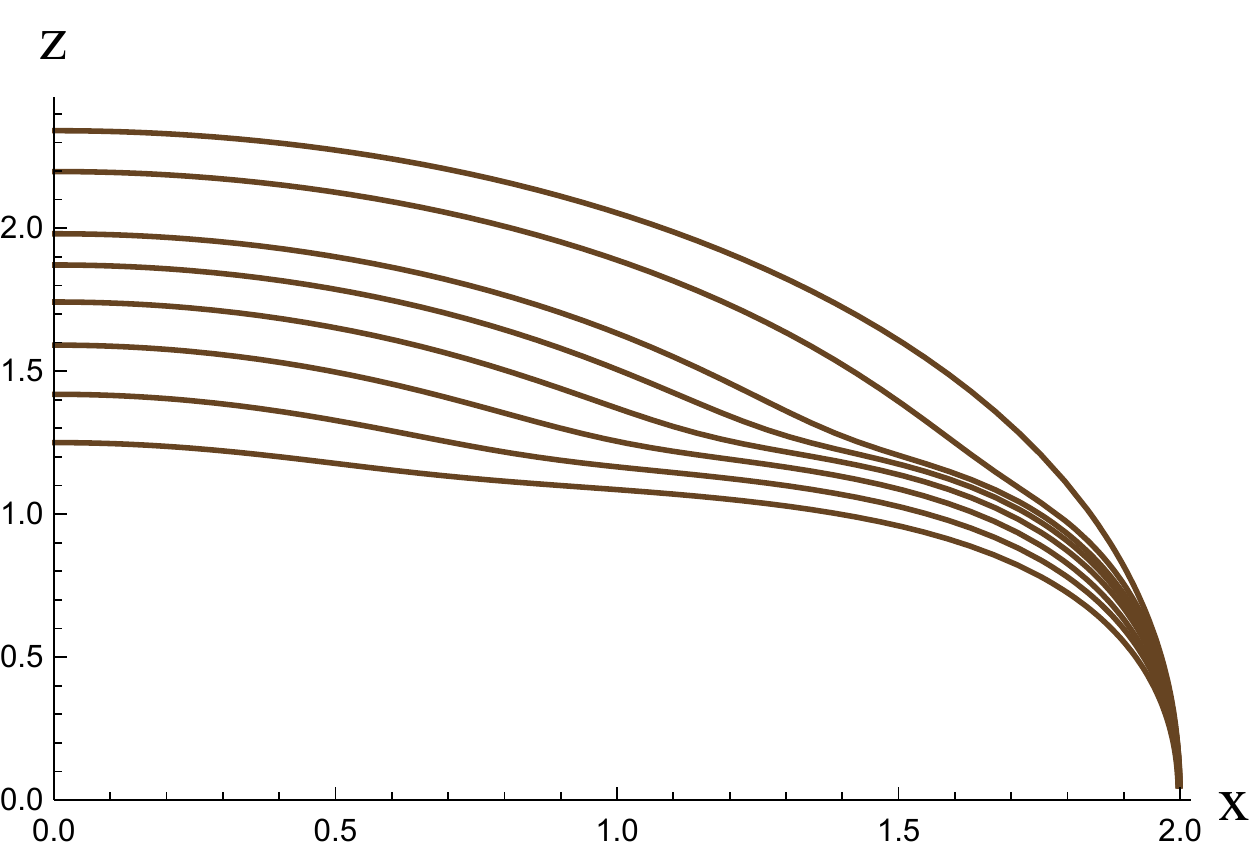}
        \caption{Profiles of the string $z(x)$ (with the boundary condition $z(2)=0$) at different moments of the boundary time  $\nu =1$ (left) and $\nu=4$ (right). In (\ref{f}) we take $m=1$.}
        \label{fig:prof-c1}
  \end{figure}

  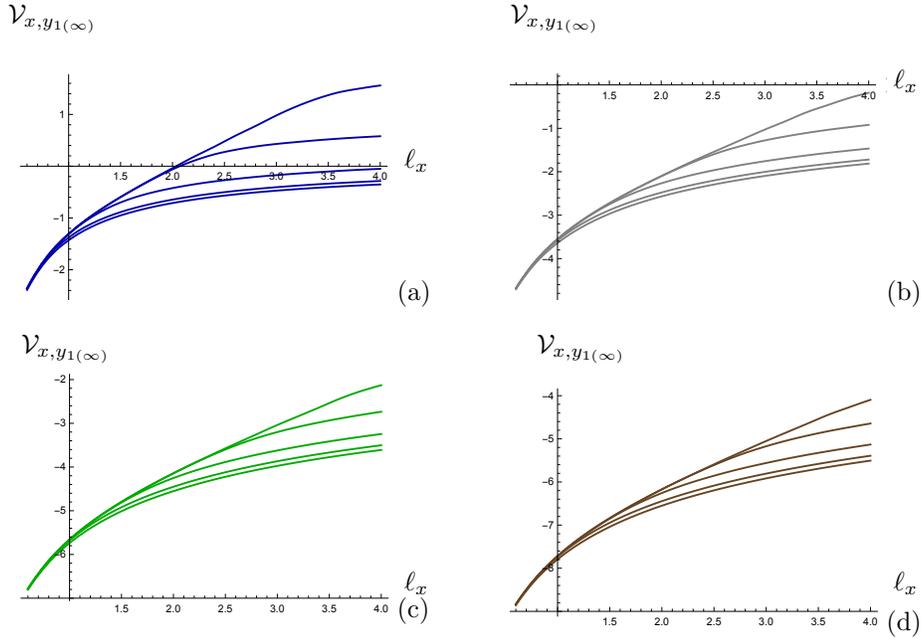
\begin{figure*}
\centering \begin{picture}(185,210)
\put(-75,140){\includegraphics[width=5.cm]{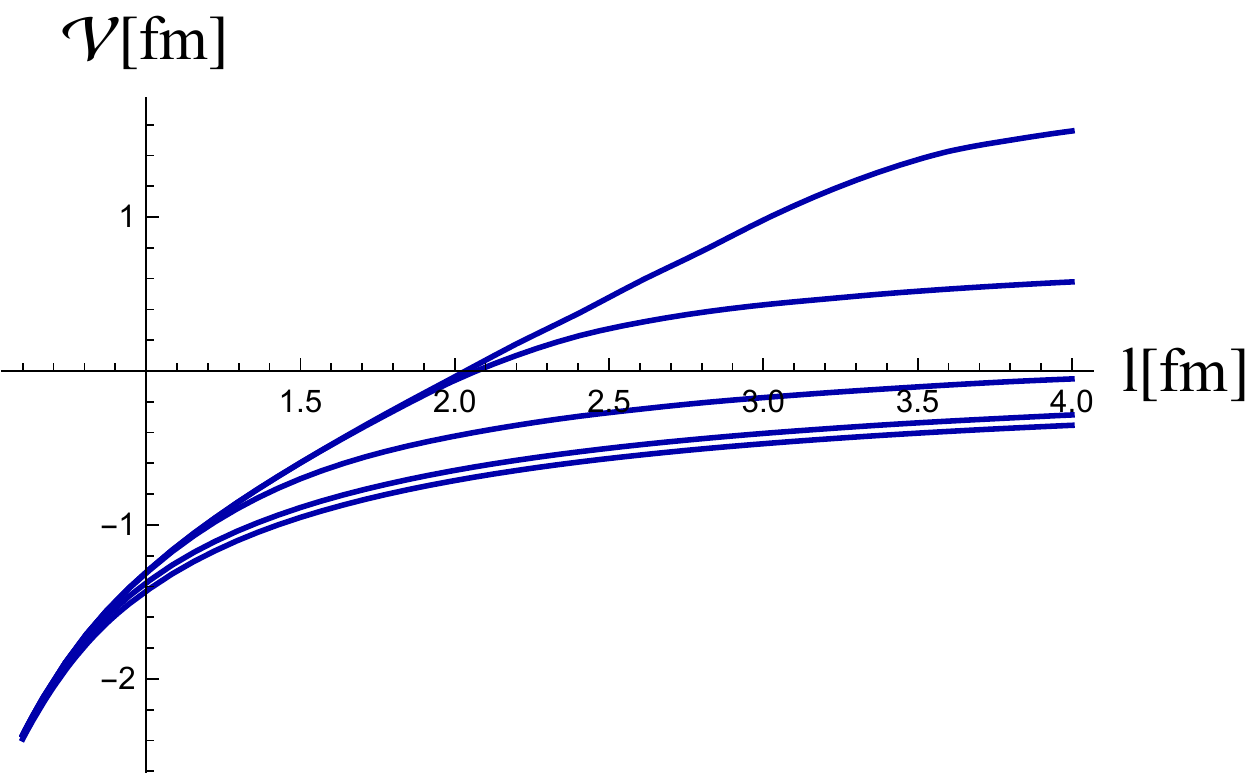}(a)}
 \put(-80,245){${\cal V}_{x,y_{1(\infty)}}$}
 \put(70,190){$\ell_x$}
\put(110,140){\includegraphics[width=5.cm]{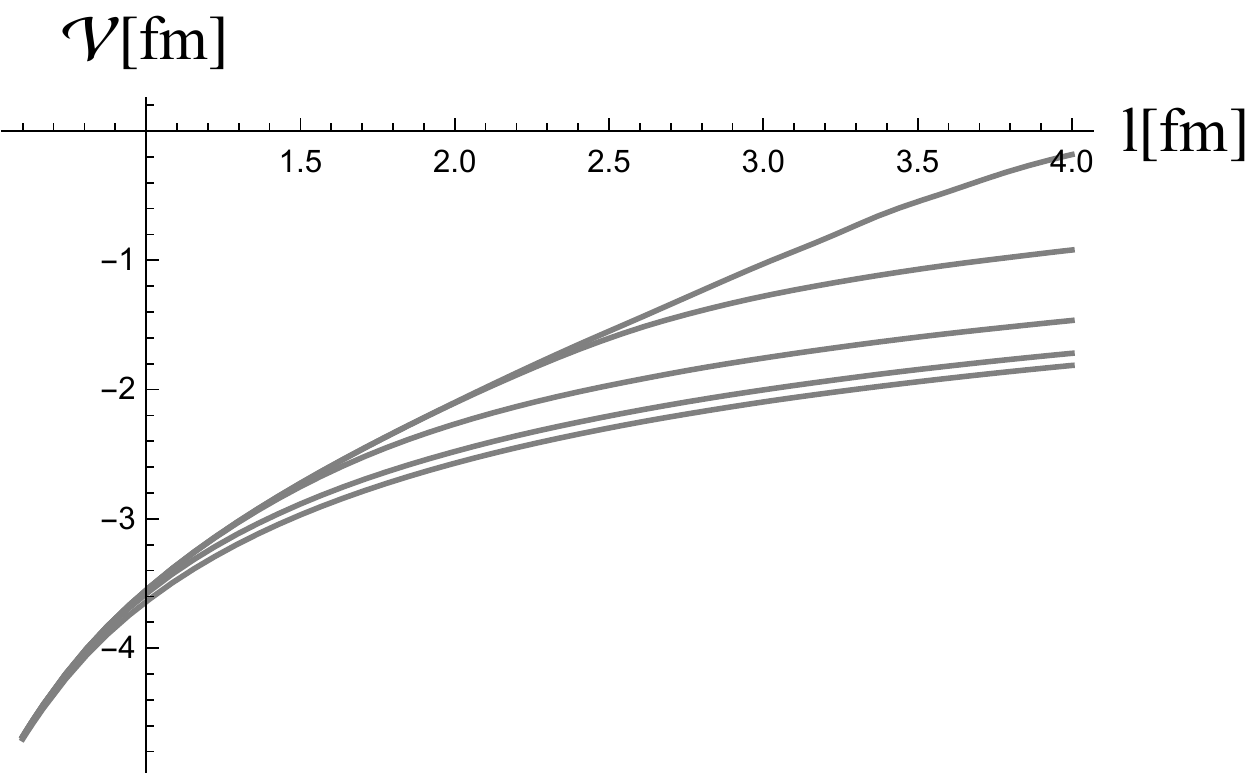}(b)}
 \put(110,245){${\cal V}_{x,y_{1(\infty)}}$}
 \put(255,220){$\ell_x$}
\put(-75,20) {\includegraphics[width=5.cm]{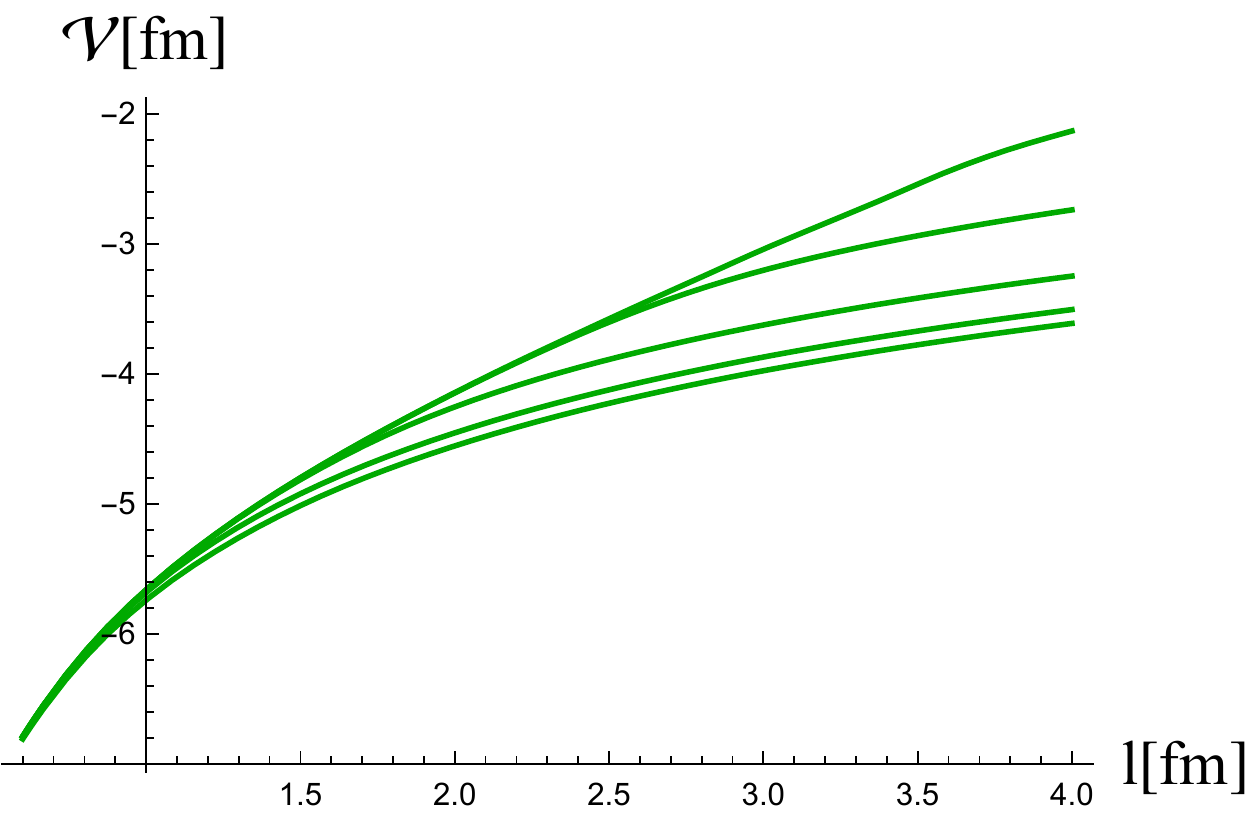}(c)}
 \put(-75,120){${\cal V}_{x,y_{1(\infty)}}$}
 \put(70,30){$\ell_x$}
\put(110,15){\includegraphics[width=5.cm]{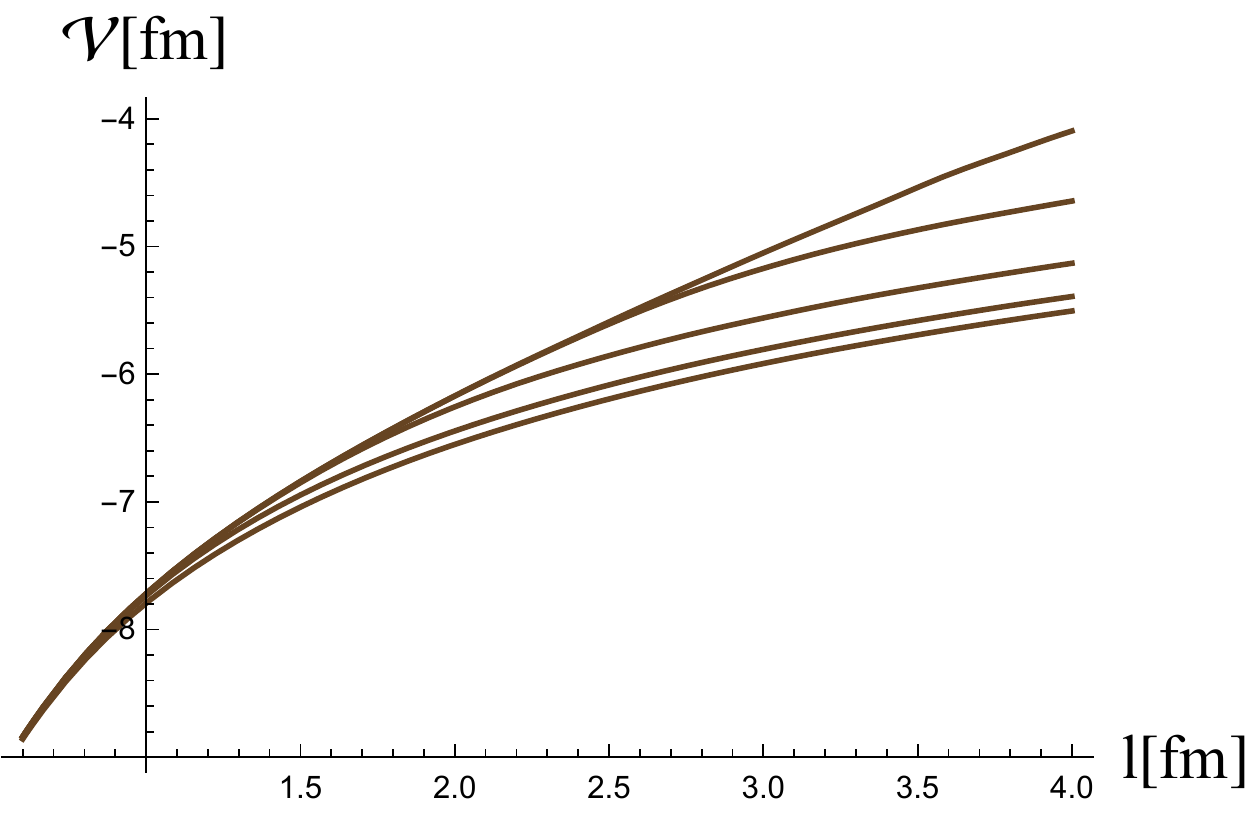}(d)}
 \put(120,120){${\cal V}_{x,y_{1(\infty)}}$}
 \put(255,30){$\ell_x$}
 \end{picture}\caption{The pseudopotential ${\cal V}_{x,y_{1(\infty)}}$ as a function of $\ell_{x}$ at fixed values of $t$ for $\nu=1,2,3,4$ ((a),(b),(c),(d), respectively). 
 Different curves correspond to time $t=0.1, 0.5, 0.9, 1.4, 2$ (from down to top, respectively). In (\ref{f}) we take $m=1$}. 
 \label{fig:pottherm1}
  \end{figure*}

\subsection{Wilson loops on the $xy_1$-plane}\label{Sect:5.2.1}

\subsubsection{Rectangular strip infinite along the $y_1$-direction}\label{Sect:5.2.1a}

{As in Sec.~\ref{Sect:WL-ind}  we start from the spatial rectangular Wilson loop on the $xy_1$-plane with the assumption that one side  of the loop
is infinite along the $y_1$-direction and the other has finite size along  the $x$-direction (see (\ref{C1xy1})).
Here we suppose the dependence  $v = v(x)$, $z = z(x)$.
The Nambu-Goto action takes the form similar to (\ref{7.2})
\begin{eqnarray}\label{7.2m}
S_{x,y_{1(\infty)}} =  L_y \int   \frac{dx}{z^{1+\frac1{\nu}}}\sqrt{1 - f(z,v) v^{\prime\,2} - 2v'z'},
\end{eqnarray}
 but with the time-dependent blackening function $f=f(z,v)$. 
The corresponding equations of motion are
\begin{eqnarray}\label{eq11}
v'' &=&  \frac{1}{2}\frac{\partial f}{\partial z}v'^{2} + \frac{(\nu + 1)}{\nu z}(1 - fv'^{2} - 2v'z'),\\\nn \label{eq12}
z''& =& - \frac{\nu+1}{\nu}\frac{f}{z} + \frac{\nu +1}{\nu}\frac{f^{2}v'^{2}}{z} - \frac{1}{2}\frac{\partial f}{\partial v}v'^{2} - \frac{1}{2} f v'^{2}\frac{\partial f}{\partial z} - v' z' \frac{\partial f}{\partial z} + 2\frac{(\nu +1)}{\nu z}f v'z',
\end{eqnarray}
 which for $\nu = 1$ coincide with  the Vaidya-AdS  equations \cite{1103.2683}.

We have to consider eqs.(\ref{eq11})  with the following boundary conditions
$z(\pm \ell_{x}/2)=0,\,\,\,\,v(\pm \ell_{x}/2)=t$,
where $\ell_x$ is the length of the Wilson loop along the $x$-direction.
To solve numerically the equations of motion (\ref{eq11}), it is convenient to  impose the initial conditions
$ z(0)=z_*, v(0) =v_*, z'(0)= 0,  v'(0)=0$.

Fig. \ref{fig:prof-c1} shows the typical behaviour of the solutions to eqs.\eqref{eq11} which satisfy the boundary conditions for different values of the critical exponent $\nu$.   In these pictures we observe the evolution of string profiles during the formation of the black brane horizon by the infalling shell with $m=1$.

For a given solution ($v(x)$, $z(x)$) to eqs.(\ref{eq11})  we can compute the functional for the Nambu-Goto action (\ref{7.2m}).
We note that the dynamical system governed by (\ref{7.2m}) has the following integral of motion
\be\label{IMC1}
\mathcal{J} = - \frac{1}{z^{1+1/\nu}\sqrt{\mathcal{R}}},
\ee
where we denote
\be\label{IMRC1}
\mathcal{R}=1 - f v'^{2} - 2v' z'.
\ee

Taking into account (\ref{IMC1})-(\ref{IMRC1}) one can represent (\ref{7.2m}) in the following form
\be
\label{SS1}
S_{x,y_{1(\infty)}}=  L_y \int^{l_{x}}_{0} \frac{dx}{z^{1+1/\nu}}\left(\frac{z_{*}}{z}\right)^{1+ 1/\nu},
\ee
where $z_{*}$ is the turning point defined from the requirements $z' = v' = 0$ and related with $\mathcal{J}$ as
$z^{1/\nu +1}_{*} = \mathcal{J}^{-1}$.

Coming to integration with respect to the $z$-variable the expression \eqref{SS1} can be represented as
\begin{eqnarray}\label{V1a}
S_{x,y_{1(\infty)},ren}= -  L_{y_1} \int^{z_{*}}_{z_{0}} \frac{\mathfrak{b}(z) }{z^{1 + 1/\nu}} dz ,
\end{eqnarray}
where $\mathfrak{b}$  is defined by
\be\label{b-c1}
\mathfrak{b}(z) = \frac{1}{z'}\left(\frac{z_{*}}{z}\right)^{1 +\frac1{\nu}}.
\ee

One can  observe that $\mathfrak{b}(z)$ tends to be $-1$ for $z \rightarrow 0$ and 
divergencies at $z=0$ are similar to the static configuration and  by this reason we put the regularization $z_0$. 
Making a substraction we come to the renormalized version of $S_{x,y_{1(\infty)}}$,  in which we can remove the regularization
\begin{eqnarray}\label{V1}
S_{x,y_{1(\infty)},ren}= -  L_{y_1} \left(\int^{z_{*}}_{z_{0}} \frac{[\mathfrak{b}(z) - \mathfrak{b}(z_{0})]}{z^{1 + 1/\nu}} dz - \nu \frac{\mathfrak{b}(z_{0})}{z^{1/\nu}_{*}}\right),
\end{eqnarray}

The pseudopotential is expressed as :
\be\label{7.4a-V-td}
{\cal V}_{x,y_{1(\infty)}}=\frac{S_{x,y_{1(\infty)},ren}}{L_{y_{1}}}.
\ee
In Fig.~\ref{fig:pottherm1}  we present the behaviour of the renormalized pseudopotential ${\cal V}_{x,y_{1(\infty)}}$ derived from the action \eqref{V1}  
as a function of $\ell_{x}$ at fixed time moments for different values of $\nu$. We see that for small distances the pseudopotential behaves  
similarly for different values of  $t$. This dependence strengthens with increasing $\nu$.
For large times we see that the pseudopotential equilibrates to its thermal value.
 The pseudopotential reaches saturation for enough large size of the strip. 
 The value of the thermalization time  grows with increasing $\ell_{x}$  for all values of $\nu$. At the same time, we observe that the saturation is reached faster for large $\nu$.

  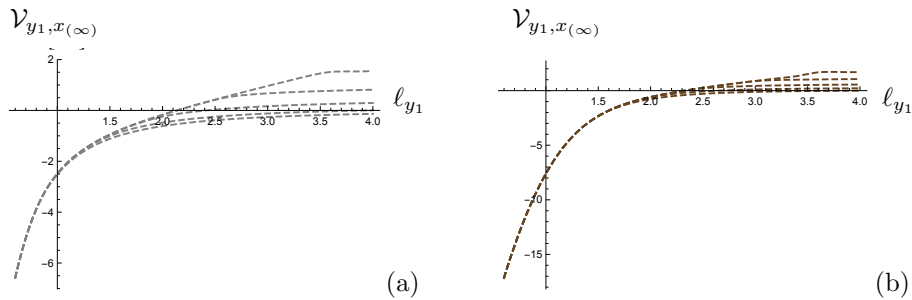
\begin{figure*}[h!]
\centering \begin{picture}(55,120)
\put(-140,0){\includegraphics[width=5cm]{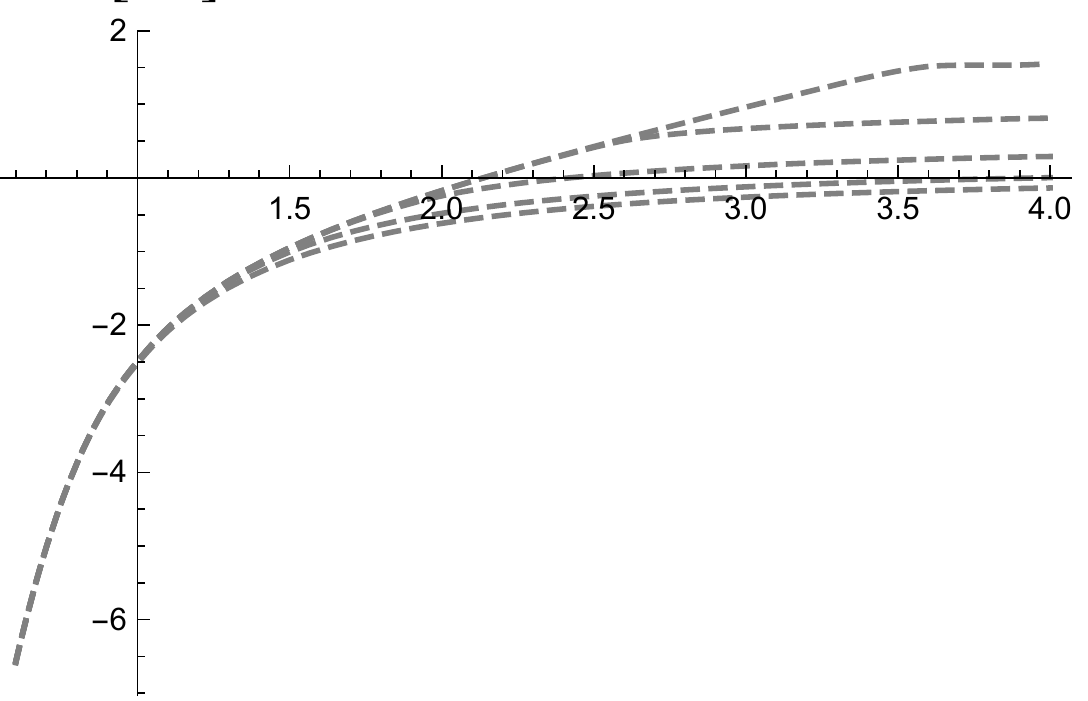}(a)}
 \put(-140,100){${\cal V}_{y_{1},x_{(\infty)}}$}
 \put(5,70){$\ell_{y_{1}}$}

\put(45,0) {\includegraphics[width=5cm]{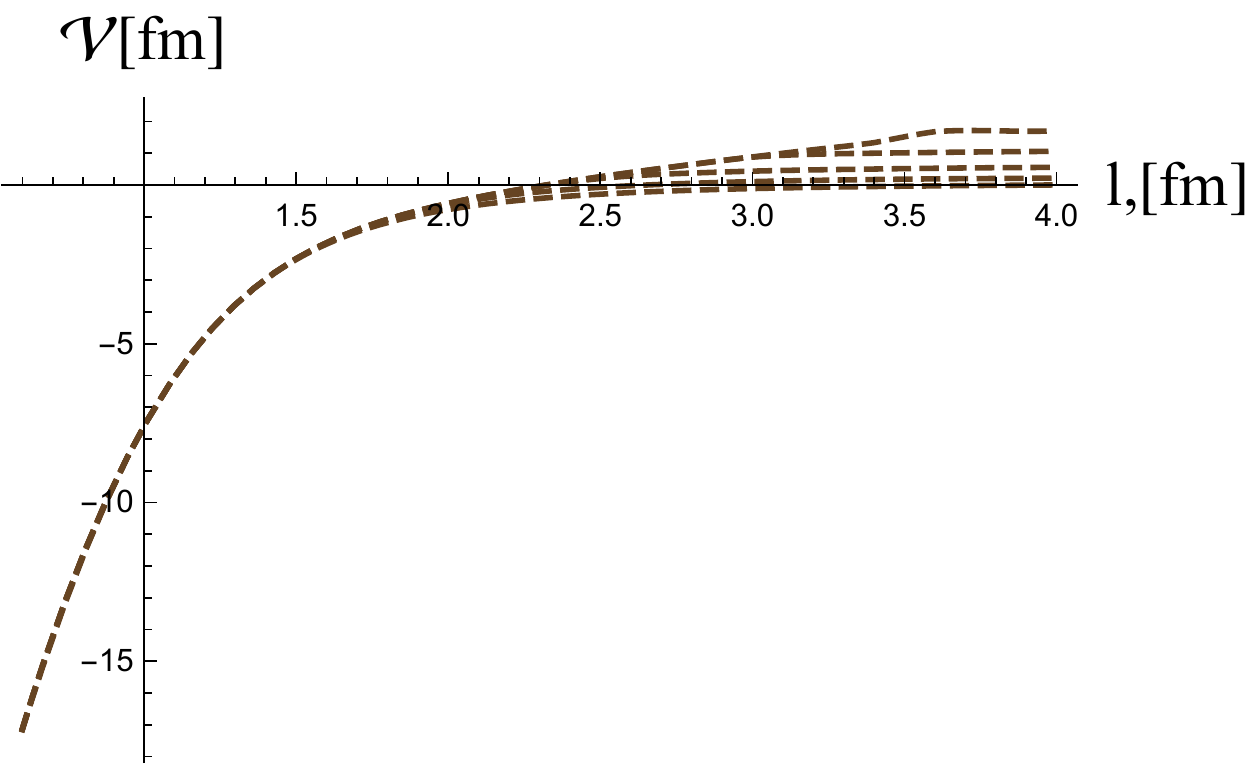}(b)}
 \put(50,100){${\cal V}_{y_{1},x_{(\infty)}}$}
  \put(190,70){$\ell_{y_{1}}$}
\end{picture}
        \caption{The pseudopotential  ${\cal V}_{y_{1},x_{(\infty)}}$ as a function of $\ell_{y_{1}}$ at fixed values of $t$ for $\nu=2,4$ ((a),(b) respectively).  The plot for $\nu=1 $ is the same as in Fig.\ref{fig:pottherm1}.a. Different curves correspond to $t=0.1, 0.5, 0.9, 1.4, 2$ from down to top. In (\ref{f}) we take $m=1$.}
        \label{fig:BHl2}
  \end{figure*}

\subsubsection{Rectangular strip infinite along the $x$-direction}\label{Sect:5.2.2}

Now we consider the rectangular Wilson loop on the $xy_1$-plane with the assumption that its contour is infinite along the $x$-direction while it has finite stretch along the $y_1$-direction (see (\ref{C2xy1})). As in the previous section, we specify this type of the strip
by the subscript $y_{1},\,x_{(\infty)}$. Thus, the corresponding Nambu-Goto action can be represented by
\begin{eqnarray}\label{7.5}
\frac{S_{y_{1},\,x_{(\infty)}}}{ L_{x}} =\int   \frac{dy_{1}}{z^2}\sqrt{\frac{1}{z^{\frac2{\nu}-2}} - f(z,v) v^{\prime 2} - 2v'z'}, 
\end{eqnarray}
with the notation  $\prime \equiv \frac{d}{d y_{1}}$.
The time-independent analogue of  (\ref{7.5}) is given by (\ref{7.5-2}).

The equations of motion following from (\ref{7.5}) are
\begin{eqnarray}
v''&= & \frac{1}{2}\frac{\partial f}{\partial z} v'^{2} + \frac{\nu +1}{\nu z}\left(z^{2 -2/\nu} - \frac{2\nu}{(1 + \nu)} fv'^{2} -2v'z' \right)\!,
\label{eqc2.1}\\\nn
z''&= & - \frac{\nu +1}{\nu}f z^{1 - 2/\nu} + \frac{2(\nu -1)z'^{2}}{\nu} + \frac{2}{\nu}\frac{f^{2}v'^{2}}{z} -\nn \frac{1}{2\nu}\frac{\partial f}{\partial v}v'^{2}\\&&\nn -  \frac{1}{2\nu}f\frac{\partial f}{\partial z}v'^{2} 
  - z'v'\frac{\partial f}{\partial z} + \frac{4}{z}fz'v'.  \label{eqc2.2}
\end{eqnarray}
It is worth to be noted that  eqs.(\ref{eqc2.1})  match with (\ref{eq11}) taken with $\nu =1$ and also reproduce those for the AdS-case.

The boundary conditions for eqs.(\ref{eqc2.1}) read
$z(\pm \ell_{y_{1}}/2)=0,\,\,\,\,v(\pm \ell_{y_{1}}/2)=t$,
where $\ell_{y_{1}}$ is the length of the Wilson loop along the $y_{1}$-direction.

 As in the previous case the action (\ref{7.5}) can be simplified on equations of motions. For this purpose we note that the dynamical system governed  by action \eqref{7.5} has
the integral of motion 
\be\label{J-C2}
\mathcal{J} = - \frac{1}{z^{2/\nu}\sqrt{\mathcal{R}}},
\ee
where
\be\label{R-TD-C2}
\mathcal{R}=\frac{1}{z^{2/\nu - 2}} - f v'^{2} - 2v' z'.
\ee
 
Taking into account (\ref{J-C2})-(\ref{R-TD-C2}) the action (\ref{7.5})  is represented in the form
\be\label{S-c2}
S_{y_{1},\,x_{(\infty)}} = L_{x} \int^{l_{y_{1}}}_{0} dy_{1} \frac{z_{*}^{1/\nu+1}}{z^{2/\nu+2}},
\ee
with the turning point  $z_{*}$   related with $\mathcal{J}$ as
$z^{1/\nu+1}_{*} = \mathcal{J}^{-1}$.

As in the previous case we present $S_{y_{1},\,x_{(\infty)}}$ as 
\begin{eqnarray}\label{SR-C2a}
S_{y_{1},\,x_{(\infty)}} = - L_{x} \int^{z_{*}}_{z_0} \frac{\mathfrak{b}(z) }{z^{2}}dz  ,
\end{eqnarray}
where 
$\mathfrak{b}$ is defined by

\be \label{frakbc2}
\mathfrak{b}(z) = \frac{1}{z'}\left(\frac{z_{*}^{1+1/\nu}}{z^{2/\nu}}\right).
\ee
Performing renormalization we come to
\begin{eqnarray}\label{SR-C2}
S_{y_{1},\,x_{(\infty)},ren} = - L_{x} \int^{z_{*}}_{0} \frac{\mathfrak{b}(z) - \mathfrak{b}(z_{0})}{z^{2}}dz  - \frac{\mathfrak{b}(z_{0})}{z_{*}}.
\end{eqnarray}

The renormalized pseudopotential derived from the action (\ref{SR-C2}) as a function of $\ell_{y_{1}}$  for different values of $t$ and $\nu$ is demonstrated in Fig.~\ref{fig:BHl2}.
Here we again observe that for small $\ell_{y_{1}}$ the behavior of the pseudopotential is similar for different values of $t$. 
As in the previous case the behavior intensifies with the increasing value of the dynamical exponent.
However, in contrast to the previous case there is no substantial dependence, on a given scale, of the thermalization time on the dynamical exponent. 
 Comparing Fig.\ref{fig:pottherm1} and Fig.\ref{fig:BHl2} one can see that for the same scale  $\ell_{y_{1}}$  the thermalization of the Wilson loop occurs faster for the configuration with a long extent in the $y_{1}$-direction.
The dependence on the dynamical exponent $\nu$ is also stronger for the latter case of the orientation.


\subsection{Wilson loop on the $y_1y_2$-plane}\label{Sect:5.2.3}
Finally, we come to the configuration  located on the $y_1y_2$-plane. 
We assume that this infinite rectangular strip is invariant  along the $y_2$-direction, see (\ref{C3y1y2}).
As in Sect.~\ref{Sect:5.1.2} we use $y_{1},y_{2,(\infty)}$ for the subscript of the action

 \begin{eqnarray}\label{7.7a}
\frac{S_{y_{1},\,y_{2,(\infty)}} }{ L_{y_{2}} }=\int  \frac{dy_{1}}{z^{1+\frac{1}{\nu}}}\sqrt{\frac{1}{z^{\frac2{\nu}-2}} - f v^{\prime 2} - 2v'z'}, 
\end{eqnarray}
where we define $\prime \equiv \frac{d}{d y_{1}}$.

The equations of motion corresponding to (\ref{7.7a}) can be written down in the following form
\bea\label{eqc3.1}
v'' &= &\frac{1}{2}\frac{\partial f}{\partial z}v'^{2} + \frac{2}{z\nu}\left(z^{2 - \frac2{\nu}} - \frac{\nu +1}{2}fv'^{2} - 2v'z' \right),\\\nn 
z''&=& -\frac{2}{\nu}fz^{1 - 2/\nu} + 2\frac{\nu -1}{\nu}\frac{z'^{2}}{z} + \frac{\nu +1}{\nu z}f^{2}v'^{2} 
 - \frac{1}{2}\frac{\partial f}{\partial v} v'^{2} - \frac{1}{2}f\frac{\partial f}{\partial z} v'^{2} \\&&\nn
- z'v'\frac{\partial f}{\partial z} + \frac{2(\nu +1)}{\nu z} f v' z'.
\eea
 One can check that eqs. (\ref{eqc3.1}) coincide with (\ref{eq11}) and (\ref{eqc2.1}) for $\nu =1$ 
as well as come to be the equations for the AdS case.
The boundary conditions to be satisfied by eqs.(\ref{eqc3.1}) have the standard form
$z(\pm \ell_{y_{1}}/2)=0,\,\,\,\,v(\pm \ell_{y_{1}}/2)=t$,
where $\ell_{y_{1}}$ is the length of the Wilson loop along the $y_{1}$-direction. }
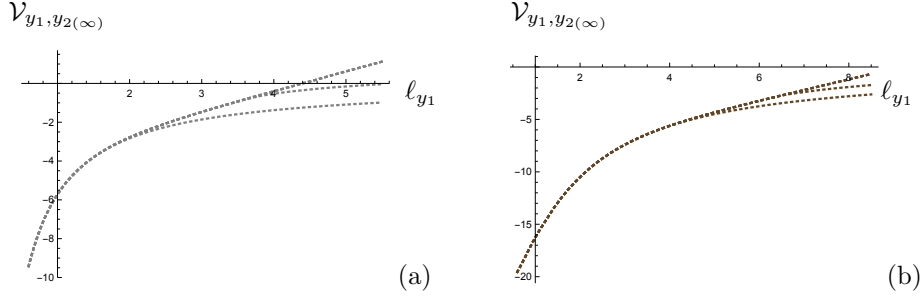
\begin{figure*}[h!]\centering
\begin{picture}(165,120)
\put(-85,0){\includegraphics[width=5cm]{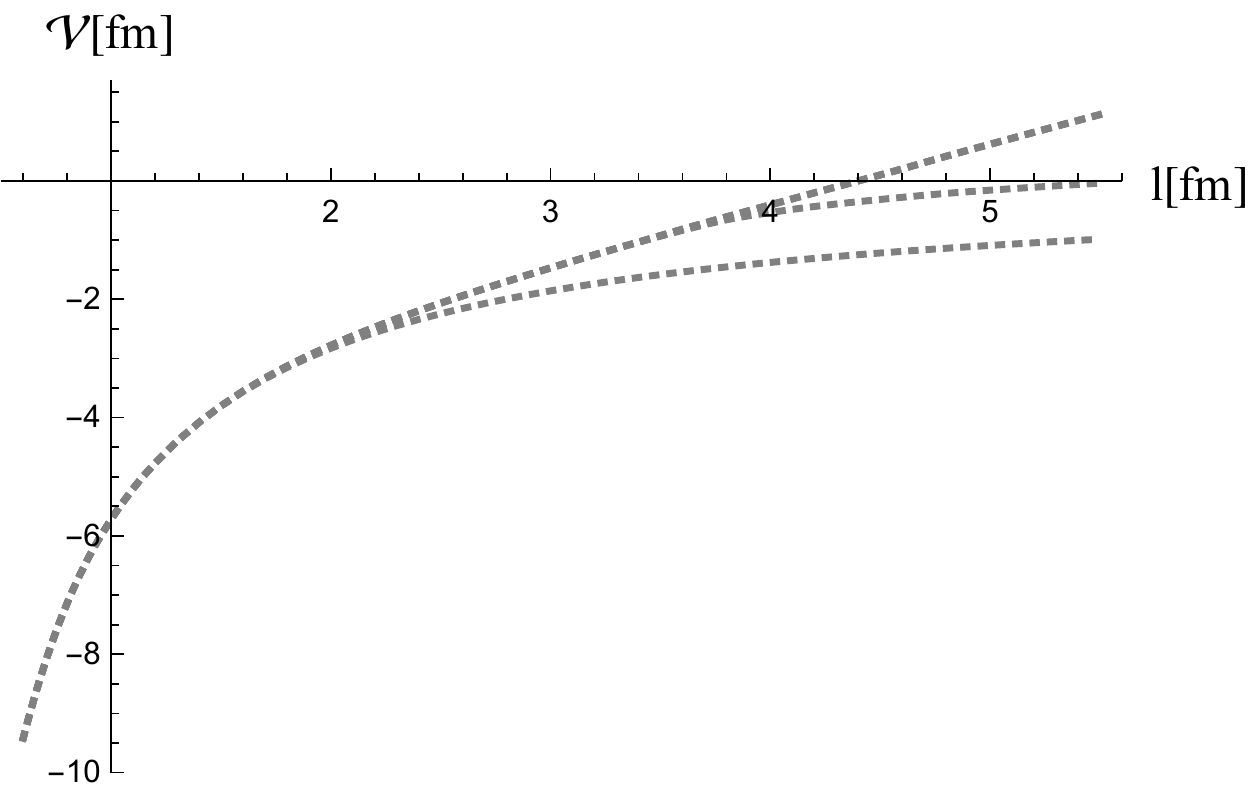}(a)}
 \put(-90,100){${\cal V}_{y_{1},y_{2(\infty)}}$}
 \put(60,70){$\ell_{y_{1}}$}

\put(100,0) {\includegraphics[width=5cm]{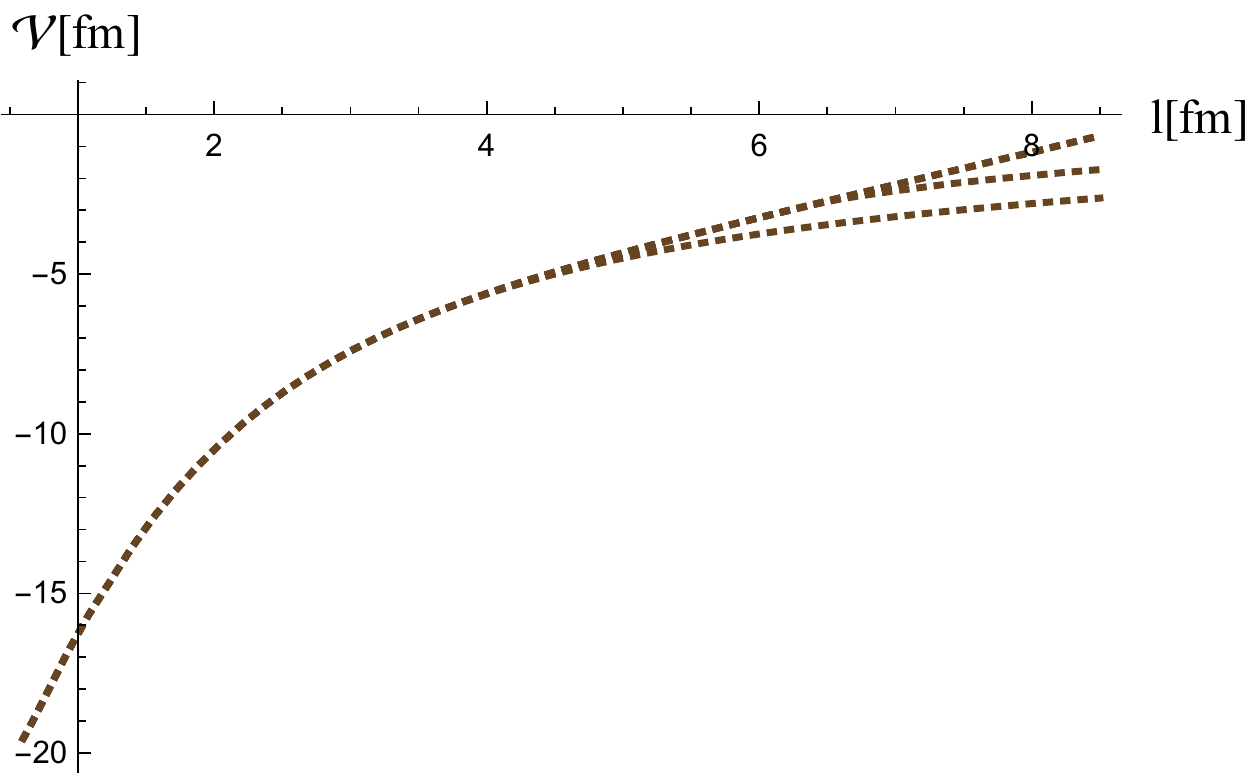}(b)}
 \put(100,100){${\cal V}_{y_{1},y_{2(\infty)}}$}
  \put(240,70){$\ell_{y_{1}}$}
\end{picture}
        \caption{The pseudopotential ${\cal V}_{y_{1},\,y_{2(\infty)}}$ as a function of the length $\ell_{y_{1}}$ at fixed values of $t$, $\nu=2,4$
        ((a),(b) respectively). (a): we take $t=0.1, 0.5, 0.9,1.4, 2$ from down to top, respectively; for plots (b): $t=0.4, 1.5, 2.5, 3.34, 4$ from down to top, respectively.  In (\ref{f}) we take $m=1$.}
        \label{fig:BHl3}
  \end{figure*}
  
Following our strategy we compute the functional (\ref{7.7a}) on a given solution to (\ref{eqc3.1}).
 We note that the integral of motion for the configuration governed by the action (\ref{7.7a}) reads
\begin{equation}\label{R-TD-C3}
\mathcal{J} = - \frac{1}{z^{\frac{3}{\nu}- 1}\sqrt{\mathcal{R}}},
\end{equation}
where we define
\begin{eqnarray}
\mathcal{R} = \frac{1}{z^{2/\nu-2}} - f (v')^{2} - 2v'z'.
\end{eqnarray}

Plugging (\ref{R-TD-C3}) into (\ref{7.7a}) we come to the following form for the functional of the Nambu-Goto action
\be\label{S-c3}
S_{y_{1},\,y_{2,(\infty)}} = L_{y_{2}} \int_0^{l_{y_{1}}} dy_1 \frac{z^{2/\nu}_{*}}{z^{4/\nu}},
\ee
where the turning point $z_{*}$ is related with $\mathcal{J}$ as $z^{2/\nu}_{*} = \mathcal{J}^{-1}$.
Changing the variable of integration from $y_1$ to $z$ we get
\begin{eqnarray}\label{SR-NG-C3m}
S_{y_{1},\,y_{2,(\infty)}}= -  L_{y_{2}} \int^{z_{*}}_{z_{0}} \frac{\mathfrak{b}(z)}{z^{1 + 1/\nu}} dz ,
\end{eqnarray}
where 
\be\label{frakbc3}
\mathfrak{b}(z) = \frac{1}{z'}\left(\frac{z_{*}^{2/\nu}}{z^{3/\nu-1}}\right).
\ee
The renormalized action in terms of the $z$-variable reads
\begin{eqnarray}
\frac{S_{y_{1},\,y_{2,(\infty)},ren}}{L_{y_{2}} }= -  \int^{z_{*}}_{z_{0}} \frac{\mathfrak{b}(z) - \mathfrak{b}(z_{0})}{z^{1 + 1/\nu}} dz + \nu \frac{\mathfrak{b}(z_{0})}{z^{\frac1{\nu}}_{*}}.\nn\\\label{SR-NG-C3}
\end{eqnarray}

Here the pseudopotential is expressed from \eqref{SR-NG-C3} as:
\be
{\mathcal V}_{y_{1},\,y_{2(\infty)}}=\frac{S_{y_{1},\,y_{2(\infty)},ren}}{L_{y_{2}}}.
\ee

The dependence of the pseudopotential ${\mathcal V}_{y_{1},\,y_{2(\infty)}}$ on the length $\ell_{y_{1}}$ is shown in Fig.\ref{fig:BHl3}.
As for the previous configurations of Wilson loop located on the $xy_{1}$-plane, the pseudopotential $\mathcal{V}_{y_{1},\,y_{2(\infty)}}$  
tends to its thermal value for large $t$. 
We note that the influence of the critical exponent on the rate of the thermalization process for ${\mathcal V}_{y_{1},\,y_{2(\infty)}}$ is  even higher than for ${\cal V}_{x,\,y_{1(\infty)}}$ and ${\cal V}_{y_1,\,x_{(\infty)}}$.
In all these cases the saturation time increases with $\ell_{y_{1}}$ and $\nu$.


\section{Thermalization times}\label{Sect:6}
\subsection{Thermalization times of spatial Wilson loops}\label{Sect:6a}

\begin{figure*}[h!]
\centering \begin{picture}(185,160)
\put(-85,20){ \includegraphics[width=4.5cm]{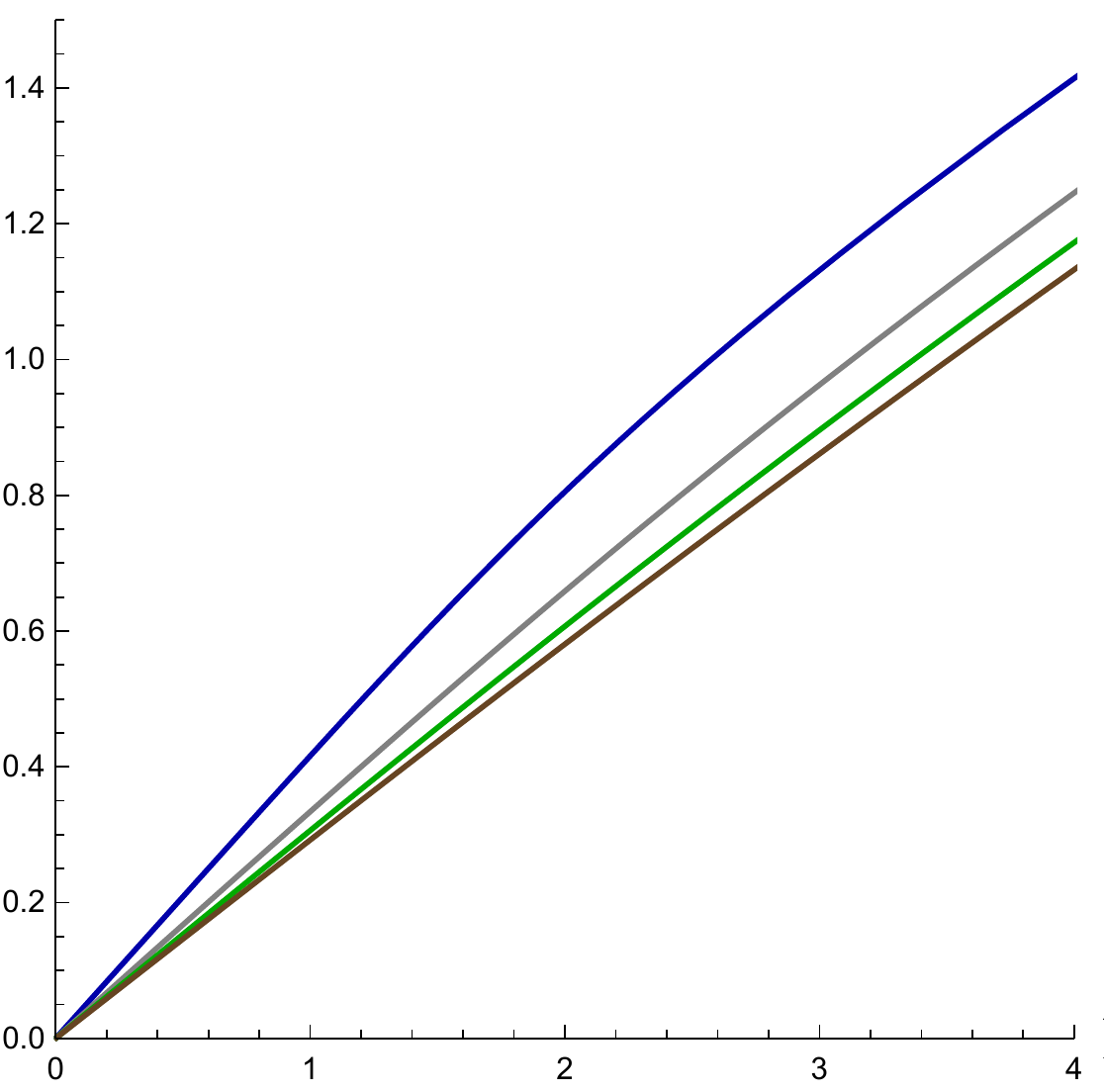}\,\,\,\,\,\,\,\,\,\,\,\,\,\,\,\,\,\,\,\,\,\,(a)}
 \put(-85,150){$t_{therm}$}
 \put(45,25){$l_{x}$}
\put(120,20){\includegraphics[width=4.5cm]{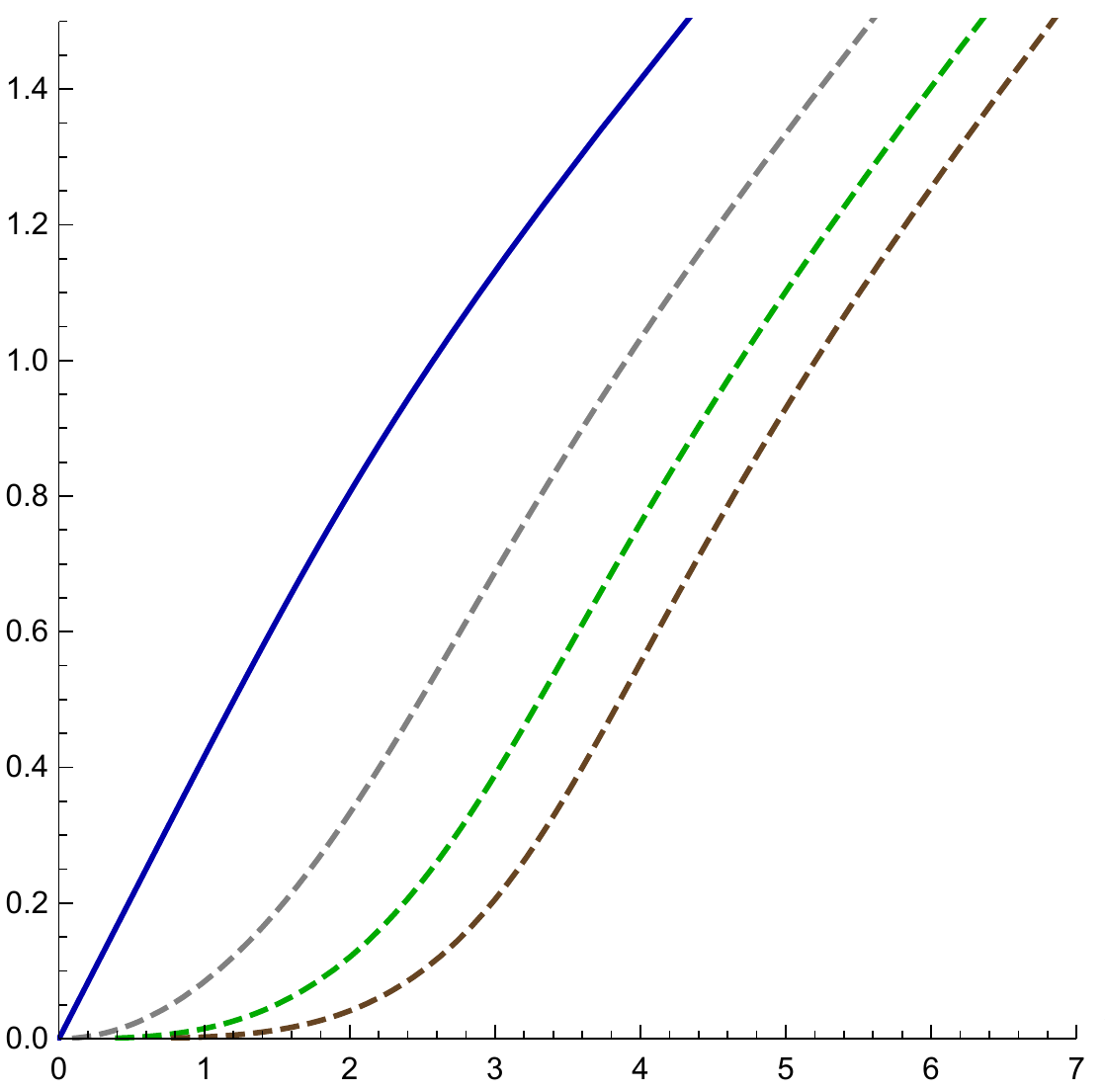}\,\,\,\,\,\,\,\,\,\,\,\,\,\,\,\,\,\,\,\,\,\,(b)}
 \put(110,150){$t_{therm}$}
 \put(250,25){$l_{y_{1}}$}
 \end{picture}
           \caption{The thermalization time for the Wilson loop in the $xy_{1}$-plane with a short extent in the $x$- and $y$-directions ((a) and (b), respectively).
            Different curves correspond to different values of $\nu=1,2,3,4$ (from top to down for each plot). }
        \label{fig:thermc1c2}
  \end{figure*}
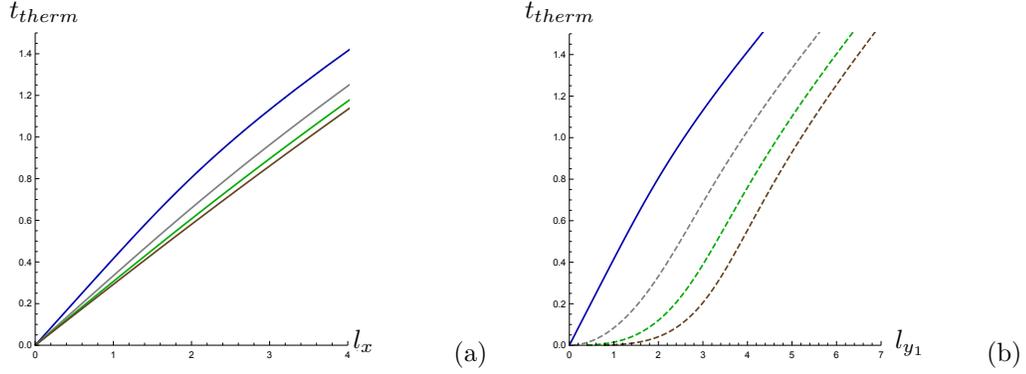

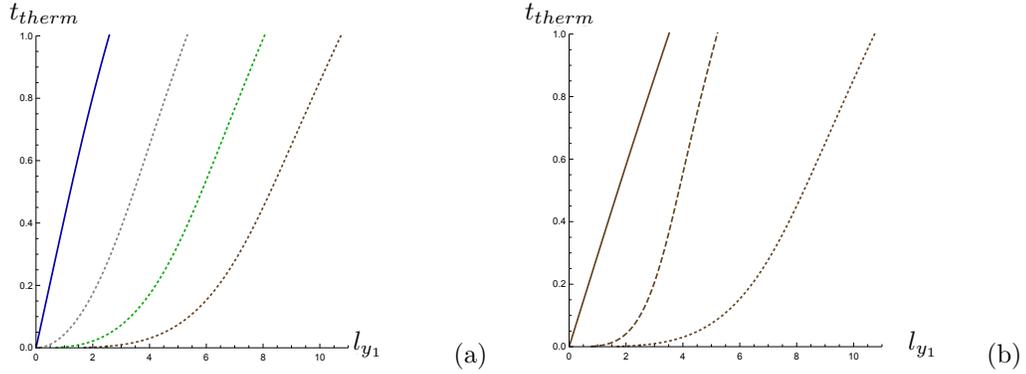
\begin{figure*}[h!]
\centering \begin{picture}(185,170)
\put(-85,20){ \includegraphics[width=4.5cm]{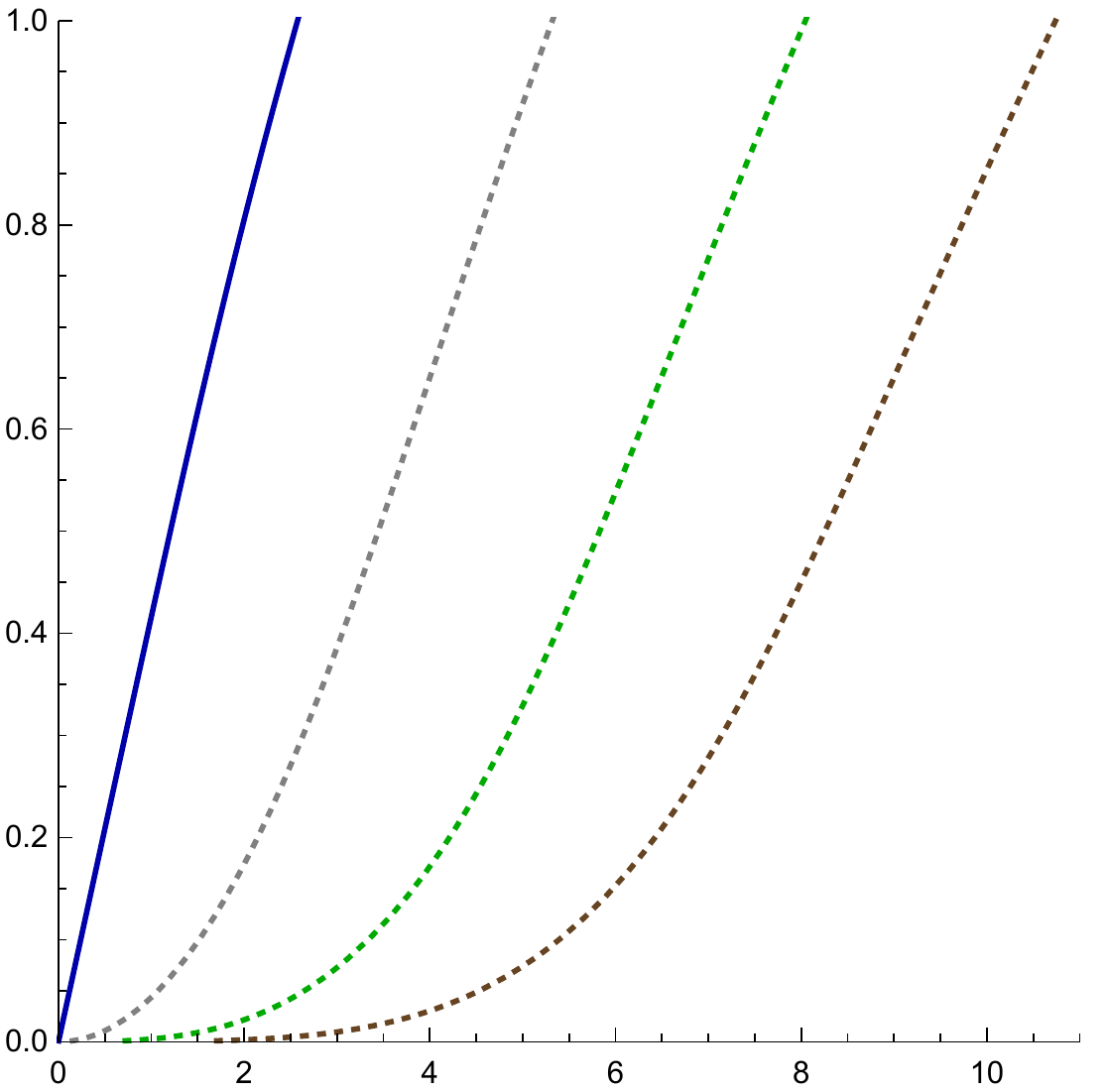}\,\,\,\,\,\,\,\,\,\,\,\,\,\,\,\,\,\,\,\,\,\,(a)}
 \put(-85,150){$t_{therm}$}
 \put(45,25){$l_{y_{1}}$}
\put(120,20){\includegraphics[width=4.5cm]{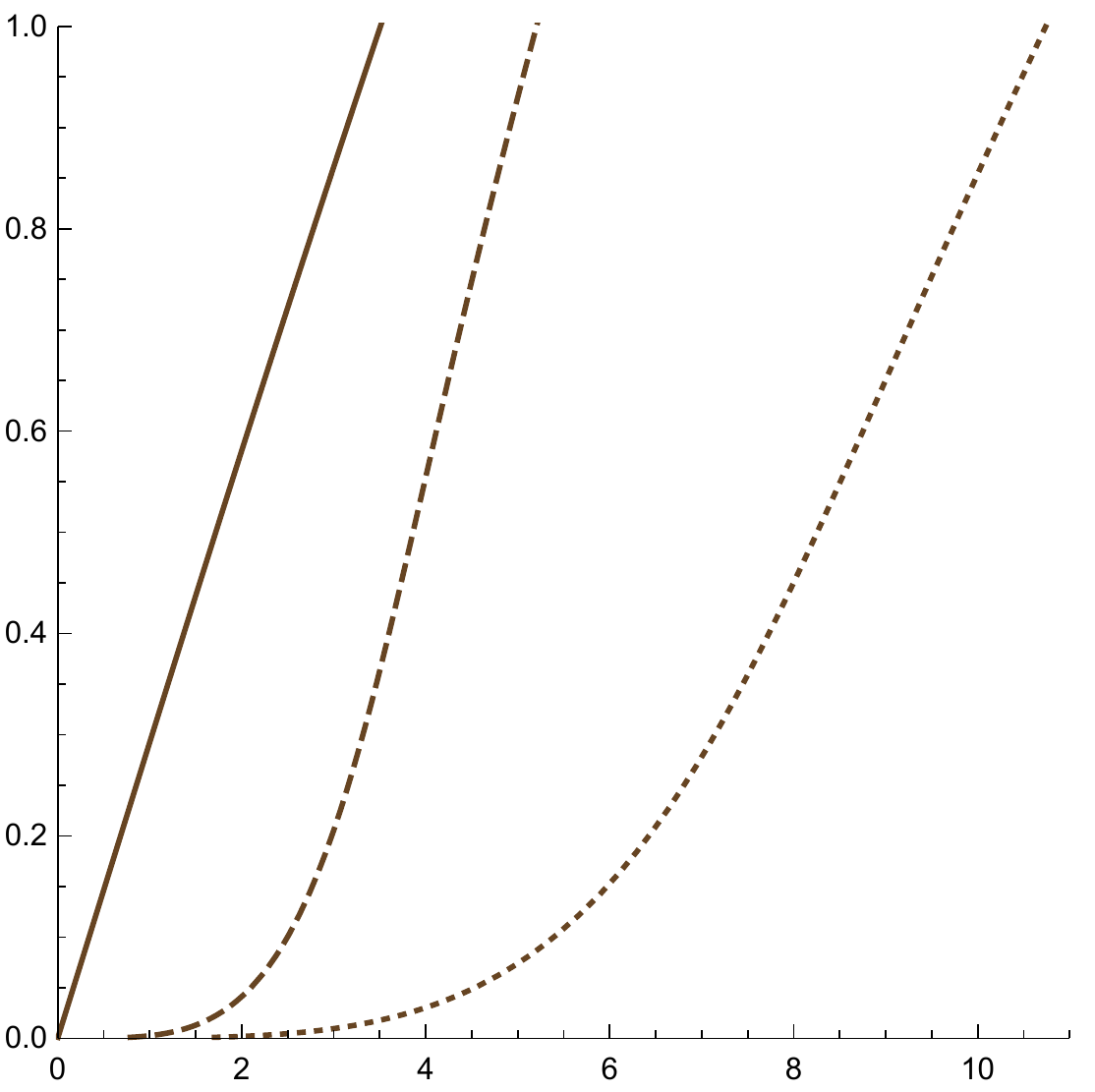}\,\,\,\,\,\,\,\,\,\,\,\,\,\,\,\,\,\,\,\,\,\,(b)}
 \put(110,150){$t_{therm}$}
 \put(255,25){$l_{y_{1}}$}
 \end{picture}
  \caption{(a)The thermalization time as a function of $\ell$ for the Wilson loop in the $y_1y_{2}$-plane, different curves correspond to different values of $\nu=1,2,3,4$ (from left to right). (b) The thermalization time as a function of $\ell$ for $\nu=4$, for Wilson loops with short extents in the $x$- and $y$- directions lying in the $xy_1$-plane and for the Wilson loop in the $y_1y_2$-plane (from left to right).}
        \label{fig:thermc3}
  \end{figure*}

In this section, we compare the thermalization time for spatial Wilson loops with different orientations and its dependence on the value of the dynamical exponent $\nu$.  
To simplify these estimations we  consider  the thin shell limit. 
We are interested in the value of the boundary time $t_{therm}$ when the string profile is totally covered by the thin shell, i.e.:

\be
t_{therm}(\ell)=\int_0^{z_*(\ell)}\frac{dz}{f(z)},
\ee
where $\ell$ is the length between the string endpoints on the boundary given by \eqref{7.3da},\eqref{7.6} or \eqref{7.7im} for different orientations.
In Fig.\ref{fig:thermc1c2} we plot the dependence on $\ell$ of  the thermalization time  for two configurations in the $xy_{1}$-plane. In Fig.\ref{fig:thermc3}(a) the behavior of the thermalization time as a function of $\ell$ for the configuration in the transverse $y_1y_{2}$-plane is presented.  One can see that the thermalization time decreases with increasing $\nu$ for all cases plotted in  Fig.~\ref{fig:thermc1c2} and  Fig.~\ref{fig:thermc3}. 
The dependence on the length $\ell$ for  the loop in the $xy_{1}$-plane with the short extent in the $x$-direction
is linear. At the same time, the dependence for the loop in the same plane, but with the short extent in the $y_1$-direction,
as well as for the loop  in $y_1y_{2}$-plane, is not linear for small $\ell$ asymptoting to the linear dependence only for large  $\ell$.  It should be noted that the deviation from linearity strengthens with increasing $\nu$.  We also see that for the configuration in the transverse plane,  the deviation of the thermalization time for the anisotropic cases from the isotropic one the thermalization time  is much stronger than for the other orientations. In Fig.\ref{fig:thermc3}(b) the comparison of thermalization times for  different orientations in the case $\nu=4$ is plotted.  This plot shows that the dependence on the orientation is crucial, varying the orientation we change the order of thermalization time. This means that characteristic scale depends on the orientation. The similar behavior of the thermalization time on $\ell$ was observed for the thermalization time of two-point correlators in \cite{AGG}.

\subsection{Thermalization times of different observables}\label{Sect:6b}

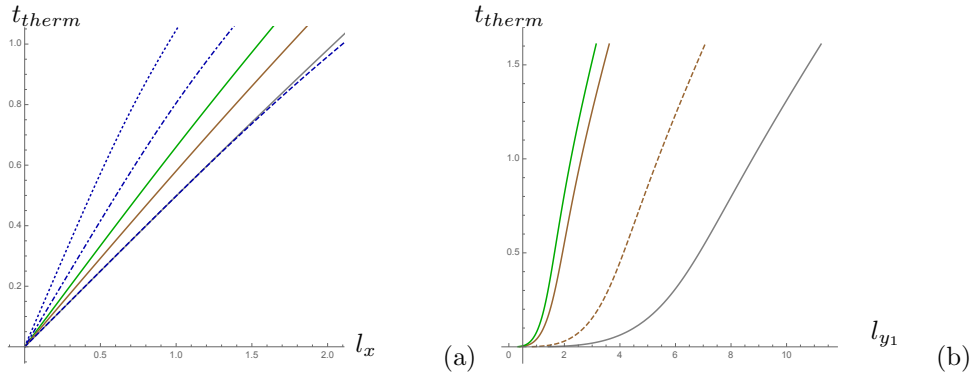
\begin{figure*}[h!]
\centering \begin{picture}(185,160)
\put(-80,20){ \includegraphics[width=4.5cm]{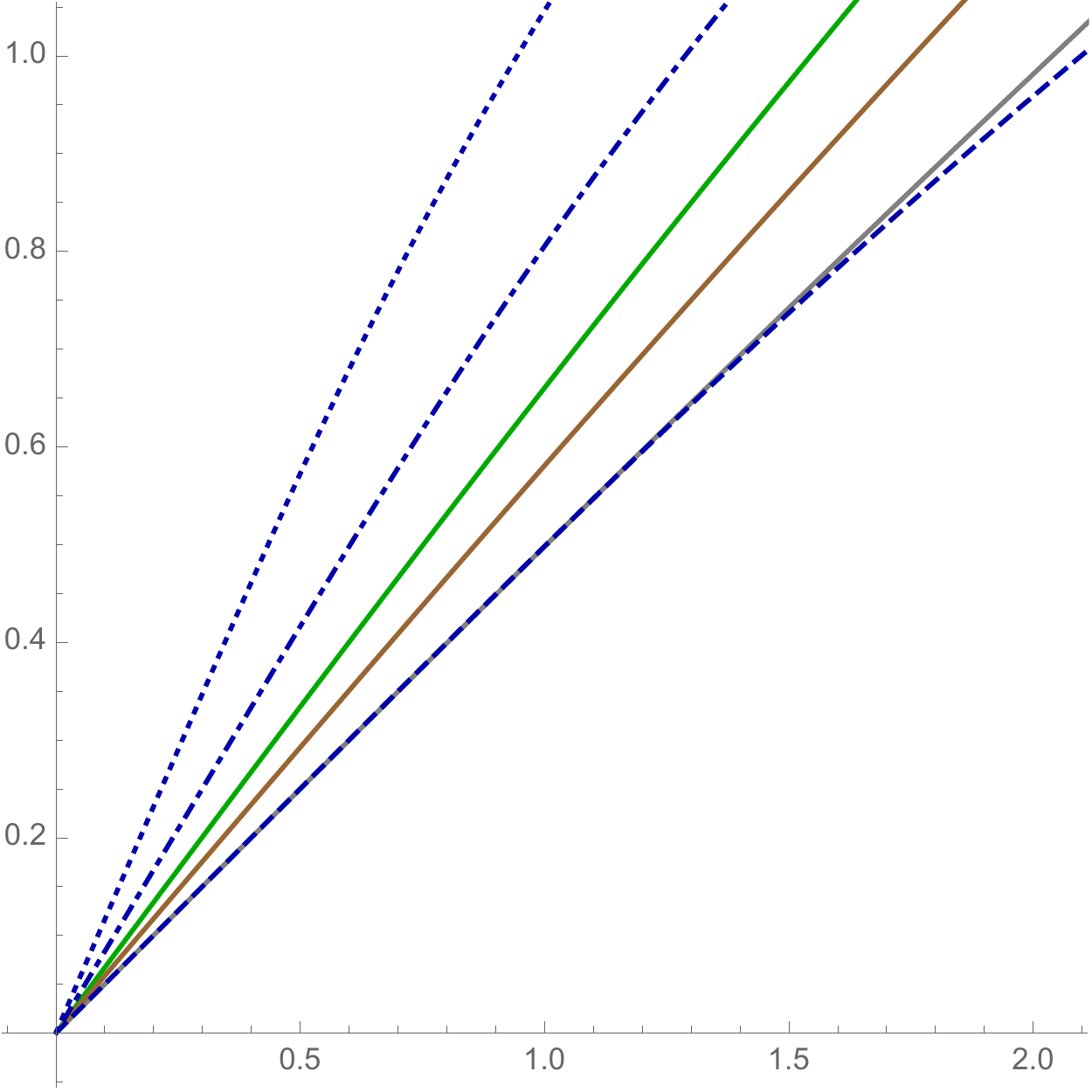}\,\,\,\,\,\,\,\,\,\,\,\,\,\,\,\,\,\,\,\,\,\,(a)}
 \put(-75,150){$t_{therm}$}
 \put(55,25){$l_{x}$}
\put(110,20){\includegraphics[width=4.5cm]{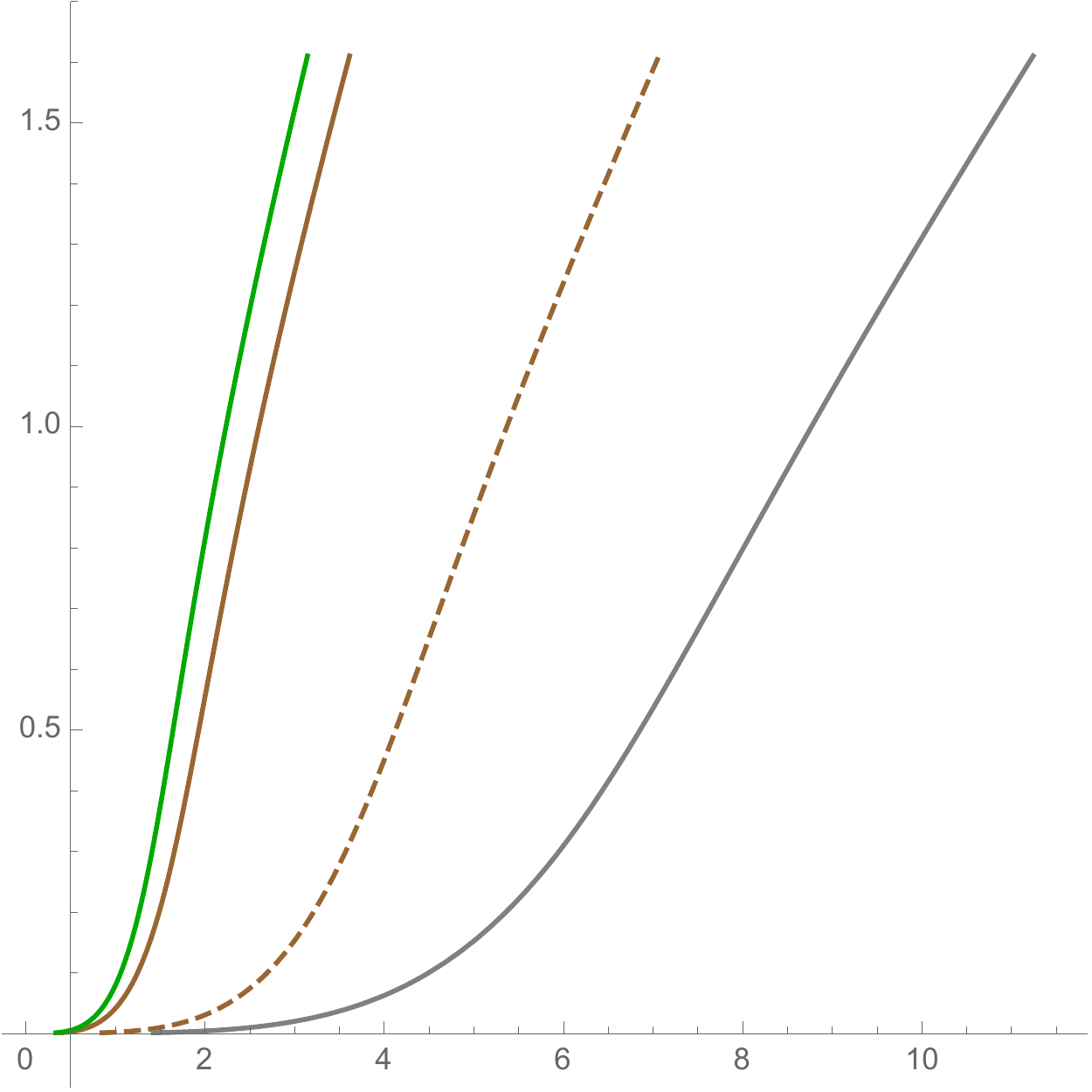}\,\,\,\,\,\,\,\,\,\,\,\,\,\,\,\,\,\,\,\,\,\,(b)}
 \put(100,150){$t_{therm}$}
 \put(250,30){$l_{y_{1}}$}
 \end{picture}
  \caption{The thermalization times of the two-point correlators, holographic entanglement entropy and Wilson loops for different configurations. (a)The solid lines (from left to right) correspond to the entropy(green), the Wilson loop (brown) and the two-point  correlator (gray) with the dependences on the longitudinal direction $x$ in the anisotropic background with $\nu=4$, while the dashed, dash-dotted and dotted lines represent the behaviour of the two-point correlator, Wilson loop and  entropy in the isotropic spacetime, respectively.
(b)The solid curves correspond to the entropy(green), the Wilson loop (brown) on the $xy_{1}$-plane and the two-point correlator (gray) with the dependences on the trasversal direction $y_{1}$. The dashed curve corresponds to the Wilson loop on the $y_{1}y_{2}$-plane. All plots in (b) are for the anisotropic background with $\nu=4$.}
        \label{fig:ttimes}
  \end{figure*}

It is interesting to compare the thermalization times of different observables. In our work \cite{AGG} we 
have studied two-point correlation functions and the holographic entanglement entropy in the Lifshitz-like backgrounds. By virtue of the spatial anisotropy of the metric we had two different configurations of the correlators and entropy with respect to the longitudinal and transversal directions. We have observed that the entanglement entropy for a subsystem
delineated in the transversal direction thermalizes faster then the two-point correlator and Wilson
loop in the longitudinal one. In \cite{AGG} we
also have calculated  thermalization times for two-point correlators.   In Appendix~\ref{App:C}  the additional computations for the thermalization time of the holographic entanglement entropy are given.
In Figure \ref{fig:ttimes} we show the comparison of the thermalization times for two point correlation functions, holographic entanglement entropy and Wilson loops for different configurations. 

We see that the order of the thermalization process  in a certain direction is similar to the case in the isotropic and ordinary Lifshitz backgrounds \cite{1110.5035}. 
The two-point correlator  is  the observable that thermalizes first, then we  observe the thermalization of the Wilson loops, and the entanglement entropy is the observable that thermalizes last. One should be noted that the thermalization process of the Wilson loop and the entanglement entropy in the anisotropic background is faster even in the longitudinal direction than thermalization of the same observables in the isotropic case. From Figure \ref{fig:ttimes} (a) it is also interesting to see that the curves for the two-point correlator in the $x$-direction in the anisotropic background and the correlator in the isotropic spacetime match.

\section{Conclusions}

In this paper, we have explored the  holographic scenario of the formation of the quark-gluon plasma using the  bulk backgrounds
\eqref{Vaidya-LL}, which possess spatially anisotropy.
To probe the formation of the quark-gluon plasma  we have used the rectangular spatial Wilson loops located on the boundary of our background
\eqref{Vaidya-LL}. We have considered three possible configurations of Wilson loops on the boundary:  the infinite rectangular strip  located on the plane including one longitudinal and one transverse directions, the $xy_{1}$-plane,  with a short extent in the $x$- or $y_1$-direction, and the infinite rectangular strip  located on the  transverse $y_1y_2$-plane.

We have analyzed Wilson loops both for static and time-dependent cases  using the static 
  black brane \eqref{Ll-bh}, and Vaidya solutions \eqref{Vaidya-LL}, respectively.
The results obtained in this paper show how the expectations of Wilson loops  are modified in the presence of anisotropy in the strong coupling limit.

We have found, that at small distances, the pseudopotential derived from the Wilson loop located in the $xy_1$-plane has a nontrivial dependence on the parameter $\nu$. Namely, for $\nu>1$ a breaking of the Coulomb phase has been observed. For Wilson loops lying in the transverse plane the Coulomb phase is unbroken, and all dependence for small $\ell$ is encoded in the $\nu$ dependent constant. At large $\ell$ all pseudopotentials are linear growing functions.
Also we have found that the magnetic string tension is also affected by the anisotropic parameter $\nu$. 
For the contour located on the transverse directions, the dependence of the string tension on the temperature is suppressed by anisotropy, 
so the magnetic string tension becomes close to a constant value.
This effect is clearly seen for large $\nu$.
We  have also observed interesting results for Wilson loops in the Vaidya backgrounds \eqref{Vaidya-LL}.
 The effect of anisotropy parameter on the thermalization time for different orientations also has been investigated. In the ${x,y_{1(\infty)}}$-case of orientation the dependence of thermalization time on scale $\ell$ is linearly growing function for all $\nu$. For other orientation the dependence is not linear. In the transverse orientation this can be seen very clear. Until some critical value the dependence  is slowly growing function, and after this critical value it shows the linear growth.

A common feature in the behavior of the pseudopotential  is the tendency of achievement of the saturation for large values of the boundary time. 
The dynamical exponent also influences to the thermalization of Wilson loops, so that the value of $\nu$ increases this behavior strengthens. 
We have seen that the thermalization is much faster than in the AdS case. It is worth noticing that the approach to the saturation also depends on the orientation of the Wilson loop. The configuration on the transversal directions the system saturates quicker than for the contours on the $xy_{1}$-plane. 

Comparing to results for  the evolution of the holographic entanglement entropy in the  backgrounds \eqref{Vaidya-LL} in \cite{AGG} we have found that the thermalization process both of the Wilson loops and the entanglement entropy is faster in the transverse direction. We have seen 
the similar behavior of the thermalization time as a function of $\ell$ for the two-point correlators in \cite{AGG}.

We also calculated the dependence of the jet quenching parameter on the orientation and on the anisotropic parameter $\nu$. We also noticed that for special orientations the string action defined the  jet quenching parameter becomes complex that leads to  a suppression of the jet quenching parameter. This phenomena requires more elaborations as well as the study of the time evolution of the  jet quenching parameter modeling by the 
light-like Wilson loop in the time-depending background \eqref{Vaidya-LL}.

It would be interesting to generalize our results to the case of the modified  backgrounds that  describe confinement/deconfinement phase transition \cite{Arefeva:2016rob}.

\section*{Acknowledgments}
We are grateful to Oleg Andreev and Oleg Teryaev for useful discussions. 

$$\,$$
\appendix

\section{Asymptotics for static pseudopotentials}\label{App:B}
Here we collect the expressions for asymptotics of static pseudopotentials.
\subsection{Rectangular strip in $xy_1$-plane infinite along the $y_1$-direction}

\begin{figure*}[tbp]
\centering \begin{picture}(185,180)
\put(-80,0){\includegraphics[width=3.5cm]{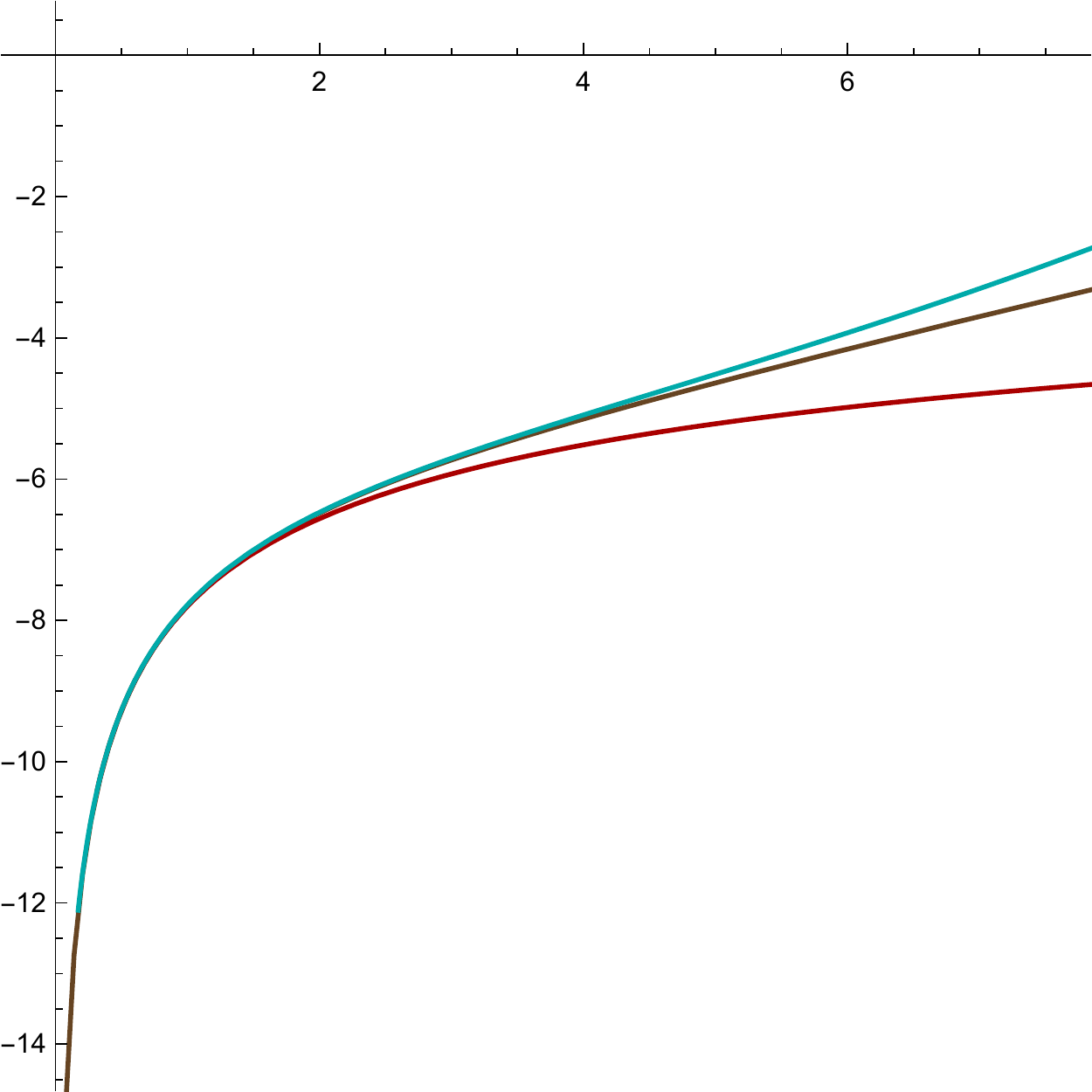}(a)}
 \put(-80,110){$\mathcal{V}_{x,y_{1(\infty)}}$}
 \put(20,100){$\ell_x$}
\put(45,0){\includegraphics[width=3.5cm]{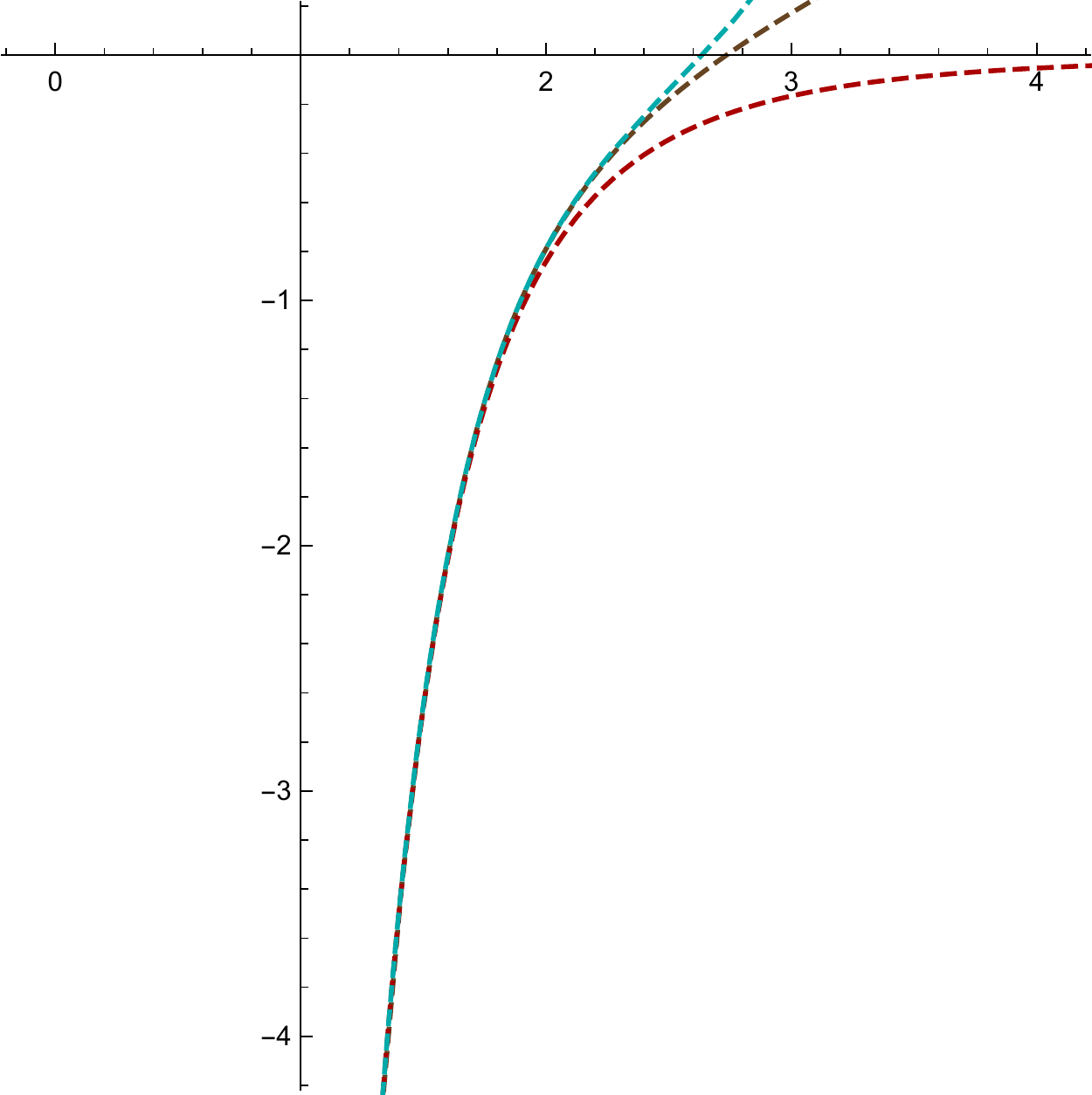}(b)}
 \put(55,110){$\mathcal{V}_{x_{(\infty)},y_{1}} $}
 \put(140,100){$\ell_{y_1}$}
\put(160,10) {\includegraphics[width=3.5cm]{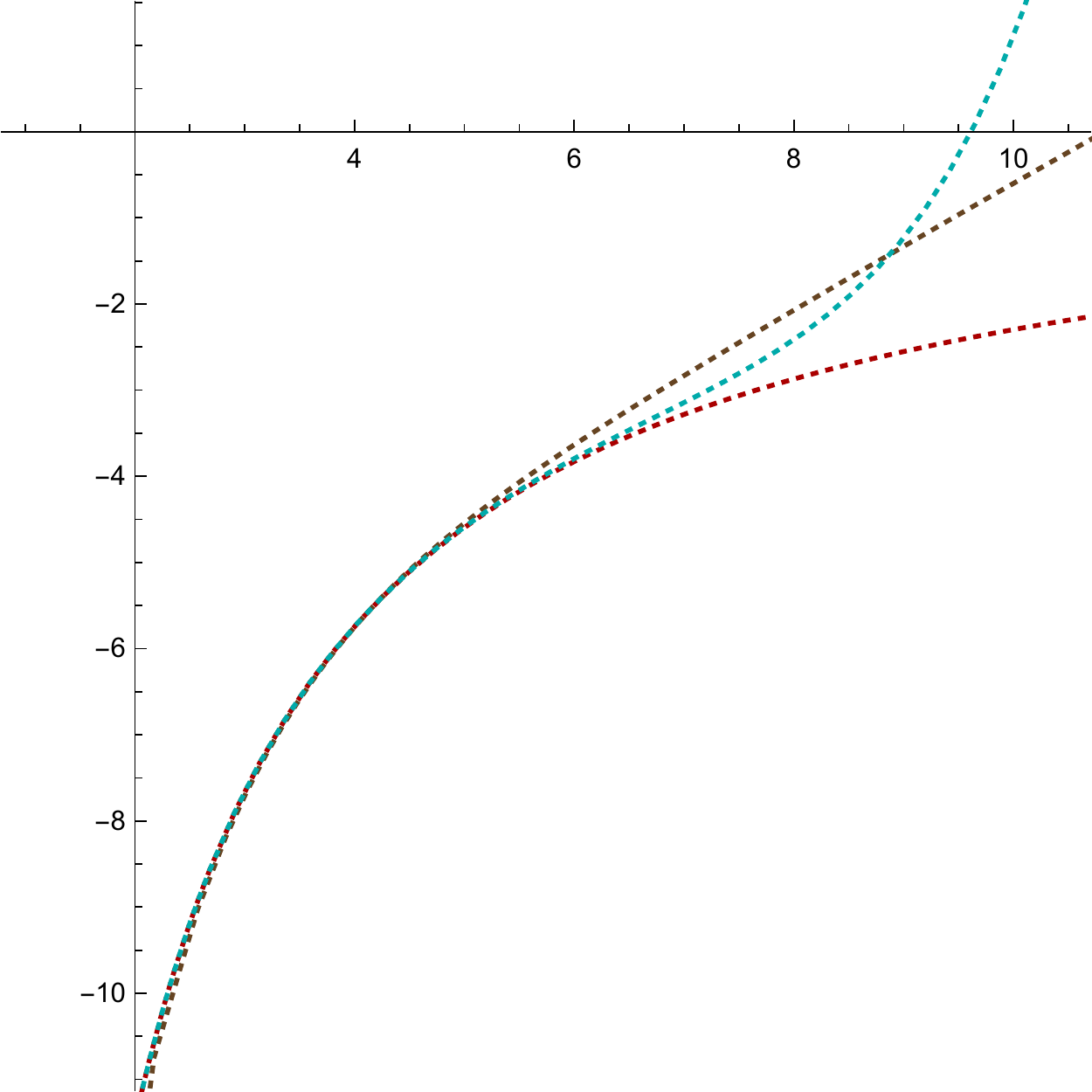}(c)}
 \put(170,120){$\mathcal{V}_{y_1,y_{2(\infty)}} $}
 \put(270,100){$\ell_{y_1}$}
 \end{picture}
\caption{a) The comparison of the pseudopotentials given by exact formula \eqref{7.4a} (the brown line), massless approximation (the red bottom line)  and the first massive correction  (the cyan top  line) \eqref{B6}. 
 b)  The comparison of the pseudopotentials given by exact formula \eqref{7.7a} (the brown dashed line), massless approximation (the red bottom dashed line)  and the first massive correction  (the cyan top dashed line)  \eqref{Vy1x}.$\,\,\,$ c)  The comparison of the pseudopotentials given by exact formula \eqref{7.7a-2} (the brown dotted line), massless approximation (the red bottom dotted line)  and the first massive correction   (the cyan top dotted line) \eqref{B25}. For all plots $m=0.2$. }
 \label{fig:approx}
\end{figure*}
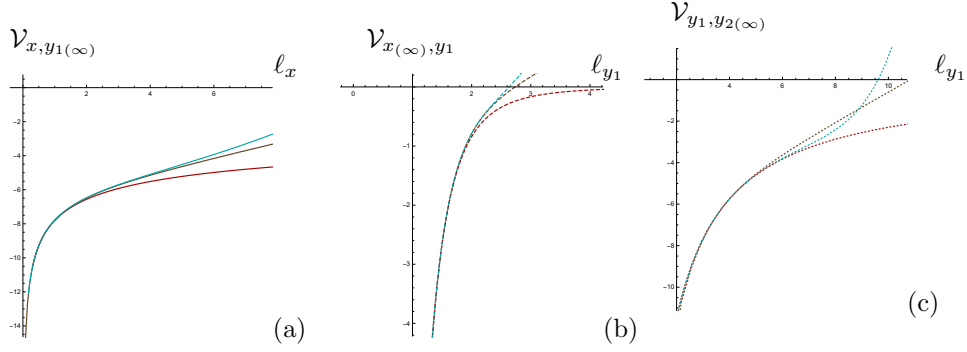

The integral in \eqref{7.3d} can be evaluated approximately

\begin{eqnarray}\label{7.3c-app}
\ell_{x}&\approx&-\frac{\sqrt{\pi } z_* \Gamma \left(-\frac{1}{2 \nu +2}\right) (6 \nu +4+ (2 \nu +1)m z_*^{\frac{2}{\nu }+2})}{2 (\nu +1) (3 \nu +2)
   \Gamma \left(\frac{\nu }{2 \nu +2}\right)}.\eea
    For $\nu=4$ we have
    \bea\nn
   84\ell_x&\approx&5.12 z_* \left(28+9 m
   z_* ^{5/2}\right)\eea
   and this gives   \bea\label{IA}
   z_*&=&\ell_x\left(0.586  - 0.0495 \ell_x^{5/2} m+{\cal O}( \ell_x^{5} m^2)\right).
 \eea
Evaluating \eqref{7.4a} for $\nu=4$ we get approximately
\bea\label{BVxy1inf}
{\cal V}_{x,y_{1(\infty)}} 
 &=& - \frac{\frac{8
   \sqrt{\pi } \Gamma \left(\frac{9}{10}\right)}{\Gamma
   \left(\frac{2}{5}\right) }+ \frac{\sqrt{\pi }  \Gamma \left(\frac{9}{10}\right)
   }{\Gamma \left(\frac{2}{5}\right)}mz_*^{5/2}+{\cal O}\left(m^2z_*^5\right)}{z_*^{1/4}} .
 \eea 
 Substituting \eqref{IA}  in (\ref{BVxy1inf}) we get   
  \bea
  \label{B6}
 {\cal V}_{x,y_{1(\infty)}} &=& -\frac{7.80}{l_x^{1/4}}\left( 1-0.012 \,m\,l_x^{5/2}
   +{\cal O}\left(m^2\,l_x^{5}\right) \right)
   \eea

 In Figure \ref{fig:approx} we present the comparison of the pseudopotentials given by exact formula \eqref{7.4a}, \eqref{7.3d} and the approximated formula \eqref{B6}.
   
   Inverting  \eqref{7.3c-app} for arbitrary $\nu$ we obtain
   \bea\nn
   \label{4xzs}
z_*&=&l_x\left(\frac{ \Gamma \left(\frac{\nu }{2 \nu +2}\right)}{2
   \sqrt{\pi } \Gamma \left(1-\frac{1}{2 (\nu
   +1)}\right)} - B_{l_x} ml_x^{\frac{2}{\nu }+2}+{\cal O}( \ell_x^{\frac{4}{\nu }+4} m^2)\right),\eea
    where \bea\nn
 B_{l_x}&=&\frac{ \nu   \left(\frac{\Gamma
   \left(\frac{\nu }{2 \nu +2}\right)}{\Gamma
   \left(1-\frac{1}{2 (\nu +1)}\right)}\right)^{\frac{2}{\nu
   }+4} \Gamma \left(2-\frac{1}{2 (\nu +1)}\right)
   }{(\nu +1)2^{\frac{2}{\nu }+5}\pi
   ^{\frac{1}{\nu }+\frac{3}{2}} \Gamma \left(\frac{5 \nu
   +4}{2 \nu +2}\right)}.
\end{eqnarray}

The pseudopotential  ${\cal V}_{x,y_{1(\infty)}}$ can be evaluated approximately 
\bea\nn
{\cal V}_{x,y_{1(\infty)}} &=&l_x^{-\frac{1}{\nu}}\left(A_{l_x}+F_{l_x} \,m\, l_x^{\frac{2}{\nu}+2}
+{\cal O}(m^2\, l_x^{\frac{4}{\nu}+4})\right),\eea
where
$$A_{l_x}=\frac{
   \nu ^2 
   \Gamma \left(-\frac{1}{2 (\nu +1)}\right) \Gamma
   \left(1-\frac{1}{2 (\nu +1)}\right)^{\frac{1}{\nu }}
   \left( \Gamma \left(\frac{\nu }{2 \nu
   +2}\right)\right){}^{-1/\nu }}{2^{2-\frac{1}{\nu }}\pi ^{\frac{\nu +1}{2 \nu }-\frac{1}{\nu }-1} (\nu +1)^2 \Gamma
   \left(\frac{3 \nu +2}{2 \nu +2}\right)},$$
   $$ F_{l_x}=\frac{
   \csc \left(\frac{\pi }{2 \nu +2}\right) \Gamma
   \left(\frac{-1}{2 (\nu +1)}\right) \Gamma
   \left(\frac{2 (\nu +1)-1}{2 (\nu +1)}\right)^{-\frac{1+3\nu}{\nu }}
   \Gamma \left(\frac{\nu }{2 \nu
   +2}\right)^{\frac{2\nu+1}{\nu }}}{2^{\frac{1}{\nu }+3} \pi ^{\frac{\nu +1}{2 \nu }-1}(\nu +2) \Gamma
   \left(\frac{1}{2 \nu +2}\right) \Gamma \left(-\frac{\nu
   +2}{2 \nu +2}\right)}$$
$$+ \frac{ \Gamma \left(\frac{-1}{2 (\nu
   +1)}\right) \Gamma \left(\frac{3 \nu +2}{2 \nu +2}\right)
    \Gamma \left(\frac{4(\nu +1)-1}{2 (\nu +1)}\right) 
   \Gamma \left(\frac{\nu }{2 \nu
   +2}\right)^{\frac{1}{\nu }+2}}{(\nu +2)2^{\frac{1}{\nu }+3} \pi
   ^{\frac{\nu +1}{2 \nu }} \Gamma
   \left(-\frac{\nu +2}{2 \nu +2}\right) \Gamma \left(\frac{5
   \nu +4}{2 \nu +2}\right)\Gamma \left(\frac{2 (\nu +1)-1}{2 (\nu +1)}\right)^{\frac{1+3\nu}{\nu
   }}}.$$
For large $\nu$  one can expand the coefficients and obtain 
\bea\nn
\frac{\ell_x^{1/\nu}}{2\nu\left(1 -
 \frac{\pi ^2}{24 \nu^2 }\right)}
{\cal V}_{x,y_{1(\infty)}} &=& - 1-+\frac{1+3\nu}{3\left(24 \nu^2 - \pi ^2\right)} \,m \,\ell_x^{\frac{2}{\nu}+2}+{\cal O}(m^2\, \ell_x^{\frac{4}{\nu}+4}).
\eea

\subsection{Rectangular strip in the $xy_1$-plane infinite along the $x$-direction}

For $\nu=4$ the integral \eqref{7.6} can be evaluate approximately \bea\nn
\frac{\ell_{y_1}}{z_*^{1/4}}&=&\frac{22\cdot 2^{4/5} \pi  \Gamma
   \left(\frac{6}{5}\right) }{5 \Gamma
   \left(\frac{1}{10}\right) \Gamma
   \left(\frac{21}{10}\right)}+\frac{6\cdot 2^{4/5} \pi   \Gamma \left(\frac{6}{5}\right)
  }{5 \Gamma \left(\frac{1}{10}\right) \Gamma
   \left(\frac{21}{10}\right)} z_*^{5/2}+{\cal O}\left(m^2z_*^{5}\right)\eea
   and inverting we get
   \bea
   \label{zsl2}
z_*&=& \ell_{y_1}^4 \left[0.0412-0.0000155
   m\,\ell_{y_1}^{10}
   +{\cal O}\left(m^2l_2{}^{20}\right) \right].
\end{eqnarray}

The pseudopotential for $\nu=4$ reads
\bea\nn
{\cal V}_{y_{1},\,x_{(\infty)}}(z_*)&=&\frac{-2\frac{\sqrt{\pi } \Gamma
   \left(\frac{3}{5}\right)}{\Gamma \left(\frac{1}{10}\right)
   }+4\frac{ \sqrt{\pi }  \Gamma \left(\frac{3}{5}\right)
   }{\Gamma
   \left(\frac{1}{10}\right)}mz_*^{5/2}+{\cal O}(m^2z_*^{5})}{z_*}.\\\label{BVy1xinf}\eea
Plugging  \eqref{zsl2} in (\ref{BVy1xinf}) we get
\bea\nn
\label{Vy1x}
{\cal V}_{y_{1},\,x_{(\infty)}}(\ell_{y_1})&=&-\frac{13.5}{\ell_{y_1}^4}\left(1-0.00031m \ell_{y_1}^{10}+{\cal{O}}(m^2\ell_{y_1}^{20} )\right).
\eea

Performing the same for arbitrary $\nu$ we have 
\begin{eqnarray}\nn
\frac{z_*}{l_{y}^{\nu } }&=&   \left(\frac{2\ \pi ^{1 /2}\Gamma
   \left(\frac{1}{2 \nu +2}\right)}{\nu  \Gamma
   \left(\frac{\nu +2}{2 \nu +2}\right)}\right)^{\nu }-
\frac{
    \nu  (\nu +2) 
    \left(\frac{\Gamma \left(\frac{1}{2 \nu
   +2}\right)}{\nu  \Gamma \left(\frac{\nu +2}{2 \nu
   +2}\right)}\right)^{3 \nu +2}}{8^{\nu+ 1}\pi ^{\frac{3 \nu }{2}+1}(2 \nu +3)}m\ell_{y_1}^{2 \nu +2}+O(m^2 \ell_{y_1}^{4\nu+4})
\end{eqnarray}
and
\begin{eqnarray}\label{sec-gennu}
{\cal V}_{y_{1},\,x_{(\infty)}}(\ell_{y_1})&=&\frac{Q_{l_y}+ C_{l_y}m\ell_{y_1}^{2+2\nu}+{\cal{O}}(m^2\ell_{y_1}^{4+4\nu})}{\ell_{y_1}^{\nu}},
\end{eqnarray}
where \bea\nn
Q_{l_y}&=&\frac{2^{\nu } \pi ^{\frac{\nu +1}{2}}
   \nu^{\nu +1} \Gamma
   \left(-\frac{\nu }{2 \nu +2}\right) \Gamma
   \left(\frac{\nu +2}{2 \nu +2}\right)^{\nu }}{(\nu
   +1)\Gamma \left(\frac{1}{2 \nu
   +2}\right)^{\nu +1}},\\\nn
C_{l_y}&=&-\frac{  \Gamma
   \left(\frac{1}{2 \nu +2}\right)^{\nu +1} \Gamma
   \left(-\frac{\nu }{2 \nu +2}\right) \Gamma
   \left(\frac{\nu +2}{2 \nu +2}\right)^{-\nu -2}}{2^{\nu +3} \nu ^{\nu } \pi ^{\frac{\nu
   }{2}+\frac{1}{2}}(2
   \nu +3)}.\eea

\,\,\,\,\,\,\,\,\,\,\,\,\,For large $\nu$ we have the expansion for $\eqref{sec-gennu}$
\bea\nn
\,\,\,\,\,\,\,\,\,\,\,\,\,\,\,&\,&\ell_{y_1}^{\nu}{\cal V}_{y_{1},\,x_{(\infty)}}(\ell_{y_1})=-2\pi^{1/2+\nu}+ \frac{ (2 \nu
   -1) e^{\frac{24 \nu -\pi ^2+12}{24 \nu }}}{4 \pi ^{\nu +1}\nu }m \ell_{y_1}^{2+2\nu}+{\cal{O}}(m^2\ell_{y_1}^{4+4\nu}).
\eea

\subsection{Rectangular strip in $y_1y_2$-plane infinite along the $y_2$-direction}

For
$\nu=4$ the integral \eqref{7.7i} can be evaluated approximately

\bea\nn
\frac{\ell_{y_1}}{z_*^{1/4}}&=&  \frac{2
   \sqrt{\pi } \Gamma \left(\frac{3}{4}\right)
   }{\Gamma \left(\frac{5}{4}\right)}+\frac{\sqrt{\pi } \Gamma \left(\frac{13}{4}\right)
   }{\Gamma \left(\frac{15}{4}\right)}mz_*^{5/2} +{\cal O}(m^2 z_*^{10})  \eea
   and this gives   \bea
   \label{zsl3}
z_*&=&\ell_{y_1}^4\left(0.00190 -2.52\,\,10^{-10} m\ell_{y_1}^{10} +{\cal O}\left(m^2\ell_{y_1}^{20}\right) \right).
   \end{eqnarray}

The pseudopotential for $\nu=4$
\bea\label{BVy1y2inf}
{\cal V}_{y_{1},\,y_{2(\infty)}}(z_*)&=&\frac{\frac{2\sqrt{\pi } \Gamma
   \left(-\frac{1}{4}\right)}{\Gamma \left(\frac{1}{4}\right)
   }+\frac{\sqrt{\pi }  \Gamma \left(\frac{9}{4}\right)
   mz_*^{5/2}}{ \Gamma
   \left(\frac{11}{4}\right)}+{\cal O}(m^2z_*^5)}{z_*^{1/4}}.\nn\\
   \eea

Substituting \eqref{zsl3} in (\ref{BVy1y2inf}) we get
\bea\label{B25}\nn
{\cal V}_{y_{1},\,y_{2(\infty)}}&=&  -\frac{23.0\left(1 - 0.741\cdot 10^{-9}m\,\ell_{y_1}^{10}+{\cal O}\left(m^2\ell_{y_1}^{20}\right)\right)}{\ell_{y_1}}.
\end{eqnarray}
For an arbitrary value of $\nu$  from \eqref{7.7i} we have
\bea
\label{zsnul3}
\,\,\,\,\,\,\,\,\,\,\,\,\,\,\,z_*=  \ell_{y_1}^{\nu }\left[C_1+C_2 m \ell_{y_1}^{2 \nu +2}+{\cal O}(m^2 \ell_{y_1}^{4+4\nu} )\right],\eea
where
\bea\nn
C_1&=&2^{-\nu } \nu ^{-\nu } \pi ^{-\nu /2}
   \left(\frac{\Gamma
   \left(\frac{1}{4}\right)}{\Gamma
   \left(\frac{3}{4}\right)}\right)^{\nu }\\\nn
C_2&=&-\frac{2^{-3 \nu -5} \nu ^{-3 \nu -1} \pi ^{-\frac{3
   \nu }{2}-1} \Gamma \left(\frac{1}{4}\right)^{3
   \nu +3}  \Gamma \left(\frac{\nu
   }{2}+\frac{5}{4}\right)}{\Gamma \left(\frac{\nu
   }{2}+\frac{7}{4}\right)\Gamma \left(\frac{3}{4}\right)^{3 \nu
   +3}}\nn
\eea
The pseudopotential for general $\nu$ can expressed as
\bea\nn
\frac{4{\cal V}_{y_{1},\,y_{2(\infty)}}(z_*)}{\sqrt{\pi } \nu  z_*^{-1/\nu }
}&=&\frac{2 \Gamma \left(-\frac{1}{4}\right)}{\Gamma \left(\frac{1}{4}\right)}+\frac{m \Gamma \left(\frac{\nu
   }{2}+\frac{1}{4}\right) z_*^{\frac{2}{\nu
   }+2}}{\Gamma \left(\frac{1}{4} (2 \nu+3)\right)}+{\cal O}(m^2z_*^{4+4/\nu}).\label{BVy1y2inf2}
   \eea
   Plugging \eqref{zsnul3}  in (\ref{BVy1y2inf2}) we get the following formula for the pseudopotential for large $\nu$
\bea
&\,&\ell_{y_1}{\cal V}_{y_{1},\,y_{2(\infty)}}(\ell_{y_1})=-\frac{4 \pi  \nu ^2
   \Gamma \left(\frac{3}{4}\right)^2}{ \Gamma
   \left(\frac{1}{4}\right)^2}+\frac{ \Gamma \left(\frac{1}{4}\right)^{2 \nu +1} \Gamma
   \left(\frac{3}{4}\right)^{-2 \nu -1}}{2^{2 \nu +\frac{5}{2}} \nu ^{2 \nu +\frac{3}{2}} \pi
   ^{\nu } } m\ell_{y_1}^{2 \nu +2}+O(m^2 \ell_{y_1}^{4+4\nu}).\nn
   \eea

\section{Thermalization times of holographic two-point correlators and entanglement entropy} \label{App:C}
\subsection{Thermalization time of two-point correlators}
Under the holographic approach one can find the thermalization time $t_{therm}$ of the two-point correlator at the scale $\ell$ using the Vaidya background.
For this,  one should consider a geodesic of a bulk particle with equal time endpoints located  at the distance $\ell$ and find the time when the geodesic covered by the shell. In the Vaidya-Lifshitz background (\ref{Vaidya-LL})-(\ref{Vaidya-f}) we should study the thermalization in both longitudinal and transversal directions.

For the thermalization in the longitudinal direction we have the following relation for the length 
\begin{eqnarray}\nn
\ell_{x} = 2z_{*}\int^{1}_{0}\frac{wdw}{\sqrt{f(z_{*}w)(1- w^{2})}},
\end{eqnarray}
where $w = z/z_{*}$ and the turning point is assumed to lie above the horizon, i.e. $z_{h} > z_{*}$.\\

The distance in the transversal direction is given by
\begin{eqnarray}\nn
\ell_{y_{1}} = 2 z^{1/\nu}_{*}\int^{1}_{0}\frac{w^{-1+2/\nu}dw}{\sqrt{f(wz_{*})(1 - w^{2/\nu})}}.
\end{eqnarray}

The thermalization time of the two-point correlator in both directions is defined by
\begin{eqnarray}\label{tthermtp}\nn
t_{therm} = z_{*}\int^{1}_{0}\frac{dw}{f(z_{*}w)}.
\end{eqnarray}
\subsection{Thermalization time of entanglement entropy}

To study the thermalization of the entanglement entropy we should also consider configurations in the longitudinal and transversal directions.\\

The longitudinal length scale is given by
\begin{eqnarray}
\ell_{x}   = 2 \int^{1}_{0} z_{*}w^{1+2/\nu} \frac{dw}{\sqrt{f(z_{*}w)(1 - w^{2(1+2/\nu)})}}.
\end{eqnarray}
The length for a subsystem delineated in the transversal direction is 
\begin{eqnarray}\nn
\ell_{y_{1}} = 2z^{1/\nu}_{*}\int^{1}_{0}\frac{w^{3/\nu}dw}{\sqrt{f(w,z_{*})(1 - w^{2(1+2/\nu)})}}. 
\end{eqnarray}
Here we are also interested in the value of the boundary  time when the surface is covered by the shell, i.e.
 the thermalization time of the entanglement entropy has the same expression as for the two-point correlator (\ref{tthermtp}).

\end{document}